\documentclass[12pt]{article}
\usepackage{amsmath,setspace}
\usepackage{graphicx}
\usepackage{enumerate}
\usepackage{natbib}
\usepackage{url} 

\usepackage{comment}

\usepackage{hyperref} 
\usepackage{cleveref} 

\hypersetup{colorlinks = true, 
	    linkcolor = blue, 
	    urlcolor = blue,
        citecolor = blue} 

\usepackage{amsmath,amssymb,amsfonts,amsthm,amscd}
\usepackage{booktabs}
\newtheorem{thm}{Theorem}
\newtheorem{prop}{Proposition}
\newtheorem{property}{Property}
\newtheorem{lemma}{Lemma}

\newtheorem{remark}{Remark}

\newtheorem{example}{Example}

\usepackage{algorithm}
\usepackage{algorithmic}
\renewcommand{\algorithmicrequire}{ \textbf{Input:}} 
\renewcommand{\algorithmicensure}{ \textbf{Output:}} 

\newcommand{\EE}{\mathbb E}
\newcommand{\PP}{\mathbb P}
\newcommand{\II}{\mathbb I}

\usepackage{helvet}
\usepackage{bm}
\usepackage{color}

\usepackage{multirow}

\usepackage{graphicx} 
\usepackage{subfig}    
\usepackage{float} 

\newcommand{\blind}{1}

\begin{document}

\bibliographystyle{chicago}

\def\spacingset#1{\renewcommand{\baselinestretch}%
{#1}\small\normalsize} \spacingset{1}


\if1\blind
{
  \title{\bf False Discovery Rate Control For Structured Multiple Testing: Asymmetric Rules And Conformal $Q$-values} 
 \author{Zinan Zhao$^{1}$ \ and \ Wenguang Sun$^{2}$}
 \date{}
  \maketitle
} \fi

\if0\blind
{
  \bigskip
  \bigskip
  \bigskip
  \begin{center}
    {\LARGE\bf False Discovery Rate Control For Structured Multiple Testing: Asymmetric Rules and Conformal $Q$-values}
\end{center}
  \medskip
} \fi

\bigskip

\begin{abstract}

The effective utilization of structural information in data while ensuring statistical validity poses a significant challenge in false discovery rate (FDR) analyses. Conformal inference provides rigorous theory for grounding complex machine learning methods without relying on strong assumptions or highly idealized models. However, existing conformal methods have limitations in handling structured multiple testing, as their validity often requires the deployment of symmetric decision rules, which assume the exchangeability of data points and permutation-invariance of fitting algorithms. To overcome these limitations, we introduce the pseudo local index of significance (PLIS) procedure, which is capable of accommodating \emph{asymmetric rules} and requires only \emph{pairwise exchangeability} between the null conformity scores. We demonstrate that PLIS offers finite-sample guarantees in FDR control and the ability to assign higher weights to relevant data points. Numerical results confirm the effectiveness and robustness of PLIS and demonstrate improvements in power compared to existing model-free methods in various scenarios.
 
\end{abstract} 

\noindent%
\vfill

\footnotetext[1]{Center for Data Science and School of Mathematical Sciences, Zhejiang University.}  

\footnotetext[2]{Center for Data Science and School of Management, Zhejiang University.}

\spacingset{1.2} 
\section{Introduction}
\label{sec:intro}

In various applications, data is commonly acquired and organized as ordered sequences or lattices, revealing informative structural patterns that play a crucial role in analysis and interpretation. For example, spatio-temporal data in econometric analyses may display serial or spatial dependence structures, while in genome-wide association studies, single-nucleotide polymorphisms (SNPs) often cluster along biological pathways, indicating functional relationships between genes. To make reliable and meaningful inferences, it is essential to account for the underlying structural patterns in such data. False discovery rate (FDR) methods \citep{BH95} are a powerful tool for identifying sparse signals from massive and complex data. 
In FDR analyses, a significant challenge lies in integrating structural information to enhance statistical power and interpretability, while simultaneously ensuring the validity of FDR control. In this section, we discuss recent advancements in addressing this challenge, as well as the limitations of existing approaches, and present our contributions.

\subsection{Model-based and model-free FDR methods}
\label{subsec:proscons}

Structural knowledge, such as the clustering patterns and dependence, can be leveraged to enhance the efficiency of existing FDR methods, as demonstrated by the works of \citet{bheller07}, \citet{sc09}, \citet{fan12}, \citet{sc15}, \citet{perrot21hmm}, and \citet{re22graph}. However, the validity of existing ``model-based'' FDR methods, which involve constructing new test statistics based on estimated model parameters, typically depend on idealized assumptions, such as correct specification of the data generating models, homogeneity and stationarity of the underlying processes, consistent estimation of unknown parameters, and asymptotic normality of the test statistics. These assumptions may not be fulfilled and may be hard to check in practice. The violation of these assumptions can have serious consequences in FDR analysis, including decreased power and inflated error rate. It remains a significant challenge to develop powerful FDR methods for structured multiple testing that are assumption-lean and provably valid.  

The framework of conformal inference \citep{vovk05} provides finite-sample uncertainty guarantees on a flexible class of off-the-shelf machine learning algorithms, only under the assumption of exchangeability. The important connection between machine learning and FDR analysis, as highlighted by \citet{yang21bonus}, \citet{mary22semi} and \citet{bates21testing},  indicates that efficient conformity scores, and hence provably valid and powerful conformal $p$-values, can be constructed based on complex learning algorithms. This insightful perspective leads to ``model-free'' approaches to FDR control, a research direction with considerable promise. Recent developments along this direction include the BONuS \citep{yang21bonus} and AdaDetect \citep{marandon22mlmeetfdr} procedures, which convincingly demonstrate that combining conformal $p$-values with the Benjamini-Hochberg (BH) algorithm enables the use of highly effective machine learning models while ensuring finite-sample FDR control, even in the presence of model misspecification. However, currently available conformal methods are not well-equipped to handle structured multiple testing. The limitation is illustrated next.

\subsection{Non-exchangeable scores and asymmetric decision rules}
\label{subsec:rules}
Suppose we are interested in testing $m$ null hypotheses $H_{0, i}$, for $1\leq i\leq m$, based on summary statistics $\mathbf{X}=(X_{i}: 1\leq i\leq m)$. Denote $\pmb{\Theta}=(\theta_{i}: 1\leq i\leq m)\in\{0,1\}^{m}$ the true states of nature, where $\theta_{i}=0/1$ indicates that $H_{0,i}$ is true/false. Let $\mathcal{H}_{0}=\{i\in[m]: \text{$H_{0,i}$ is true}\}$. 
The decisions are represented by a binary vector $\pmb{\delta}(\mathbf{X})=(\delta_i: 1\leq i\leq m)\in\{0,1\}^{m}$, where $\delta_i=1$ indicates that $H_{0,i}$ is rejected and $\delta_i=0$ otherwise. We allow $\delta_i$ to depend on the entire data set $\mathbf{X}$. We call $\pmb{\delta}$ a \emph{symmetric decision rule} \citep{copas74rule} if $\pmb{\delta}\{\Pi(\mathbf{X})\}=\Pi\{\pmb{\delta}(\mathbf{X})\}$ for all permutation operators $\Pi$.  

To provide context, consider a hidden Markov model (HMM) where the underlying states $\pmb{\Theta}$  are unknown, and the observations are conditionally independent given $\pmb{\Theta}$. In this toy example, we assume that the hidden states form a binary Markov chain and the non-null cases tend to appear in clusters. We further assume that the null and non-null distributions are respectively given by $(X_{i}|\theta_{i}=0)\sim \mathcal{N}(0, 1)$ and $(X_{i}|\theta_{i}=1)\sim \mathcal{N}(2, 1).$
Suppose we have observed $X_{j}=X_{k}=2.5$, $j\neq k$, where $X_{j}$ is surrounded by small observations and $X_{k}$ is surrounded by large observations. Intuitively, $X_{k}$ is more likely to be a non-null case compared to $X_{j}$ due to the clustering effects from the Markovian structure; hence symmetric rules are inappropriate. \citet{sc09} demonstrated that the optimal FDR procedure in HMMs is an asymmetric rule that relies on thresholding the local index of significance (LIS), whose value depends on the entire sequence, with higher weights given to neighboring locations. 

The presence of structured patterns within the data and the use of asymmetric rules pose a significant challenge to the conformal inference framework. To see this, we tentatively consider the following definition, which generalizes the conformal $p$-value in \citet{bates21testing} in a nuanced (and possibly improper) way:
\begin{equation}\label{def:confpv}
    p_{i}\equiv p_{i}(X_{i})=\frac{1}{1+|\mathcal{D}^{cal}|} \left[1+\sum_{j\in \mathcal{D}^{cal}}\mathbb{I}\{s_{i}(X_{i})>s_{j}(Y_{j})\} \right], \; i\in[m]=\{1, \cdots, m\}.
\end{equation}
Here $s_i(\cdot)$ is the conformity score function that is allowed to vary across $i$, $\mathcal D^{cal}$ is the index set for calibration data containing null samples $Y_{j}$, with $|\cdot|$ denoting the cardinality of a set. \cite{bates21testing} showed that if the score functions are identical and permutation-invariant, and the data points $\{X_i, i \in \mathcal H_0; Y_j, j \in D^{cal}\}$ are jointly exchangeable, then the conformal $p$-values in \eqref{def:confpv} are super-uniform and PRDS \footnote{$(p_{1},\cdots,p_{m})$ have the \emph{positive regression dependency on each subset} (PRDS) property on $\mathcal{H}_{0}$ if $\PP\left\{(p_{j},j\in[m])\in D|p_{i}=u\right\}$ is non-decreasing in $u$ for any $i\in\mathcal{H}_{0}$ and any non-decreasing set $D\subset\mathbb{R}^{m}$.}. Hence, according to \cite{by01} and \cite{sarkar02}, applying BH with these conformal $p$-values is valid for FDR control. However, score functions that leverage structural patterns are typically not permutation-invariant, leading to a violation of the joint exchangeability assumption. Consequently, the conformal $p$-values defined by \eqref{def:confpv} may become improper and compromise the PRDS property, making it problematic to apply the BH procedure.

\subsection{A preview of our method and contributions}\label{subsec:overview}


To develop a provably valid FDR procedure that can effectively leverage structural information and accommodate asymmetric rules, we propose the pseudo local index of significance (PLIS) procedure that consists of three steps:  
\begin{itemize}
    \item[1.] constructing \emph{baseline data} that preserve useful structural patterns;\vspace{-0.25cm} 
    \item[2.] calculating conformity scores based on user-specified \emph{working models};\vspace{-0.25cm} 
    \item[3.] constructing a \emph{mirror process} that emulates the true process for decision-making. 
\end{itemize}
The proposed algorithm features two critical components: an innovative algorithm for calculating the conformity score through a working model, which captures useful structural patterns and enhances the efficiency of the FDR analysis; a generic framework for constructing a mirror process for decision-making, which bypasses the $p$-value inference framework, eliminates the need for joint exchangeability assumptions, and establishes finite-sample FDR theory for asymmetric rules.

The proposed research makes several contributions. Firstly, PLIS provides a provably valid model-free approach to multiple testing. Compared to model-based FDR procedures, it achieves comparable power when the underlying model is correctly specified and  accurately estimated, while maintaining finite-sample FDR control  in scenarios where the model is mis-specified or estimated poorly. Secondly, PLIS demonstrates superior power compared to existing conformal methods due to its ability to leverage asymmetric rules that exploit structural information. Finally, we have developed a novel theoretical framework that relies only on the \emph{pairwise exchangeability} between $s_{i}(X_{i})$ and $s_{i}(Y_{i})$ under the null, which is much weaker than the \emph{joint exchangeability} requirement stipulated in existing theories for conformal inference.

\subsection{Organization}

The article is organized as follows. Section \ref{sec:model} introduces the PLIS procedure for structured multiple testing and establishes its theoretical properties. Section \ref{sec:further} explores extensions of PLIS and its relationship to existing concepts. Section \ref{sec:simu} presents simulation results to assess the numerical performance of PLIS and to compare it with existing methods. An illustration of the proposed method is provided in Section \ref{sec:application} through an application. The Supplementary Material contains proofs, extensions and additional numerical results.

\section{Structured Multiple Testing: Conformal Inference with Asymmetric Rules} \label{sec:model}

In this section, we first present the structured probabilistic model (Section \ref{subsec:spm}) and the corresponding problem formulation (Section \ref{subsec:formulation}). Next, we introduce the PLIS procedure (Section \ref{subsec:method}) and establish its theoretical properties (Section \ref{subsec:theory}). Concrete examples and guidelines are provided in Section \ref{subsec:eg} to illustrate the PLIS framework. Section \ref{subsec:semi-supervised-plis} and Section \ref{subsec:conformalqv} respectively discuss the semi-supervised PLIS algorithm and the conformal $q$-value notion.  

\subsection{A class of structured probabilistic models}\label{subsec:spm}



Consider a multiple testing problem with binary-valued unknown states $\pmb{\Theta}=(\theta_{i}: i\in \mathcal{G})$ that form a graph $\mathcal{G}$. Let $m=|\mathcal{G}|$ be the number of nodes/hypotheses, where $\theta_{i}=0$ indicates that node $i$ is a null case, and $\theta_i=1$ otherwise.
Our study focuses on a class of structured probabilistic models (e.g. \citealp{Goodfellow-et-al-2016}), where the correlations between random variables $\mathbf{X}=(X_{i}: i\in\mathcal{G})$ are captured by the interdependence structure between corresponding latent states $\pmb{\Theta}=(\theta_{i}: i\in \mathcal{G})$. The inference units may be conceptualized as the nodes of the graph, and the interdependence structures between test statistics are encoded as edges that connect the nodes. 
The observations $(X_{i}: i\in\mathcal{G})$ are conditionally independent given $\pmb{\Theta}$, obeying
\begin{equation}\label{model:con.ind.}
 \PP\left(\mathbf{X}, \pmb{\Theta} \right)=\PP\left(\pmb{\Theta}\right)\prod_{i\in\mathcal{G}}f\left(X_{i}|\theta_{i}\right), \quad X_{i}|\theta_{i} { \sim } (1-\theta_{i})f_{0}(x)+\theta_{i}f_{1i}(x),
\end{equation}
where $f_{0}$ and $f_{1i}$ are the null and non-null densities, respectively. Although the model assumes conditional independence, it still remains highly flexible, as we do not impose any distributional constraints on the unknown states $\pmb{\Theta}$, and allow $f_{1i}$ to vary across different inference units. Conditional independence is closely related to the widely employed exchangeability assumption in the conformal inference literature \citep{schervish2012theory, barber2023finettis}. As established by de Finetti's Theorem \citep{heath76finetti, definetti80, durrett2019probability}, if a set of random elements is exchangeable, there must exist a latent $\eta$ such that the random elements are independent and identically distributed (i.i.d.) conditional on $\eta$. For a more detailed and in-depth exploration of the generalization of model \eqref{model:con.ind.} and the exchangeability assumption, please refer to Section \ref{subsec:semi-supervised-plis} and Section C of the Supplementary Material.

We make two assumptions without loss of generality: (a) $f_{0}(-x)=f_{0}(x)$; and (b) a larger magnitude of $|X_{i}|$ provides stronger evidence against the null hypothesis. In situations where the assumptions do not hold, we may transform $X_i$ into $z$-values using the formula $z_{i}=\Phi^{-1}\{F_{0}(X_{i})\}$, where $F_{0}$ represents the cumulative distribution function (CDF) of $X_i$ under the null, and $\Phi$ represents the CDF of a standard Gaussian variable.

We discuss several special cases of Model \eqref{model:con.ind.}. First, we examine the scenario where the latent states $\theta_{i}$ are independent and identically distributed (i.i.d.) and the non-null densities $f_{1i}$ are identical (and hence denoted as $f_{1}$). Under this assumption, Model (\ref{model:con.ind.}) reduces to Efron's two-group model \citep{efron01}:
\begin{equation}\label{model:two-group}
X_{i}\overset{i.i.d.}{\sim}(1-\pi)f_{0}(x)+\pi f_{1}(x),
\end{equation}
where $\pi=\PP(\theta_{i}=1)$ is the proportion of non-null cases. While the independent case is not the primary focus of this work, it is worth noting that our proposed methodology remains applicable in this scenario. In Section \ref{subsec:eg}, we present a tailored version of our proposal for Model \eqref{model:two-group}, and discuss its connection to the AdaDetect procedure \citep{marandon22mlmeetfdr}.

When the latent states $\pmb{\Theta}$ form a homogeneous binary Markov chain, a two-state hidden Markov model (HMM, \citealp{rabiner89hmm}) can be recovered from (\ref{model:con.ind.}). HMMs have been extensively studied in the context of multiple testing (e.g. \citealp{sc09, sesia19knockoffhmm, perrot21hmm}). By properly accounting for the dependence between test statistics and leveraging the HMM structure, these methods can effectively control the FDR and provide more accurate inference. Section \ref{subsec:eg} presents a customized algorithm for HMMs based on our proposal and illustrates its superiority over existing methods.

The class of models in \eqref{model:con.ind.} also includes useful models such as the Ising model \citep{onsager44Crystal} and conditional random field model \citep{lafferty01field}, which have been applied in multiple testing for climate change analysis and network analysis \citep{liang10graph, shu15Multiple, liu16graph, re22graph}. In contrast to HMMs which assume homogeneous transition probabilities and identical emission probabilities, model \eqref{model:con.ind.} offers greater flexibility to accommodate deviations from these assumptions. Moreover, in contrast to the stationary hidden states model proposed by \citet{wwb08}, Model (\ref{model:con.ind.}) allows the underlying process for $\pmb{\Theta}$ to be non-stationary. As many real-world phenomena exhibit non-stationary behaviors, Model (\ref{model:con.ind.}) is better equipped to provide a more accurate representation of the underlying dynamics.

\subsection{Problem formulation and basic setup} \label{subsec:formulation}




We consider a multiple testing problem that arises from Model \eqref{model:con.ind.}, which can be complex in nature and difficult to learn accurately from data. The hypotheses of interest are
$H_{0,i}:\, \theta_{i}=0 \; vs.\; H_{1,i}:\,\theta_{i}=1,$ for $i\in\mathcal{G}$.  A key feature of the models in \eqref{model:con.ind.} is the exchangeability of null observations $(X_{i}: i\in\mathcal G, \theta_i=0)$, which remains true regardless of the model complexity. This important characteristic provides a strong foundation for conformal inference and has been widely assumed. If we make the additional assumption that our working model has learned a score function  that is permutation-invariant, then we can properly define the conformal $p$-values in \eqref{def:confpv} for FDR analysis. 

As previously indicated, we want to employ a score tailored specifically to index $i$. The score  does not require permutation invariance regarding $i$. This attribute provides a substantial degree of flexibility, enabling the construction of powerful scores that can effectively leverage useful structural patterns. However, the violation of permutation invariance poses a significant challenge to the conformal inference framework, as the $p$-value is no longer super-uniform under the null. To address this issue, Section \ref{subsec:method} develops a new framework that eliminates the need of defining $p$-values as required by existing methods. 

In conformal inference, the calibration dataset, denoted as $\mathbf{Y}$, plays a key role in assessing the relative significance of hypotheses. Various approaches can be employed to obtain the calibration data $\mathbf{Y}$. In the classical setting of multiple testing, $\mathbf{Y}$ can be sampled from the known null distribution $F_0$. Conversely, in certain machine learning applications, such as the outlier detection problem, $\mathbf{Y}$ may be sampled from a set of labeled null observations. For the present, we assume that the null distribution $F_{0}$ is known. In Section \ref{subsec:semi-supervised-plis}, we turn to scenarios where $F_{0}$ is unknown, yet labeled null samples are available. 


Let $s_{i}^{X} = s_{i}(X_{i})$ denote the \emph{test score} of unit $i$. The superscript $X$ is used to distinguish $s_i^X$ from the \emph{calibration score}, denoted as $s_i^{Y}$, which are computed based on the calibration data. Let $\mathbf{S}_{X}=\{s_i^X: i\in\mathcal{G}\}$ and $\mathbf{S}_{Y}=\{s_i^Y: i\in\mathcal{G}\}$. In constructing the scores, it is crucial to guarantee the pairwise exchangeability between $s_i^X$ and $s_i^Y$ for $i\in\mathcal H_0$. This important notion is rigorously defined in Section \ref{subsec:theory}.


We denote the decision for test unit $i$ as $\delta_{i}\in\{0, 1\}$, where $\delta_{i}=1$ signifies the rejection of $H_{0,i}$ and $\delta_{i}=0$ otherwise. Denote $\pmb\delta=(\delta_i: 1\leq i\leq m)$. 
The false discovery proportion (FDP) and true discovery proportion (TDP) are respectively defined as 
\begin{equation}\label{def:FDP} 
    \mathrm{FDP}(\pmb\delta)=\frac{\sum_{j\in\mathcal G} (1-\theta_j)\delta_j}{(\sum_{j\in\mathcal{G}}\delta_j)\vee 1}, \quad \mathrm{TDP}(\pmb\delta)=\frac{\sum_{j\in\mathcal G} \theta_j\delta_j}{(\sum_{j\in\mathcal{G}}\theta_{j})\vee 1}.
\end{equation}
The (frequentist) FDR and average power (AP) can be defined accordingly: $\mathrm{FDR}=\mathbb{E}[\mathrm{FDP}(\pmb\delta)] \text{ and } \mathrm{AP}=\mathbb{E}[\mathrm{TDP}(\pmb\delta)],$
where the expectation operator is taken over the observed and calibration data while holding the hidden states $\pmb\Theta$ fixed. The objective is to develop a decision rule that effectively controls the FDR while maximizing the AP. 
\subsection{The PLIS procedure for FDR control} 
\label{subsec:method}

The conformity score that best captures structural patterns from probabilistic models \eqref{model:con.ind.} is $\PP(\theta_{i}=0|\mathbf{X})$, assuming perfect knowledge of the model structure. However, accurately estimating this score is challenging. To address this, we propose to compute a score through a user-specified working model that efficiently extracts structural information and predicts unknown states. Specifically, we choose a working model $\mathcal{M}$ and corresponding algorithm $\mathcal{A}$ to compute pseudo scores $\mathbf S_X=\{s_i^X: i\in\mathcal G\}$, bearing in mind that the working model may deviate from the complex data-generating model.

To effectively leverage the structural information, the size of the calibration set needs to match that of the test set, ensuring that each test unit is paired with a corresponding calibration data point. Suppose the test and calibration data points have been paired and denoted as $\{(X_i, Y_i), i\in\mathcal G\}$. In our methodology, the initial step involves constructing the baseline dataset, denoted as $\mathbf W=(W_i\equiv h(X_i, Y_i): i\in\mathcal G)$, where 
\begin{equation}\label{hxy}
    h(x,y)=\left\{\begin{matrix}
  x& \text{if $|x|\geq |y|$}\\
  y& \text{otherwise} 
    \end{matrix}\right. .
\end{equation}
The construction of $\mathbf W$ ensures that $W_i$ depends equally on $X_i$ and $Y_i$, while satisfying the flipping-coin property, i.e., $\PP(W_{i}=X_{i})=\PP(W_{i}=Y_{i})=1/2$ if $i\in\mathcal H_0$. 

\begin{remark}\rm{
Any symmetric function satisfying $h(x,y)=h(y,x)$ can be employed to generate the baseline data $\mathbf{W}$. However, our choice to utilize \eqref{hxy} enables us to retain data with large effect sizes, thereby preserving the structural patterns among the non-null cases. This is particularly advantageous for data sets that display local dependency structures. Section E.7 of the Supplement shows that \eqref{hxy} leads to substantial power gain compared to alternative choices such as $h^\prime(x,y)=x+y$.} 
\end{remark}

\begin{figure}[!htbp]
    \centering
    \includegraphics[width=1\linewidth]{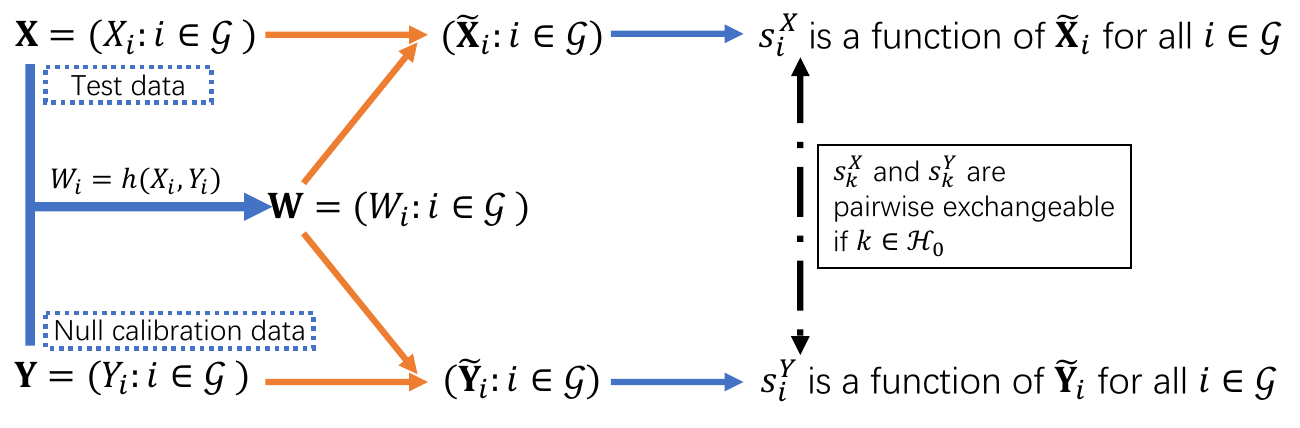}
    \caption{A graphical illustration of constructing the baseline data and calculating the scores. To create $\tilde{\mathbf{X}}_{i}$ and $\tilde{\mathbf{Y}}_{i}$, each $W_{i}$ in $\mathbf{W}$ is replaced by $X_{i}$ and $Y_{i}$ in turn, for $i\in\mathcal{G}$, with the values in remaining nodes unchanged. The same function $s_{k}(\cdot)$ is used for computing $s_{k}^{X}$ and $s_{k}^{Y}$, ensuring that the two scores are pairwise exchangeable if $k\in\mathcal{H}_{0}$. }
\label{pic:plis}
\end{figure}

We proceed to generate two parallel datasets $\{\tilde{\mathbf{X}}_{i}:i\in\mathcal{G}\}$ and $\{\tilde{\mathbf{Y}}_{i}:i\in\mathcal{G}\}$ by respectively substituting $X_i$ and $Y_i$ for $W_i$ in $\mathbf{W}$. For example, if $\mathbf X$ is an ordered sequence $(X_1, \cdots, X_m)$, then
$\tilde{\mathbf X}_i=(W_1, \cdots, W_{i-1}, X_i, W_{i+1}, \cdots, W_m)$. Next, we calculate the pseudo score $s_{i}^{X}\equiv\PP_{\mathcal{M,A}}(\theta_{i}=0|\tilde{\mathbf{X}}_{i})$, referred to as the pseudo local index of significance (PLIS). The corresponding calibration score is given by $s_{i}^{Y}\equiv\PP_{\mathcal{M,A}}(\theta_{i}=0|\tilde{\mathbf{Y}}_{i})$, applying the same working model $\mathcal M$ and computational algorithm $\mathcal A$ to the calibrated data $\tilde{\mathbf{Y}}_{i}$. Figure \ref{pic:plis} visually illustrates the process of data generation and score computation, which guarantees the pairwise exchangeability between $s_{i}^{X}$ and $s_{i}^{Y}$, a key property that will be rigorously defined and established in Section \ref{subsec:theory}. 

Suppose that a smaller value of $s_{i}^{X}$ indicates stronger evidence against the null hypothesis. Intuitively, the rejection of $H_{0,i}$ requires two pieces of evidence: firstly, $s_{i}^{X}$ must be sufficiently small, and secondly, $s_{i}^{X}$ should be smaller than its null counterpart $s_{i}^{Y}$. Therefore, we only select small $s_i^X$ from the \emph{candidate rejection set} $\mathcal{G}^{r}=\{i\in\mathcal{G}:s_{i}^{X}<s_{i}^{Y}\}$. Consider a class of thresholding rules $\pmb\delta=(\delta_{i}: i\in\mathcal{G})$, where $\delta_{i}\equiv 0$ for $i\notin\mathcal{G}^{r}$, and $\delta_{i}=\II(s_{i}^{X}\leq t)$ for $i\in\mathcal{G}^{r}$.
Correspondingly, we introduce the \emph{calibration set} $\mathcal{G}^{c}=\{i\in\mathcal{G}:s_{i}^{Y}<s_{i}^{X}\}$, which mirrors $\mathcal{G}^{r}$. This mirror set of null scores $\{s_i^Y: i\in \mathcal{G}^{c}\}$ allows us to characterize the joint behavior of their counterparts $\{s_{i}^{X}: i\in\mathcal{G}^{r}\}$ under the null. It follows that the true FDP process and its mirror process, denoted $Q(t)$, are respectively given by
\begin{equation}\label{qv}
\mbox{FDP}(t) =\frac{\sum_{j\in\mathcal{G}^{r}\cap\mathcal H_0}\mathbb{I}\left\{s_{j}^{X}\leq t\right\}}{(\sum_{j\in\mathcal{G}^{r}}\mathbb{I}\left\{s_{j}^{X}\leq t\right\})\vee 1 }, \quad    Q(t) =\frac{1+\sum_{j\in\mathcal{G}^{c}}\mathbb{I}\left\{s_{j}^{Y}\leq t\right\}}{{(\sum_{j\in\mathcal{G}^{r}}\mathbb{I}\left\{s_{j}^{X}\leq t\right\})\vee 1 } }, \quad t>0.
\end{equation}

To gain a better understanding of our methodology, it is helpful to note that $Q(t)$ provides a conservative estimate of $\mbox{FDP}(t)$. This conservativeness arises from two factors: (a) $\sum_{j\in\mathcal{G}^{c}}\mathbb{I}\{s_{j}^{Y}\leq t\}$ is always greater than or equal to $\sum_{j\in \mathcal{G}^{c}\cap\mathcal{H}_{0}}\mathbb{I}\{s_{j}^{Y}\leq t\}$; and (b) $\sum_{j\in \mathcal{G}^{c}\cap\mathcal{H}_{0}}\mathbb{I}\{s_{j}^{Y}\leq t\}$ can be seen as a mirror process that resembles the unobserved process $\sum_{j\in \mathcal{G}^{r}\cap\mathcal{H}_{0}}\mathbb{I}\{s_{j}^{X}\leq t\}$ since, as will be established in the next subsection (Property \ref{prop:pairwise}), $s_{j}^{X}$ and $s_{j}^{Y}$ are pairwise exchangeable for $j\in\mathcal H_0$. We choose the largest threshold, denoted as $\tau$, such that the conservative estimate of the FDP falls below $\alpha$: 
$$
\tau=\sup\left\{t\in \mathbf{S}_{X} \cup \mathbf{S}_{Y}: Q(t)\leq\alpha\right\}.
$$
The proposed PLIS procedure, summarized in Algorithm \ref{algo:MP}, rejects $H_{0,i}$ if $\delta_i\equiv\II(i\in\mathcal{G}^{r}, s_{i}^{X}\leq\tau)=1$. By mathematical conventions, the supremum of an empty set is defined as $-\infty$. Therefore, if the set $\{t\in \mathbf{S}_{X} \cup \mathbf{S}_{Y}:Q(t)\leq\alpha\}$ is empty, then $\tau$ is set to $-\infty$, and Algorithm \ref{algo:MP} does not reject any hypotheses.

\begin{algorithm}
\renewcommand{\algorithmicrequire}{\textbf{Input : }}
\renewcommand{\algorithmicensure}{\textbf{Output : }}
\caption{The PLIS procedure} \label{algo:MP}
\begin{algorithmic}[1]
\REQUIRE The observations $\mathbf{X}=(X_{i}: i\in\mathcal{G})$ , the null distribution $F_{0}$, a pre-specified FDR level $\alpha$, a working model $\mathcal {M}$ and a computational algorithm $\mathcal{A}$.
\ENSURE A decision rule $\pmb{\delta}=(\delta_{i}: i\in\mathcal{G})$.
\STATE Generate $\mathbf{Y}=(Y_{i}:i\in\mathcal{G})$, where $Y_i$ are i.i.d. observations from $F_{0}$. 
\STATE Create $\mathbf{W}=(W_{i}:i\in\mathcal{G})$, where $W_{i}=h(X_{i},Y_{i})$ and $h(x,y)$ is given by \eqref{hxy}. Construct $\tilde{\mathbf{X}}_{i}$ and $\tilde{\mathbf{Y}}_{i}$ by respectively  substituting $X_{i}$ and $Y_{i}$ in place of $W_{i}$ in $\mathbf{W}$ for $i\in\mathcal{G}$.

\STATE Calculate $s_{i}^{X}\equiv{\PP}_{\mathcal{M,A}}(\theta_{i}=0|\tilde{\mathbf{X}}_{i})$ and $s_{i}^{Y}\equiv{\PP}_{\mathcal{M,A}}(\theta_{i}=0|\tilde{\mathbf{Y}}_{i})$ for $i\in\mathcal{G}$ with the user-specified working model $\mathcal {M}$ and algorithm $\mathcal{A}$. Let $\mathcal{G}^{r}=\{i:s_{i}^{X}<s_{i}^{Y}\}$.

\STATE Let $\tau=\sup\{t\in\mathbf{S}_{X} \cup \mathbf{S}_{Y}:Q(t)\leq\alpha\}$, where $Q(t)$ is defined in \eqref{qv}. 

\STATE Let $\delta_{i}=\mathbb{I}\{s_{i}^{X}\leq\tau\}$ for $i\in\mathcal{G}^{r}$, otherwise $\delta_{i}=0$.

\STATE \textbf{Return} $\pmb{\delta}=(\delta_{i}: i\in\mathcal{G})$.
\end{algorithmic}
\end{algorithm}

We conclude this subsection with several remarks. First, in order to ensure pairwise exchangeability between $s_i^X$ and $s_i^Y$ under the null, it is crucial to estimate the working model $\mathcal{M}$ from $\mathbf{W}$ before calculating the conformity scores. Detailed explanations and practical guidelines for the deployment of PLIS under two widely-used working models can be found in Section \ref{subsec:eg}. Second, we can define an alternative  estimate of $\mbox{FDP}(t)$, denoted as $Q^\prime(t)$, by substituting $\mathcal{G}$ in place of both $\mathcal{G}^{r}$ and $\mathcal{G}^{c}$ in equation \eqref{qv}. However, employing $Q^\prime(t)$ tends to be overly conservative; further elaborations on this are provided in Sections \ref{ap:knockoff}.
Third, Algorithm \ref{algo:MP} closely resembles the operation of the conformal BH, albeit with several modifications. Finally, as revealed by an insightful reviewer, $Q(t)$ in \eqref{qv} is closely related to the Selective SeqStep+ algorithm \citep{barber15knockoff}. Further discussions on these connections can be found in Section D of the Supplement.

\subsection{Theory}\label{subsec:theory}

We start by establishing several exchangeability properties, first for the data points and subsequently for the conformity scores. The random elements $\{Z_{i}: i\in[K]\}$ are said to be (jointly) exchangeable if the distribution of $(Z_1, \cdots, Z_K)$ is the same as that of $(Z_{\Pi_1}, \cdots, Z_{\Pi_K})$ for any permutation $(\Pi_1, \cdots, \Pi_K)$ of the indices $\{1, \cdots, K\}$. We write 
$
(Z_1, \cdots, Z_K)\overset{d}{=}(Z_{\Pi_1}, \cdots, Z_{\Pi_K}),
$ 
where $\overset{d}{=}$ denotes equality in distribution. We will soon introduce a weaker notion, the pairwise exchangeability, which is more relevant to our theory. The following exchangeability properties are proven in Section A of the supplement. 

The first property can be easily deduced from the definition of Model \eqref{model:con.ind.} and the construction of $\mathbf{W}$ outlined in Algorithm \ref{algo:MP}.

\begin{property}[Exchangeability and conditional independence between data points] \label{prop:points} 
Suppose $\mathbf{X}=(X_{i}:i\in\mathcal{G})$ are observations from Model \eqref{model:con.ind.}, and $(Y_{i}:i\in\mathcal{G})$ are randomly drawn from the null distribution $F_{0}$. Then we have:
(a) The random variables $\left\{Y_{i}, i\in\mathcal{G}; X_{j}, j\in\mathcal{H}_{0}\right\}$ are jointly exchangeable conditional on $\pmb\Theta$; (b) $\{Z_{j}=\{X_{j},Y_{j}\}:j\in\mathcal{G}\}$ are conditionally independent given $\mathbf{W}$ and $\pmb{\Theta}$; 
(c) For $i\in\mathcal H_0$, $X_{i}$ and $Y_{i}$ are pairwise exchangeable conditional on $\mathbf{W}$ and $\pmb{\Theta}$, i.e.,
$(X_{i},Y_{i}|\mathbf{W},\pmb{\Theta})\overset{d}{=}(Y_{i},X_{i}|\mathbf{W},\pmb{\Theta}). $

\end{property}

The next property characterizes the pairwise exchangeability between $\tilde{\mathbf{X}}_{j}$ and $\tilde{\mathbf{Y}}_{j}$. 

\begin{property}[Pairwise exchangeability and conditional independence between data sets]\label{prop:sequence} Consider $\tilde{\mathbf{X}}_{j}$ and $\tilde{\mathbf{Y}}_{j}$ as generated according to Algorithm \ref{algo:MP}. Then we have: (a) $\{\mathcal{D}_{j}=\{\tilde{\mathbf{X}}_{j}, \tilde{\mathbf{Y}}_{j}\} :j\in\mathcal{G}\}$ are conditionally independent given $\mathbf{W}$ and $\pmb{\Theta}$; 
(b) For $i\in\mathcal{H}_{0}$, $\tilde{\mathbf{X}}_{i}$ and $\tilde{\mathbf{Y}}_{i}$ are pairwise exchangeable, i.e. $(\tilde{\mathbf{X}}_{i}, \tilde{\mathbf{Y}}_{i}| \mathbf{W},\pmb{\Theta}) \overset{d}{=} (\tilde{\mathbf{Y}}_{i}, \tilde{\mathbf{X}}_{i}| \mathbf{W},\pmb{\Theta})$.

\end{property}

The next property is concerned with the conditional independence and pairwise exchangeability between null scores.

\begin{property}[Pairwise exchangeability and conditional independence between scores]\label{prop:pairwise} Consider scores $s_j^X$ and $s_j^Y$ computed according to Algorithm \ref{algo:MP}, denote $\mathbf s_j=\{s_{j}^{X},s_{j}^{Y}\}$ for $j\in\mathcal{G}$. Then we have:
(a) $\{\mathbf{s}_{j}:j\in\mathcal{G}\}$ are conditionally independent given $\mathbf{W}$ and $\pmb{\Theta}$;
(b) Denote $\mathbf S_{-i}=\cup_{j\in\mathcal{G},j\neq i}\mathbf{s}_j$. For $i\in\mathcal{H}_{0}$, $s_{i}^{X}$ and $s_{i}^{Y}$ are pairwise exchangeable given $\mathbf S_{-i}$ and $\pmb\Theta$: \vspace{-0.25cm}
\begin{equation}\label{prop:pairwise-scores}
(s_{i}^{X}, s_{i}^{Y}| \mathbf{S}_{-i}, \pmb{\Theta} ) \overset{d}{=} (s_{i}^{Y}, s_{i}^{X}| \mathbf{S}_{-i}, \pmb{\Theta}),\text{ or equivalently }(s_{i}^{X}, s_{i}^{Y}, \mathbf{S}_{-i} | \pmb{\Theta} ) \overset{d}{=} (s_{i}^{Y}, s_{i}^{X}, \mathbf{S}_{-i}| \pmb{\Theta}).
\end{equation} 
\end{property}
  
The pairwise exchangeability \eqref{prop:pairwise-scores}, which is crucial in proving the FDR theory for PLIS, is a weaker assumption than the joint exchangeability among all null scores $\{s_{i}^{X}: i\in\mathcal{H}_{0}\}\cup\{ s_{i}^{Y}: i\in\mathcal{G}\}$ required in most existing conformal methods. Now we state our main theorem that establishes the validity of PLIS for controlling the FDR.

\begin{thm}\label{thm:fdr}
Consider Model \eqref{model:con.ind.} and the PLIS procedure outlined in Algorithm \ref{algo:MP}. Then Properties 1-3 hold and PLIS controls the FDR at level $\alpha$.
\end{thm}

Our approach sets itself apart from current theories in conformal inference that rely on joint exchangeability \citep{yang21bonus,mary22semi,marandon22mlmeetfdr,bates21testing}. In contrast, we introduce a framework that utilizes only pairwise exchangeability and establishes our theory grounded in the Selective SeqStep+ algorithm \citep{barber15knockoff}.

\subsection{Practical guidelines and examples}\label{subsec:eg}

PLIS utilizes resampling techniques to generate a mirror process that quantifies the FDP in multiple testing. At its core, PLIS can be situated within the conformal inference framework, which facilitates the development of valid FDR rules that exhibit enhanced robustness to model misspecification. This section presents examples that contextualize PLIS as a valuable tool for ``conformalizing'' some well-known FDR procedures, producing inferences that possess similar operational characteristics as state-of-the-art  conformal methods \citep{yang21bonus,marandon22mlmeetfdr, liang2022integrative, bates21testing}. 


\noindent\textbf{Example 1. Conformalizing the LIS procedure under HMMs.} 

A variety of real-world applications involve data with HMM-type structures. Several works  \citep{sc09, perrot21hmm} have demonstrated that exploiting the HMM structure can greatly enhance the power of FDR analysis. However, their theoretical analysis necessitates certain conditions to be satisfied, such as homogeneous transition probabilities and identical non-null distributions, which may not be strictly met in practical scenarios. The HMMs are characterized by a set of parameters, denoted as ${\vartheta}=(\mathbf A, f_0, f_1)$, where $\mathbf{A}$ is the transition probability matrix, and $f_0$ and $f_1$ are the emission densities for the null and non-null observations, respectively. 

In our setup, we use $\mathcal {M}_{\rm HM}$ as the HMM working model, where $\vartheta$ is estimated using the EM algorithm on $\mathbf{W}$. To implement the PLIS procedure, we utilize the forward-backward algorithm, denoted by $\mathcal{A}_{\rm FB}$, to compute the scores $s_{i}^{X}=\PP_{\mathcal {M}_{\rm HM},\mathcal{A}_{\rm FB}}(\theta_{i}=0|\tilde{\mathbf{X}}_{i})$ and $s_{i}^{Y}=\PP_{\mathcal {M}_{\rm HM},\mathcal{A}_{\rm FB}}(\theta_{i}=0|\tilde{\mathbf{Y}}_{i})$ for $i\in\mathcal{G}$. The decision-making process is based on Algorithm \ref{algo:MP}. The PLIS procedure can be regarded as a conformalized adaptation of the LIS procedure proposed by \citet{sc09}, computed using $\mathcal {M}_{\rm HM}$ and algorithm $\mathcal{A}_{\rm FB}$.   PLIS  differs from LIS  in that it can handle a variety of data structures that deviate from HMMs. This ability ensures valid FDR control in finite-samples, even when the working model is misspecified or the model parameters are estimated poorly.  \qed

\medskip
\noindent\textbf{Example 2. Conformalizing the adaptive $z$-value procedure under the two-group model \eqref{model:two-group} with independent observations.}
\medskip

In the context where observations are assumed to be independent and Efron's two-group model \eqref{model:two-group}, denoted $\mathcal M_{\rm TG}$, is adopted, \citet{sc07} showed that the density ratio (DR) $r(x)=\frac{f_{0}(x)}{{f}(x)}$ or the local false discovery rate ${\mathrm{Lfdr}}(x)=(1-\pi)r(x)$ serves as the optimal building block for FDR analysis. The PLIS framework treats DR and Lfdr equivalently since $(1-\pi)$ is a constant across all study units and only the relative ranking contributes to the operation of PLIS. Using the working model $\mathcal M_{\rm TG}$, the conformity score $s_i(\cdot)$ can be taken as the  DR function $\hat r(\cdot)=f_0(\cdot)/\hat f(\cdot)$, where $f_0$ is known, and $\hat f(\cdot)$ is the standard kernel density estimator constructed based on $\mathbf{W}$. The PLIS procedure operates by employing $s_{i}^{X}\equiv\hat{r}(X_{i})$ and $s_{i}^{Y}\equiv\hat{r}(Y_{i})$ in Algorithm \ref{algo:MP}. As carefully explained in Section D.7 of the Supplement, although DR is permutation-invariant with respect to $\mathbf{W}$, the null scores are not jointly exchangeable. Nevertheless, PLIS still works since $(s_{i}^{X}, s_{i}^{Y})$ are pairwise exchangeable given $\mathbf{W}$ and $\pmb{\Theta}$. 

\citet{sc07} proposed a class of adaptive $z$-value (AZ) procedures, and 
showed that AZ outperforms $p$-value based methods by adapting to the shape of the alternative distribution. PLIS may be viewed as a conformalized adaptation of AZ, which effectively incorporates the structural information into inference while controlling the FDR in finite samples. This represents a notable advantage over the AZ procedure, which only ensures asymptotic FDR control. We note that the AdaDetect method \citep{marandon22mlmeetfdr}, which also utilizes the two-group model to estimate a permutation-invariant score function, exhibits similar advantages to PLIS. However, when encountering score functions that lack permutation-invariance, particularly when utilizing the HMM as the working model, PLIS significantly deviates from AdaDetect.
 \qed 
  
Our customized PLIS procedures remain valid as long as the underlying model belongs to the broad class  \eqref{model:con.ind.}, regardless of the working models or algorithms used. Developing conformity scores using sensible models and computational algorithms is essential for improving the practicality and effectiveness of PLIS. Finally, instances of numerical issues such as significant reductions in power and convergence failures frequently serve as warnings for practitioners, highlighting the unsuitability of the current model for the specific application and emphasizing the need for a new working model.  

\subsection{Semi-supervised PLIS} \label{subsec:semi-supervised-plis}

This section considers the scenario in which the null distribution is unknown, and users only have access to $n$ labeled null data points $\mathbf{U}=\{U_{1},\cdots,U_{n}\}$ drawn from that distribution. The objective is to assess whether new data points $\{X_i: i\in\mathcal G\}$ follow the same unknown distribution as the reference dataset $\mathbf{U}$. This scenario is known as the semi-supervised multiple testing problem \citep{mary22semi}, or the outlier detection problem in the conformal inference literature \citep{marandon22mlmeetfdr,bates21testing}. We introduce a semi-supervised PLIS (Algorithm \ref{algo:pu_plis}) to solve this problem. 

The algorithm first partitions the labeled null data $\mathbf{U}$ into two subsets: the calibration set $\mathbf{Y}=(Y_{i}:i\in\mathcal{G})$ and the training set $\mathbf{U}^{tr}=\mathbf{U}\setminus\mathbf{Y}$. 
The test data $\mathbf{X}$ and calibration data $\mathbf{Y}$ are then utilized to construct the baseline data $\mathbf{W}$ by following the same steps in Algorithm \ref{algo:MP}.  We move on to the estimation issue. In the case of an HMM, an estimator for the emission distribution $f_{0}$ can be learned from the labeled null data $\mathbf{U}^{tr}$. Subsequently, we utilize the baseline data $\mathbf{W}$ to estimate $f_1$ and transition probabilities via the EM algorithm, and then calculate the conformity scores via the forward-backward procedure. If the working model is a two-group model, the conformity scores correspond to the density ratios, which can be estimated using both $\mathbf{U}^{tr}$ and $\mathbf{W}$ via the positive unlabeled (PU) learning algorithms. This approach has also been suggested by \citet{marandon22mlmeetfdr}. The above steps are summarized in Algorithm \ref{algo:pu_plis}.

\begin{algorithm}
\renewcommand{\algorithmicrequire}{\textbf{Input : }}
\renewcommand{\algorithmicensure}{\textbf{Output : }}
\caption{The semi-supervised PLIS procedure} \label{algo:pu_plis}
\begin{algorithmic}[1]
\REQUIRE The test data $\mathbf{X}=(X_{i}: i\in\mathcal{G})$ , the null samples $\mathbf{U}=\{U_{1},\cdots,U_{n}\}$, a pre-specified FDR level $\alpha$, a working model $\mathcal {M}$ and a computational algorithm $\mathcal{A}$.
\ENSURE A decision rule $\pmb{\delta}=(\delta_{i}: i\in\mathcal{G})$.
\STATE Split $\mathbf{U}$ into the calibration data $\mathbf{Y}=(Y_{i}:i\in\mathcal{G})$ and the training data $\mathbf{U}^{tr}=\mathbf{Z}\setminus\mathbf{Y}$.
\STATE Create $\mathbf{W}=(W_{i}:i\in\mathcal{G})$, where $W_{i}=h(X_{i},Y_{i})$ and $h(x,y)$ is given by \eqref{hxy}. Construct $\tilde{\mathbf{X}}_{i}$ and $\tilde{\mathbf{Y}}_{i}$ by respectively  substituting $X_{i}$ and $Y_{i}$ in place of $W_{i}$ in $\mathbf{W}$ for $i\in\mathcal{G}$.

\STATE Calculate $s_{i}^{X}\equiv{\PP}_{\mathcal{M,A}}(\theta_{i}=0|\tilde{\mathbf{X}}_{i})$ and $s_{i}^{Y}\equiv{\PP}_{\mathcal{M,A}}(\theta_{i}=0|\tilde{\mathbf{Y}}_{i})$ for $i\in\mathcal{G}$ with the user-specified working model $\mathcal {M}$ and algorithm $\mathcal{A}$ based on $\mathbf{W}$ and $\mathbf{U}^{tr}$.

\STATE Let $\tau=\sup\{t\in\mathbf{S}_{X} \cup \mathbf{S}_{Y} :Q(t)\leq\alpha\}$, where $Q(t)$ is defined in \eqref{qv}. 

\STATE Let $\delta_{i}=\mathbb{I}\{s_{i}^{X}\leq\tau\}$ for $i\in\mathcal{G}^{r}=\{j\in\mathcal{G}:s_{j}^{X}<s_{j}^{Y}\}$, otherwise $\delta_{i}=0$.

\STATE \textbf{Return} $\pmb{\delta}=(\delta_{i}: i\in\mathcal{G})$.
\end{algorithmic}
\end{algorithm}

\begin{remark}\rm{
Here we have assigned exactly one null data point to each test unit, but in situations where there is an abundance of null samples, it may be appropriate to employ the de-randomization idea  in Section \ref{subsec:deran} to improve stability. In situations where the size of the null dataset is smaller than $m$, one may conduct an independent screening procedure to narrow down the focus before using PLIS on a smaller subset of hypotheses. These represent interesting issues for future investigation.}
\end{remark}

The next theorem, established using de Finetti's theorem (cf. Section C in the Supplement) and techniques employed in proving Properties \ref{prop:points}-\ref{prop:pairwise}, generalizes Theorem \ref{thm:fdr} in two ways. Firstly, it considers the semi-supervised setup, where we have access to labeled null data instead of explicit knowledge of the null distribution. Secondly, it relaxes the assumption of conditional independence given $\pmb\Theta$ in Model \eqref{model:con.ind.} by allowing for (exchangeable) correlated noise in the data generation process.  

\begin{thm}\label{thm:puplis}
    Consider Algorithm \ref{algo:pu_plis}, if the null data $\{U_{1},\cdots,U_{n},X_{i},i\in\mathcal{H}_{0}\}$ are exchangeable conditional on the non-null data $\{X_{i}:i\notin\mathcal{H}_{0}\}$ and the true states, then 
        (a) for $i\in\mathcal{H}_{0}$, $s_{i}^{X}$ and $s_{i}^{Y}$ are pairwise exchangeable, satisfying \eqref{prop:pairwise-scores}; 
        (b) Algorithm \ref{algo:pu_plis} controls FDR at $\alpha$.
\end{thm}

In Section E.4 of the Supplement, we provide numerical results to corroborate Theorem \ref{thm:puplis} by considering the semi-supervised setup in which correlated noises are introduced to the observations generated from Model \eqref{model:con.ind.}.

\subsection{Conformal $q$-values}\label{subsec:conformalqv}

When the joint exchangeability condition fails to hold, conformal $p$-values can no longer be properly defined. To address this issue, we introduce the concept of conformal $q$-value as a significance index to measure the risk associated with individual decisions. 

We begin by examining $Q(t)$ in \eqref{qv}, which offers a conservative estimate of the FDP. Consider the minimum FDR level $\alpha$ at which $H_{0,i}$ can be ``just’’ rejected:
\begin{equation}\label{conf-qv}
    q_{i} \equiv \min_{t\in\mathbf{S}_{X}\cup\mathbf{S}_{Y},t \geq s_{i}^{X}} Q(t), \text{ for } i\in\mathcal{G}^{r}; \quad q_{i} \equiv 1, \text{ for } i\in\mathcal{G}\setminus\mathcal{G}^{r}.
\end{equation} 
The adjusted $q_i$ is referred to as the conformal $q$-value corresponding to $H_{0,i}$, owing to its resemblance to the $q$-value idea introduced by \cite{storey03qvalue}. While Storey's $q$-value is constructed based on the empirical distribution of $p$-values, our conformal $q$-value is derived from a resampling method and a carefully designed mirror process.

The conformal $q$-value is a valid and user-friendly significance index that provides clear interpretability for individual decisions, and practitioners can directly use it for decision-making by comparing them with a pre-specified $\alpha$. 
Theorem \ref{thm:fdr} and Proposition \ref{thm:qv} below together establish the validity of using the conformal $q$-value \eqref{conf-qv} in FDR analysis.  
 
\begin{prop}\label{thm:qv} 
Consider the PLIS procedure as outlined in Algorithm \ref{algo:MP} and the conformal $q$-value defined by equation \eqref{conf-qv}. Then the decision $\delta_i$ presented in Algorithm \ref{algo:MP} is equivalent to rejecting $H_{0,i}$ if $q_{i}\leq\alpha$. 
\end{prop}

\section{Connections to Existing Works and Extensions}\label{sec:further}

In this section, we first establish the connection between PLIS and the e-BH method (Section \ref{subsec:evalue}), then introduce several ensuing extensions: derandomized PLIS (Section \ref{subsec:deran}), as well as ${\rm PLIS}_{\rm cbh}$ and ${\rm PLIS}_{\rm sym}$ (Sections \ref{ap:knockoff}). Finally, we discuss the distinctions of PLIS from related works (Section \ref{subsec:comparison}).

\subsection{Connection to the e-BH procedure}\label{subsec:evalue}

In hypothesis testing, an e-value \citep{vovk21evalues} is defined as the observed value of a non-negative random variable $E$ that satisfies the condition $\mathbb{E}[E]\leq 1$ under the null hypothesis. E-values can be constructed using betting scores \citep{shafer21}, likelihood ratios, and stopped super-martingales \citep{grunwald20safe}. This section demonstrates how the PLIS framework can be utilized to construct robust and powerful e-values.

Let $(e_{i}: 1\leq i\leq m)$ be the e-values for testing $(H_{0,i}: 1\leq i\leq m)$. \citet{wang22ebh} proposed the e-BH procedure, which involves first ordering the e-values as $e_{(1)}\geq e_{(2)}\geq\cdots\geq e_{(m)}$, and then choosing a cutoff along the ranking using the following step-wise algorithm.
Let $ \hat{k}=\max\left\{i:(i/m)e_{(i)}\geq(1/\alpha)\right\}$, then reject 
hypotheses in the set $\mathcal{R}_{ebh}=\{1\leq j\leq m : e_{j}\geq e_{(\hat{k})}\}$. We call $\{e_{j}: 1\leq j\leq m\}$ a set of generalized e-values if 
\begin{equation}\label{cond:generalev}
    \mathbb{E} \left( \textstyle\sum_{j \in \mathcal{H}_{0} }e_{j} \right)\leq m.
\end{equation}
Condition (\ref{cond:generalev}) is strictly weaker than the condition that $\mathbb{E}[e_{j}]\leq1$ for all $j \in \mathcal{H}_{0}$. \citet{wang22ebh} proved that if  $\{e_{j}: 1\leq j\leq m\}$ are a set of generalized e-values, then the e-BH procedure controls FDR at $\alpha$ under arbitrary dependence. We can construct a set of generalized e-values based on the PLIS framework. Specifically, define 
\begin{equation}\label{newev}
    e_{j} = \frac{m \delta_{j} }{1+\sum_{i\in\mathcal{G}^{c}} \mathbb{I}\{ s_{i}^{Y}\leq \tau\}}, \quad \mbox{for $j\in\mathcal G$},
\end{equation}
where $\delta_{j}$, $\mathcal{G}^{c}$ and $\tau$ are defined by Algorithm \ref{algo:MP}. The following theorem can be proved using similar techniques for proving Theorem \ref{thm:fdr}.
\begin{thm}\label{thm:ev}
    The variables $e_{j}$ defined in (\ref{newev}) constitute a set of generalized e-values. 
\end{thm}
The following proposition demonstrates that the PLIS procedure is equivalent to implementing the e-BH procedure with generalized e-values as defined by \eqref{newev}. The connection to e-BH offers valuable insights into our theory, providing a new perspective on why PLIS performs robustly in the face of general dependence and model misspecification.

\begin{prop}\label{prop:ev}
 If we implement the e-BH procedure with e-values in \eqref{newev}, then the rejection set $\mathcal{R}_{ebh}=\mathcal{R}$, where $\mathcal{R}=\{i\in\mathcal{G}:\delta_{i}=1\}$ is the set of rejected hypotheses by Algorithm \ref{algo:MP}.
\end{prop}

\subsection{Derandomized PLIS}\label{subsec:deran}

While the PLIS framework employs resampling methods, which can introduce additional uncertainties  and may be viewed as undesirable by practitioners, the fact that the average of e-values is still an e-value \citep{vovk21evalues} offers motivation to consider a principled derandomization method. This involves aggregating multiple random samples to reduce variability. Similar ideas have been utilized in contemporaneous works, including \citet{ren22derandomized} and \citet{sesia2023}.

The derandomized PLIS procedure involves running the PLIS procedure repeatedly for $N$ times and averaging the results for decision-making. Specifically, for the $k$-th replication, $k\in[N]$, we generate $\{Y_{j}^{(k)}: j\in\mathcal G\}\overset{i.i.d.}{\sim}f_{0}$, and construct $\tilde{\mathbf{X}}_{i}^{(k)}$ and $\tilde{\mathbf{Y}}_{i}^{(k)}$ by respectively substituting $X_{i}^{(k)}$ and $Y_{i}^{(k)}$ in place of $W_{i}^{(k)}$ in $\mathbf{W}^{(k)}$, where $\mathbf{W}^{(k)}$ is created by combining $\mathbf{X}$ and $\mathbf{Y}^{(k)}=(Y_{i}^{(k)}:i\in\mathcal{G})$ via \eqref{hxy}. We then compute the scores $\{s_{i}^{X(k)},s_{i}^{Y(k)}:i\in\mathcal{G}\}$ and threshold $\tau^{(k)}$ based on Algorithm \ref{algo:MP}. The corresponding e-values $\{e_{i}^{(k)}: i\in \mathcal G\}$ for replication $k$ can be calculated using \eqref{newev}. This process is repeated for $k=1, \cdots, N$. Finally, we average the individual e-values $e_{i}^{(k)}$ to compute the summary e-values $\bar e_{i}=\frac{1}{N} \sum_{k=1}^{N} e_i^{(k)}$ for all $i\in\mathcal G$, and apply the e-BH procedure with $\{\bar e_i: i\in \mathcal G\}$. The proposed derandomized PLIS procedure is summarized in Algorithm \ref{algo:derandomPLIS}. The theoretical FDR control of derandomized PLIS follows from the validity of the e-BH procedure and Theorem \ref{thm:ev}.

\begin{remark}\rm{
Algorithm \ref{algo:derandomPLIS} has allowed different replications to have varied $(\alpha_{k})_{k=1}^{N}$. As noted by \citet{ren22derandomized}, derandomization could lead to higher power if the e-values are carefully aggregated through powerful weighting schemes. We have investigated this issue with some preliminary results provided in Section E.6 in the Supplement.  The development of optimal weights is an interesting  direction for future research.}
\end{remark}

\begin{algorithm}
\renewcommand{\algorithmicrequire}{\textbf{Input : }}
\renewcommand{\algorithmicensure}{\textbf{Output : }}
\caption{The Derandomized PLIS procedure} \label{algo:derandomPLIS}
\begin{algorithmic}[1]
\REQUIRE Observations $\mathbf{X}=(X_{i}: i\in\mathcal{G})$, the null distribution $F_{0}$, working model $\mathcal {M}$, algorithm $\mathcal{A}$, number of replications $N$, weights $(\alpha_{k})_{k=1}^{N}$, nominal FDR level $\alpha$. 
\ENSURE Decisions $\pmb{\delta}=(\delta_{i}:i\in\mathcal{G})$. 


\FORALL{$k=1,2,\cdots,N$} 

\STATE Generate $\{Y_{j}^{(k)}\}_{j=1}^{m}\overset{i.i.d.}{\sim}F_{0}$. Denote $\mathbf{Y}^{(k)}=(Y_{j}^{(k)}:j\in\mathcal{G})$.

\STATE Create $\mathbf{W}^{(k)}=(W_{i}^{(k)}:i\in\mathcal{G})$ by combining $\mathbf{X}$ and $\mathbf{Y}^{(k)}$ via \eqref{hxy}. Construct $\tilde{\mathbf{X}}_{i}^{(k)}$ and $\tilde{\mathbf{Y}}_{i}^{(k)}$ by respectively substituting $X_{i}$ and $Y_{i}^{(k)}$ in place of $W_{i}^{(k)}$ in $\mathbf{W}^{(k)}$ for $i\in\mathcal{G}$. 

\STATE Calculate $s_{i}^{X(k)}\equiv{\PP}_{\mathcal{M,A}}(\theta_{i}=0|\tilde{\mathbf{X}}_{i}^{(k)})$ and $s_{i}^{Y(k)}\equiv{\PP}_{\mathcal{M,A}}(\theta_{i}=0|\tilde{\mathbf{Y}}_{i}^{(k)})$ for $i\in\mathcal{G}$ with user-specified working model $\mathcal {M}$ and algorithm $\mathcal{A}$. Let $\mathcal{G}^{r(k)}=\{i\in\mathcal{G}:s_{i}^{X(k)}<s_{i}^{Y(k)}\}$, $\mathcal{G}^{c(k)}=\{i\in\mathcal{G}:s_{i}^{Y(k)}<s_{i}^{X(k)}\}$, and $\mathbf{S}_{X}^{(k)}=\{s_{i}^{X(k)}\}_{i\in\mathcal{G}}$, $\mathbf{S}_{Y}^{(k)}=\{s_{i}^{Y(k)}\}_{i\in\mathcal{G}}$.

\STATE Let $Q^{(k)}(t)=\frac{1+\sum_{i\in\mathcal{G}^{c(k)}}\mathbb{I}\{s_{i}^{Y(k)}\leq t\}}{1\vee(\sum_{i\in\mathcal{G}^{r(k)}}\mathbb{I}\{s_{i}^{X(k)}\leq t\}) }$ and $\tau^{(k)}=\sup\{t \in\mathbf{S}_{X}^{(k)} \cup \mathbf{S}_{Y}^{(k)}:Q^{(k)}(t)\leq\alpha_{k}\}$.

\STATE Calculate $e_{i}^{(k)} = \frac{m \mathbb{I}\{ s_{i}^{X(k)}\leq \tau^{(k)}\} }{1+\sum_{j\in\mathcal{G}^{c(k)}} \mathbb{I}\{ s_{j}^{Y(k)}\leq \tau^{(k)}\}}$ for $i\in\mathcal{G}^{r(k)}$, otherwise $e_{i}^{(k)}=0$.

\ENDFOR

\STATE Let $\bar{e}_{i}=\frac{1}{N}\sum_{k=1}^{N}e_{i}^{(k)}$ for $i\in\mathcal{G}$. Denote the ordered statistics by $\bar{e}_{(1)}\geq \bar{e}_{(2)}\geq\cdots\geq \bar{e}_{(m)}$. Let $\hat{k}=\max\{i:({i\bar{e}_{(i)}}/{m})\geq({1}/{\alpha})\}$.

\STATE Let $\delta_{i}=\mathbb{I}\{\bar{e}_{i}\geq \bar{e}_{(\hat{k})}\}$ for $i\in\mathcal{G}$.

 \textbf{Return} $\pmb{\delta}=(\delta_{i}:i\in\mathcal{G})$
\end{algorithmic}
\end{algorithm}

\subsection{Comparison with conformal and knockoff methods}
\label{ap:knockoff}

PLIS is developed based on the principles of conformal inference, which involve the utilization of machine learning algorithms to calculate conformity scores and the use of calibration data to quantify uncertainties in decision-making. Inspired by an insightful comment from a reviewer, Section D.4 of the Supplement shows that the mirror process \eqref{qv}, originally adapted from the conformal BH algorithm (see Section D.3), falls into the class of Selective SeqStep+ algorithms \citep{barber15knockoff}. Our innovative construction of baseline data $\mathbf{W}$,  development of pairwise exchangeable scores, and  utilization of the new mirror process offer several enhancements to existing conformal methods, enabling the extension of the knockoff framework beyond regression settings \citep{barber15knockoff,candes18konckoff,ren23sideinfo}.

This section explores two variations of PLIS, emphasizing its connection to existing works and providing insights into its advantages.

The first variation replaces both $\mathcal{G}^{r}$ and $\mathcal{G}^c$ with $\mathcal{G}$ in Algorithm \ref{algo:MP}. This variation, denoted as ${\rm PLIS}_{\rm cbh}$, essentially involves applying the conformal BH algorithm \citep{mary22semi,bates21testing} to the conformity scores $\{(s_i^X, s_i^Y):i\in\mathcal{G}\}$ generated by Algorithm \ref{algo:MP}. However, implementing the conformal BH algorithm with pairwise exchangeable scores may result in overly conservative FDR levels. Additionally, using the adaptive BH algorithm \citep{marandon22mlmeetfdr,bates21testing} to overcome this conservativeness is inappropriate, as conformal $p$-values computed from pairwise exchangeable scores do not satisfy the super-uniformity and PRDS conditions.

The second variation involves applying symmetrization techniques, which have shown success in regression contexts \citep{barber15knockoff, candes18konckoff,ren23sideinfo,du2021sda}. Its key challenge lies in identifying efffective anti-symmetric functions that fulfill $T_{j}(a,b)=-T_{j}(b,a)$ to transform pairwise exchangeable scores into symmetrized test statistics. In Section D.4 of the Supplement, we demonstrate that the mirror process \eqref{qv} corresponds to utilizing the Selective SeqStep+ algorithm proposed by \cite{barber15knockoff} with a carefully designed anti-symmetric function. To underscore the significance of the symmetrization process, we consider a popular choice of anti-symmetric function $T_{j}\equiv s_{j}^{Y}-s_{j}^{X}$, as suggested in \cite{candes18konckoff} and their later works. This approach, labeled as ${\rm PLIS}_{\rm sym}$, rejects $H_{0,j}$ if $T_{j}=s_{j}^{Y}-s_{j}^{X}\geq\tau_{\rm sym}$, where $\tau_{\rm sym}$ is determined by the following FDP process: $\tau_{\rm sym} = \inf \left\{ t\in\{|T_{j}|:j\in\mathcal{G}\}: \frac{1+\sum_{j\in\mathcal{G}} \mathbb{I}\{T_{j} \leq -t\} }{\sum_{j\in\mathcal{G}} \mathbb{I}\{T_{j} \geq t\}}\leq \alpha \right\}.$ However, utilizing $T_{j}=s_{j}^{Y}-s_{j}^{X}$ may lead to information loss because the contrast has a tendency to cancel out the inherent clustering structure  among non-null effects. 

Section E.8 presents numerical results that illustrate the advantages of PLIS over its two variations ${\rm PLIS}_{\rm cbh}$ and ${\rm PLIS}_{\rm sym}$.

\subsection{Comparison with covariate-adaptive methods}\label{subsec:comparison}

Structured multiple testing is a subject of extensive research that has attracted significant interest. Our PLIS procedure distinguishes itself from existing covariate-adaptive methods in several  aspects. Firstly, in the works of \citet{lei18adapt}, \citet{Ignatiadis21IHW} and \citet{ren23sideinfo}, the structural information is represented by a covariate sequence and is incorporated through techniques such as weighted $p$-values or covariate-adaptive thresholds. By contrast, our approach leverages structural knowledge about the underlying data generation process and integrates it into inference via a user-specified working model. Secondly, the theory in existing methods is developed based on group-wise cross-weighting \citep{Ignatiadis21IHW} or covariate modulation through data masking \citep{lei18adapt, ren23sideinfo}. In contrast, PLIS operates within the theoretical framework of conformal inference, harnessing the flexibility of machine learning algorithms to learn powerful conformity scores, with associated decision risk being assessed by employing carefully calibrated null scores. Thus, PLIS not only facilitates the effective incorporation of structural information but also guarantees the validity of inference, even in scenarios where the machine learning models are mis-specified. Finally, PLIS offers an intuitive, fast, and user-friendly computational algorithm that efficiently computes scores from the baseline data. This differs from the more computationally intensive algorithms proposed in \citet{lei18adapt} and \citet{ren23sideinfo}, which involve intricate techniques such as adaptive unmasking and boundary updating.

\section{Simulation}
\label{sec:simu}

This section presents simulation results that compare PLIS, which is implemented as described in the examples of Section \ref{subsec:eg}, with several competing methods. 
For all of the numerical experiments, we set $m=2000$ and $\alpha=0.05$. The observations are generated following this model: $X_{i}|\theta_{i}\overset{ind.}{\sim}(1-\theta_{i})\mathcal{N}(0,1)+\theta_{i}\mathcal{N}(\mu,1)$. The latent states $\pmb\Theta$ obey various dependence structures in different experiments. 
The following methods are considered in our analysis: (a) $\rm PLIS_{HM}$\footnote{In subsequent simulations, $\rm PLIS_{HM}$ is referred to as PLIS if $\rm PLIS_{TG}$ is not involved.}, implemented via Algorithm \ref{algo:MP}, utilizing the working model $\mathcal{M}_{\rm HM}$ as described in Section \ref{subsec:eg}; (b) BH \citep{BH95}; (c) {AdaDetect} \citep{marandon22mlmeetfdr}, where the density ratio is estimated by $\hat r(\cdot)=f_0(\cdot)/\hat f(\cdot)$, with $f_0$ being known and $\hat f(\cdot)$ being the standard kernel estimator based on $\{ \mathbf{X}, \mathbf{Y}\}$. (d) $\rm PLIS_{TG}$, implemented via Algorithm \ref{algo:MP} using the working model $\mathcal{M}_{\rm TG}$ as described in Section \ref{subsec:eg}. (e) AdaPT \citep{lei18adapt}, implemented using the \textbf{R} package \texttt{adaptMT}. Throughout all the simulations conducted, the FDR and Average Power (AP) levels are computed by averaging the results obtained from 200 replications.

\subsection{Comparisons with symmetric rules}\label{simu:sym}

We first consider HMMs in which $\pmb{\Theta}=(\theta_{i})_{i=1}^{m}$ is a binary Markov chain with transition matrix $\mathbf{A}=(a_{ij})_{i,j=0,1}=\left(\PP(\theta_{t+1}=j|\theta_{t}=i)\right)_{i,j=0,1}$. Here, we fix $a_{00}=0.95$, and the initial state of the latent chain is set to be $\theta_{1}=0$.  

\begin{figure}[htbp]
    \centering
    \includegraphics[width=1\linewidth]{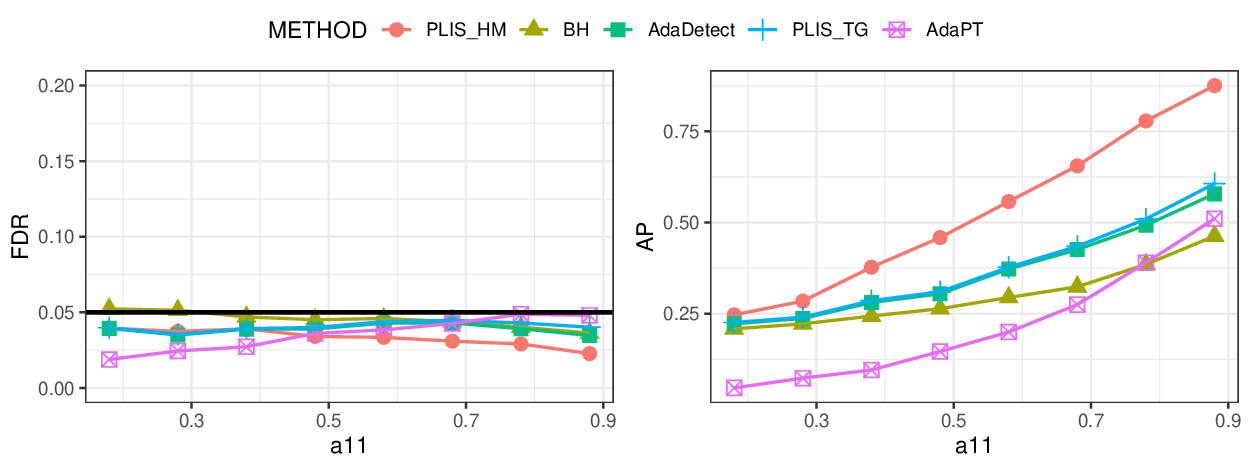}
    \caption{The comparison of FDR and AP for HMMs with $\mu=2.6$ and varying $a_{11}$. }\label{pic:hmma11}
\end{figure}

We apply BH, {AdaDetect}, AdaPT, $\rm PLIS_{HM}$ and $\rm PLIS_{TG}$ to the simulated data, and summarize the results in Figure \ref{pic:hmma11}. Our results demonstrate that all of the methods under consideration effectively control the FDR at the nominal level. Both the $\rm PLIS_{TG}$ and AdaDetect exhibit similar performance and outperform BH in terms of power. Notably, $\rm PLIS_{HM}$ exhibits the highest power among all five methods in most scenarios. This can be attributed to the use of asymmetric rules in $\rm PLIS_{HM}$. AdaPT outperforms/underperforms BH when $a_{11}$ becomes large/small. Additional simulation results for HMMs are presented in Section E.2 of the Supplement. 


\subsection{Comparisons beyond HMMs}\label{simu:beyondHMM}

\begin{figure}[!htbp]
    \centering
    \includegraphics[width=1\linewidth,height=0.32\linewidth]{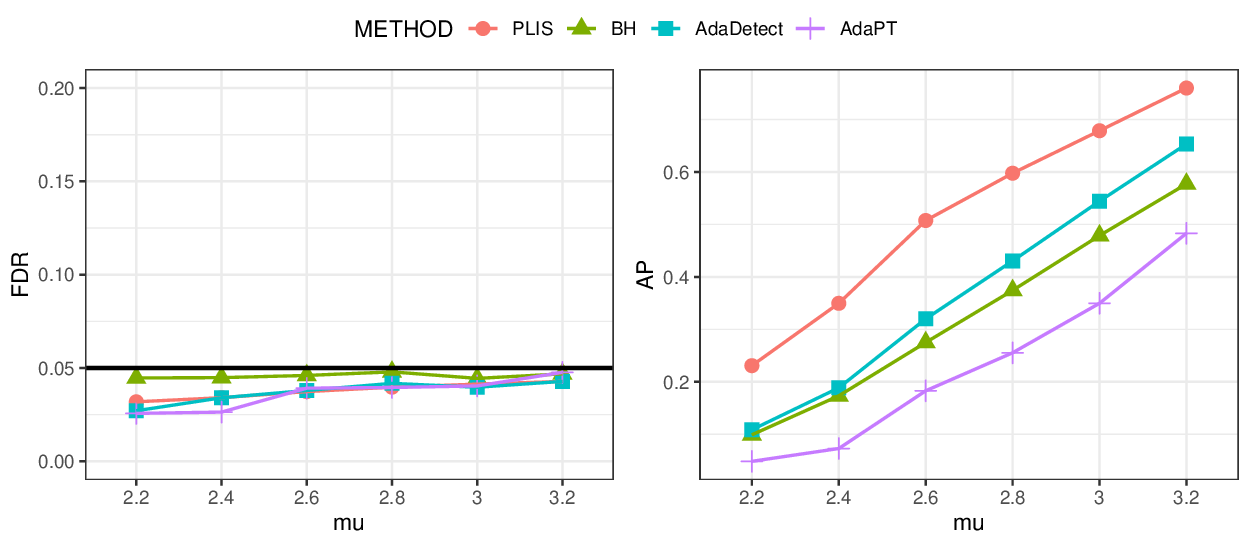}
    \caption{FDR and AP comparison for the heterogeneous HMM with exponentially vanishing $a_{11}$.}\label{pic:exp}
\end{figure}

This section examines the case where the true transition matrices change over time, which is commonly referred to as a heterogeneous HMM. Such models are particularly useful since the structure of real-world data is often non-stationary. Specifically, we define $\mathbf{A}^{(k)}=\left(a_{ij}^{(k)}\right)_{i,j=0,1}$ as the transition probability matrix for the $k$-th transition, where $a_{ij}^{(k)}=\PP(\theta_{k}=j|\theta_{k-1}=i)$, and $a_{00}^{(k)}=0.95$ is fixed for all $k$. The conditional distributions remain the same as in the preceding simulations. We consider a scenario in which $a_{11}^{(k)}=0.9e^{-k/1000}$ decreases with $k$ exponentially. This decreasing trend indicates that the stochastic system will gradually stabilize over time, with outliers becoming increasingly rare within the clusters. Since the data in this scenario possess a complex structure, accurately modeling them can be challenging. However, these data do exhibit patterns resembling HMMs, with moderate deviations from homogeneity conditions. Hence, we adopt a homogeneous HMM as a working model to capture the underlying structural patterns in the data and subsequently utilize the PLIS framework for inference. To assess the performance of PLIS and to compare it with other methods, specifically BH, AdaDetect, and AdaPT, we vary the parameter $\mu$ from 2.2 to 3.2 and apply these methods to simulated data. The results are summarized in Figure \ref{pic:exp}.

Our results demonstrate that all methods are effective in controlling the FDR, with PLIS exhibiting higher average power compared to other methods. This highlights the effectiveness of PLIS in leveraging structural information, even when wrong models and algorithms are employed. Additionally, our analysis of more general data structures, such as two-layer dynamic models and structured models generated from a renewal process, demonstrates that PLIS remains more powerful than the other methods in most cases (cf. Figure 8-12 in Section E of the Supplement). These findings underscore the versatility and robustness of PLIS in a variety of settings, suggesting its considerable potential for analyzing complex data structures in practice.

\subsection{Comparisons under a covariate-adaptive model}
\label{app:simu-cov}

Let $\{S_i: i\in\mathcal G\}$ denote the covariates that encode side information. Consider a covariate-adaptive model that can be described using a hierarchical approach:   
$
\theta_i|(S_i=s) \overset{ind.}{\sim} \text{Bernoulli}(\pi_s),     X_i|(S_i=s,\theta_i=0) \overset{i.i.d.}{\sim} F_0(x),  X_i|(S_i=s,\theta_i=1) \overset{ind.}{\sim} F_{1s}(x),
$
where $\mathbb{P}(\theta_i=1|S_i=s) = \pi_s$ represents the local sparsity level. An equivalent model, motivated from an empirical Bayes perspective, is given by $X_i|(S_i=s) \overset{ind.}{\sim} F_s(x) = (1-\pi_s)F_0(x) + \pi_s F_{1s}(x).$

Let $S_i=i$ for $i\in[3000]$ and fix $F_0=\mathcal{N}(0,1)$. We consider the following scenarios: 
(i) $F_{1s}=\mathcal{N}(\mu+0.2\sin(0.6s),1)$; $\pi_{s}=0.4(1+\sin(0.2s))$ for $s\in[201,500]\cup[801,1100]\cup[1501,1800]\cup[2101,2400]$, and $\pi_{s}=0.02$ otherwise; 
(ii) $F_{1s}=\mathcal{N}(2.8,1)$; $\pi_{s}=2\pi$ for $s\in[201,350]\cup[1501,1650]$, $\pi_{s}=\pi$ for $s\in[801,1000]\cup[2101,2300]$, and $\pi_{s}=0.02$ otherwise. Hence the external covariate $S_i$ captures the local dependence structure, with $\pi_s$ and $F_{1s}$ exhibiting similar values when $S_i$ values are comparable.

\begin{figure}[!htbp]
        \centering
        \includegraphics[width=1\linewidth]{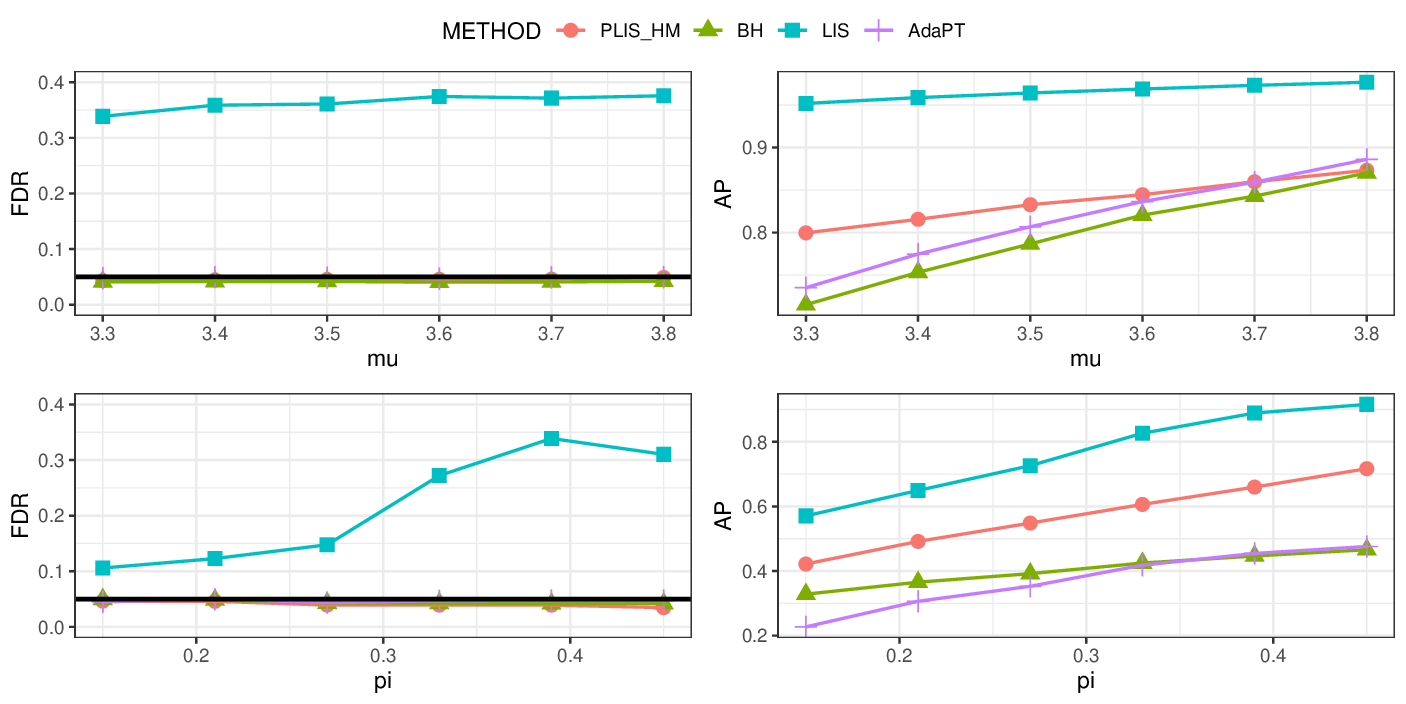}
        \caption{FDR and AP comparison under covariate-adaptive models.}
        \label{fig:PLIS_covariate}
    \end{figure}

 We apply PLIS (employing the HMM as its working model), LIS (\citealp{sc09}; using the HMM as its underlying true model), BH and AdaPT (\citealp{lei18adapt}; using the \textbf{R} package \texttt{adaptMT} with $(S_i)_{i=1}^m$ as its covariate sequence) at FDR level $\alpha=0.05$. The results are presented in Figure \ref{fig:PLIS_covariate}. The following patterns can be observed. First, AdaPT demonstrates increased efficiency as $S_i$ becomes more informative and may outperform both BH and PLIS in certain settings. Second, LIS, which mistakenly assumes that the data are generated from an HMM, suffers from severe FDR inflation. In contrast, PLIS, despite utilizing the same EM algorithm as LIS, remains effective in FDR control even when the true model significantly diverges from the working model.

\section{Application}
\label{sec:application}

The identification of eye states has wide-ranging applications across various fields. By monitoring changes in eye states, such as eyelid closures and eye movements, systems can alert drivers when they exhibit signs of drowsiness, thereby preventing potential accidents. In mental health diagnosis, analyzing eye states can offer valuable insights into conditions such as schizophrenia or autism spectrum disorder. Eye state information can also effectively improve facial recognition performance and human-computer interactions. 

This section illustrates semi-supervised PLIS for outlier detection using the Electroencephalogram (EEG) Eye State dataset, which is accessible at the UCI Machine Learning Repository (\url{https://archive.ics.uci.edu}). 
By conceptualizing the eye state sequence in a visual task as a Markovian process, the HMM provides a useful tool for modeling eye movement patterns, primarily due to its ability in capturing temporal dependencies within nearby latent states. In recent works such as \cite{chuk2020Eye} and \cite{hsiao2022Eye}, 
it is shown that HMMs can effectively represent the dynamic characteristics of eye movement behaviors during a range of cognitive activities, offering valuable insights into the interplay between eye movement patterns and individual differences in cognitive styles or abilities. Furthermore, these studies indicate that the HMM is an appropriate model for describing the local dependence structure within eye states.

During the data preparation process, the eye states of one individual were continuously recorded in chronological order, with detections and manual labeling denoted as 0/1 to signify the eye-open/eye-closed states. The analysis utilizes data collected from an experiment focusing on measuring the Frontal left Hemisphere (F7). The primary goal is to pinpoint any eye-closed states (outliers) within a sequence of 1500 consecutive records (test samples). 
After collecting the test dataset, the remaining samples of this measurement with label ``0" serve as the null samples. Concretely, we set aside the test samples and proceeded to generate the calibration samples (consisting of 1500 data points) and training samples by splitting the remaining data points with a label of ``0". Subsequently, we applied semi-supervised/conformal multiple testing methods, employing the null samples in both training and calibrating stages. The training data is also used to evaluate the null distribution and calculate $p$-values when implementing BH and AdaPT.


For the anomaly detection task, we implement various methods at the nominal FDR level of 0.05. The methods being considered include BH, AdaPT, AdaDetect, as well as the proposed semi-supervised PLIS (Algorithm \ref{algo:pu_plis}) with two distinct working models: the HMM and Efron's two-group model, respectively denoted as $\rm PLIS_{HM}$ and $\rm PLIS_{TG}$. The counts of discovered outliers and corresponding FDPs, computed based on the true labels of test samples and presented in parentheses, are provided for each method below: $\rm PLIS_{HM}$: 69 (0.014), AdaPT: 38 (0), AdaDetect/$\rm PLIS_{TG}$: 37 (0.054), and BH: 12 (0.083).  
Notably, the detected outliers are identical for AdaDetect and $\rm PLIS_{TG}$, corroborating the concluding remark in Example 2 of Section \ref{subsec:eg} regarding the similarity of AdaDetect and $\rm PLIS_{TG}$ when employing Efron's two-group model as the working model.

\begin{figure}[!htbp]
    \centering
    \includegraphics[width=1\linewidth]{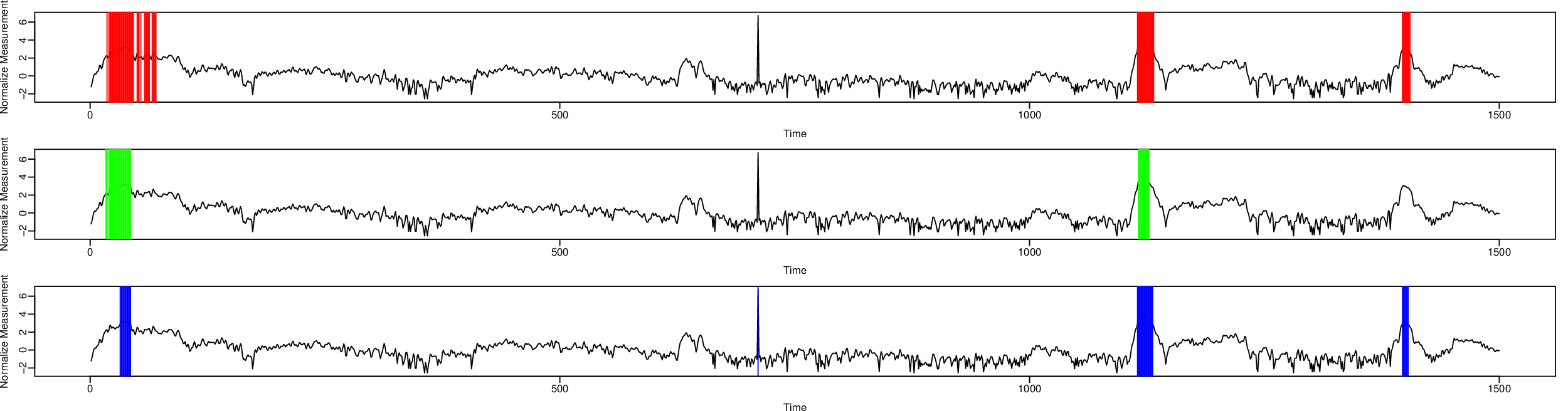}
    \caption{The discoveries of $\rm PLIS_{HM}$ (red, top), AdaPT (green, middle) and AdaDetect (blue, bottom) in detecting eye-closed states in 1500 consecutive records according to F7 measurements.}\label{pic:eye}
\end{figure}

To further examine the comparison results, we visualize the outliers identified by $\rm PLIS_{HM}$, AdaPT, and AdaDetect in Figure \ref{pic:eye}. The following patterns can be observed. First, AdaPT shows notable efficacy in detecting large outlier clusters compared to AdaDetect, which may be attributed to its successful integration of structural patterns, as encoded in the external covariate ``Time'', in the decision-making process. Second, AdaPT exhibits reduced sensitivity in cases involving smaller clusters, leading to the omission of the final segment where structural information is relatively weak. 
Third, AdaDetect seems to demonstrate superior effectiveness in identifying isolated outliers. However, it turns out that the discovery at the single location of 711 is indeed a false positive, with the true label being ``0’’. This aligns with our intuition: if a person closes the eyes, the closed state will persist for a certain period of time. Conversely, by leveraging the structural information, $\rm PLIS_{HM}$ and AdaPT successfully filter out this seemingly plausible signal. Finally,  $\rm PLIS_{HM}$ outperforms both AdaPT and AdaDetect by detecting the highest number of outliers while keeping the FDP well below the nominal level. This is made possible through the utilization of the HMM working model, which effectively exploits the informative structures within the data.


\singlespacing
\bibliography{BB}



\begin{center}\bf\Large
	Online Supplementary Material for ``False Discovery Rate Control For Structured Multiple Testing: Asymmetric Rules And Conformal $Q$-values''
\end{center}

\bigskip\bigskip

\renewcommand{\appendixname}{Appendix~\Alph{section}}

\appendix

	\renewcommand{\theequation}{\thesection.\arabic{equation}}

\setcounter{page}{1}
 \setcounter{equation}{0}
 \setcounter{figure}{5}

This supplement contains the proofs of theorems and propositions (Sections \ref{ap:proof} and \ref{ap:proof2}), some extensions of the proposed PLIS procedure (Sections \ref{ap:exch}), a comparative review of PLIS and related methods (Section \ref{ap:existing_procedure}), and additional numerical results (Section \ref{ap:simu}). 

\section{Proofs For Results in Section 2}\label{ap:proof}
In this section, proofs for theories in Section 2.4 and Section 2.7 are provided. And we further prove Theorem 2 for semi-supervised PLIS in Appendix \ref{ap:exch} after introducing de Finetti's Theorem (Lemma \ref{thm:definetti}).

\subsection{A general theory on pairwise exchangeability}

To begin, we introduce some notation. 

\begin{itemize}

\item The observed data $\mathbf{X}=(X_{j}:j\in\mathcal{G})$ and corresponding latent states $\pmb{\Theta}=(\theta_{j}:j\in\mathcal{G})$ are generated from Model (2). 

\item Denote the null samples $\mathbf{Y}=(Y_{j}:j\in\mathcal{G})$, where $Y_{j}\overset{i.i.d.}{\sim}f_{0}$. 

\item Let $\mathbf{W}=(W_{j}: j\in\mathcal{G})$ be a structured data set on the same graph $\mathcal{G}$. 

\item We use $\mathbf{W}$ as the baseline data and construct two new data sets for every $i\in\mathcal G$: $\tilde{\mathbf{X}}_{i}$ and $\tilde{\mathbf{Y}}_{i}$, by substituting $X_i$ and $Y_i$ in place of $W_i$ in $\mathbf{W}$, respectively.  

\item  Denote $s_{i}^{X}$($s_{i}^{Y}$) the scores computed based on $\tilde{\mathbf{X}}_{i}$ and $\tilde{\mathbf{Y}}_{i}$: $s_{i}^{X}=g_{i}(\tilde{\mathbf{X}}_{i})$ and $s_{i}^{Y}=g_{i}(\tilde{\mathbf{Y}}_{i})$ for $i\in\mathcal{G}$, where $\{g_{i}(\cdot):i\in\mathcal{G}\}$ are a class of functions that are $\mathbf{W}$-measurable (including non-random functions). 

\item Let $\mathbf{S}_{-i}=\{s_{j}^{X},s_{j}^{Y}: j\neq i, j\in\mathcal G\}$. 

\item Let $Z_{i}=\{X_{i},Y_{i}\}$ denote a set with two elements and $Z_{i}'=(X_{i},Y_{i})$ denote a 2-dimensional vector.

\item Denote $T_{i}=\{X_{i},Y_{i},W_{i}\}$ a set of variables with three elements and  $T_{i}'=(X_{i},Y_{i},W_{i})$ a 3-dimensional vector.

\end{itemize}

The following lemma provides a useful result on pairwise exchangeability.

\begin{lemma}\label{lemma1}
Suppose the following two conditions hold: (i) $\{T_{j}:j\in\mathcal{G}\}$ are mutually independent conditional on $\pmb{\Theta}$, and (ii) for $i\in\mathcal{H}_{0}$, $(X_{i},Y_{i},W_{i}) \overset{d}{=} (Y_{i},X_{i},W_{i})$. Then  
\begin{enumerate}[(a)]
    \item $\{Z_{j}=\{X_{j},Y_{j}\}:j\in\mathcal{G}\}$ are conditionally independent given $\mathbf{W}$ and $\pmb{\Theta}$, and for $i\in\mathcal{H}_{0}$, $(X_{i},Y_{i}|\mathbf{W},\pmb{\Theta})\overset{d}{=}(Y_{i},X_{i}|\mathbf{W},\pmb{\Theta})$.
    \item $\{\mathcal{D}_{j}=\{\tilde{\mathbf{X}}_{j},\tilde{\mathbf{Y}}_{j}\}:j\in\mathcal{G}\}$ are conditionally independent given $\mathbf{W}$ and $\pmb{\Theta}$, and for $i\in\mathcal{H}_{0}$, $(\tilde{\mathbf{X}}_{i}, \tilde{\mathbf{Y}}_{i}| \mathbf{W},\pmb{\Theta}) \overset{d}{=} (\tilde{\mathbf{Y}}_{i}, \tilde{\mathbf{X}}_{i}| \mathbf{W},\pmb{\Theta})$.
    \item Denote $\mathbf s_j=\{s_{j}^{X},s_{j}^{Y}\}$ for $j\in\mathcal{G}$. Then $\{\mathbf{s}_{j}:j\in\mathcal{G}\}$ are conditionally independent given $\mathbf{W}$ and $\pmb{\Theta}$, and for $i\in\mathcal{H}_{0}$, $s_{i}^{X}$ and $s_{i}^{Y}$ are exchangeable conditional on other scores and the latent states, i.e., $(s_{i}^{X},s_{i}^{Y}|\mathbf{S}_{-i},\pmb{\Theta}) \overset{d}{=} (s_{i}^{Y},s_{i}^{X}|\mathbf{S}_{-i},\pmb{\Theta})$, or equivalently, $(s_{i}^{X},s_{i}^{Y},\mathbf{S}_{-i}) \overset{d}{=} (s_{i}^{Y},s_{i}^{X},\mathbf{S}_{-i})$ given $\pmb{\Theta}$.
\end{enumerate}

\end{lemma}

\subsection{Proof of Lemma \ref{lemma1}}

Note that $T_{i}'=(Z_{i}',W_{i})$. Given Condition (i), we can conclude that $\{T_{j}':j\in\mathcal{G}\}$ are mutually independent conditional on $\pmb{\Theta}$. For $j\in\mathcal{G}$, let $A_{j}$ denote a Borel set on $\mathbb{R}^2$, and let $B_{j}$ denote a Borel set on $\mathbb{R}$. Additionally, let $B^{(m)}$ denote the Cartesian product of $\{B_{j}, j\in\mathcal G\}$. Then we have
\begin{equation*}
\begin{split}
   &  \PP (Z_{i}'\in A_{i},i\in\mathcal{G}|\mathbf{W}\in B^{(m)},\pmb{\Theta}) \\
    = & \frac{1}{\PP(\mathbf{W}\in B^{(m)}|\pmb{\Theta})} \PP (Z_{i}'\in A_{i},i\in\mathcal{G},\mathbf{W}\in B^{(m)}|\pmb{\Theta}) \\
    = &\frac{1}{\PP(\mathbf{W}\in B^{(m)}|\pmb{\Theta})} \PP (T_{i}'\in A_{i}\times B_{i},i\in\mathcal{G}|\pmb{\Theta}) \\
    = &\frac{1}{\PP(\mathbf{W}\in B^{(m)}|\pmb{\Theta})} \prod_{i\in\mathcal{G}} \PP (T_{i}'\in A_{i}\times B_{i}|\pmb{\Theta}) \\
    = &\frac{1}{\PP(\mathbf{W}\in B^{(m)}|\pmb{\Theta})} \prod_{i\in\mathcal{G}} \left[ \PP (Z_{i}'\in A_{i}|W_{i}\in B_{i},\pmb{\Theta}) \PP (W_{i}\in B_{i}|\pmb{\Theta}) \right]\\
    = &\frac{1}{\PP(\mathbf{W}\in B^{(m)}|\pmb{\Theta})} \left(\prod_{i\in\mathcal{G}}\PP (W_{i}\in B_{i}|\pmb{\Theta})\right) \left( \prod_{i\in\mathcal{G}} \PP (Z_{i}'\in A_{i}|\mathbf{W}\in B^{(m)},\pmb{\Theta})  \right)\\
    = &\prod_{i\in\mathcal{G}} \PP (Z_{i}'\in A_{i}|\mathbf{W}\in B^{(m)},\pmb{\Theta}).
\end{split}
\end{equation*}
Note that the preceding calculations apply to any Borel sets $A_{j}$ and $B_{j}$, which implies that $\{Z_{i}':i\in\mathcal{G}\}$ are conditionally independent given $\mathbf{W}$ and $\pmb{\Theta}$. This result leads to the conclusion that $\{Z_{i}:i\in\mathcal{G}\}$ are conditionally independent given $\mathbf{W}$ and $\pmb{\Theta}$. 

For $i\in\mathcal{H}_{0}$, since $(X_{i},Y_{i},W_{i})\overset{d}{=}(Y_{i},X_{i},W_{i})$, we have  
$$(X_{i},Y_{i}|\mathbf{W},\pmb{\Theta})\overset{d}{=}(X_{i},Y_{i}|W_{i},\pmb{\Theta})\overset{d}{=}(Y_{i},X_{i}|W_{i},\pmb{\Theta})\overset{d}{=}
(Y_{i},X_{i}|\mathbf{W},\pmb{\Theta}),
$$
where the first and last steps follow from Condition (i). 

Let $\mathcal{D}_{i}=\{\tilde{\mathbf{X}}_{i},\tilde{\mathbf{Y}}_{i}\}$, $\mathcal{D}=\{\mathcal{D}_{i}: i\in\mathcal{G} \}$, and $\mathcal{D}_{-i}=\mathcal{D}\backslash\mathcal{D}_{i}$. We assert that $\mathcal{D}_{i}$ is conditionally independent of $\mathcal{D}_{-i}$ given $\mathbf{W}$ and $\pmb{\Theta}$. This is due to the fact that, given $\mathbf{W}$ and $\pmb{\Theta}$, the randomness of the components in $\mathcal{D}$ solely arises from $\{Z_{i}:i\in\mathcal{G}\}$. Moreover, the elements $Z_{i}$ are conditionally independent, thereby establishing the conditional independence between $\mathcal{D}_{i}$ and $\mathcal{D}_{-i}$.

Now, let us examine the scores $s_{i}^{X}\equiv g_{i}(\tilde{\mathbf{X}}_{i})$ and $s_{i}^{Y}\equiv g_{i}(\tilde{\mathbf{Y}}_{i})$, $i \in\mathcal{G}$. Let $\mathbf{s}_{i}= \{s_{i}^{X},s_{i}^{Y}\}$. Based on the conditional independence between $\mathcal{D}_{i}$ and $\mathcal{D}_{-i}$ given $\mathbf{W}$ and $\pmb{\Theta}$, we infer that
\begin{equation} \label{s-ind}
\mbox{$m$ components of $\{\mathbf{s}_{i}:i\in\mathcal{G}\}$ are mutually independent conditional on $\mathbf{W}$ and $\pmb{\Theta}$.} 
\end{equation}
This result is due to the fact that $\mathbf{s}_{i}$ is a function of $\mathcal{D}_{i}$, and $g_{i}(\cdot)$ is $\mathbf{W}$-measurable.

For $i\in\mathcal{H}_{0}$, it holds that $(X_{i},Y_{i}|\mathbf{W},\pmb{\Theta})\overset{d}{=}(Y_{i},X_{i}|\mathbf{W},\pmb{\Theta})$. Consequently, we can deduce that $(\tilde{\mathbf{X}}_{i},\tilde{\mathbf{Y}}_{i}|\mathbf{W},\pmb{\Theta}) \overset{d}{=} (\tilde{\mathbf{Y}}_{i},\tilde{\mathbf{X}}_{i}|\mathbf{W},\pmb{\Theta})$. As a result, the scores $s_{i}^{X}$ and $s_{i}^{Y}$ must satisfy
\begin{equation}\label{sxy}
(s_{i}^{X}|\mathbf{W}, \pmb{\Theta})\overset{d}{=}(s_{i}^{Y}|\mathbf{W}, \pmb{\Theta}). 
\end{equation}
Let $\mathbf{S}_{-i}=\{\mathbf{s}_{j}:j\in\mathcal{G},j\neq i\}$. It can be shown that 
\begin{equation*}
    (s_{i}^{X},s_{i}^{Y}|\mathbf{S}_{-i},\mathbf{W},\pmb{\Theta}) \overset{d}{=} (s_{i}^{X},s_{i}^{Y}|\mathbf{W},\pmb{\Theta}) \overset{d}{=} (s_{i}^{Y},s_{i}^{X}|\mathbf{W},\pmb{\Theta}) \overset{d}{=} (s_{i}^{Y},s_{i}^{X}|\mathbf{S}_{-i},\mathbf{W},\pmb{\Theta}),
\end{equation*}
where the first and last steps follow from \eqref{s-ind}, and the second step follows from \eqref{sxy}.

After integrating out $\mathbf{W}$, we can obtain $(s_{i}^{X},s_{i}^{Y}|\mathbf{S}_{-i},\pmb{\Theta}) \overset{d}{=} (s_{i}^{Y},s_{i}^{X}|\mathbf{S}_{-i},\pmb{\Theta})$, which establishes the desired result on pairwise exchangeability:
$$
\mbox{$(s_{i}^{X},s_{i}^{Y},\mathbf{S}_{-i}|\pmb{\Theta}) \overset{d}{=} (s_{i}^{Y},s_{i}^{X},\mathbf{S}_{-i}|\pmb{\Theta})$ for $i\in\mathcal{H}_{0}$.}
$$

\subsection{Justifications of Properties 1-3}

Properties 1-3 are intermediate conclusions through the proof of Lemma \ref{lemma1}. Note that the scores $s_{i}^{X}$ and $s_{i}^{Y}$ defined in Algorithm 1 are specific examples of the scores defined in Lemma \ref{lemma1}. Specifically, Algorithm 1 takes $g_{i}(\cdot)=\mathbb{P}_{\mathcal{M,A}}(\theta_{i}=0|\cdot)$ for $i\in\mathcal{G}$. Hence the properties can be established by verifying the two conditions of Lemma \ref{lemma1}, for $W_i$ constructed based on a symmetric function  $h(x,y)=h(y,x)$ in Algorithm 1. 

To establish Condition (i), we consider any bivariate function $h$ and assume that $W_{i}=h(X_{i},Y_{i})$ for $i\in\mathcal{G}$. In this case, the conditional independence assumption of $\{T_{i}=\{X_{i},Y_{i},W_{i}\}:i\in\mathcal{G}\}$ holds trivially. This is due to the fact that $\{Z_{i}=\{X_{i},Y_{i}\}:i\in\mathcal{G}\}$ are mutually independent given $\pmb{\Theta}$, and $W_{i}$ is  a function of $Z_{i}$.
To establish Condition (ii), it should be noted that the function $h(x,y)$ defined by (5) is used to construct $\mathbf{W}$ in the PLIS procedure (Algorithm 1). As $h(x,y)$ is a symmetric function and $X_{i}$ and $Y_{i}$ are i.i.d. for any $i\in\mathcal{H}_{0}$, it follows that
\begin{equation*}
(X_{i},Y_{i},W_{i})=(X_{i},Y_{i},h(X_{i},Y_{i}))\overset{d}{=}(Y_{i},X_{i},h(Y_{i},X_{i})) = (Y_{i},X_{i},h(X_{i},Y_{i}))=(Y_{i},X_{i},W_{i}),
\end{equation*}
establishing Condition (ii).
Finally, Properties 1-3 on exchangeability can be justified by following the arguments in the proof of Lemma \ref{lemma1}.

\subsection{Proof of Theorem 1}

PLIS utilizes the threshold $\tau=\sup\{t \in\mathbf{S}_{X}\cup\mathbf{S}_{Y}:Q(t)\leq\alpha\}$, which guarantees that $Q(\tau)\leq \alpha$ holds true at all times. It follows that
\begin{equation*}
\begin{split}
    \mathrm{FDP}(\tau) &= \frac{\sum_{j\in\mathcal{G}^{r}\cap\mathcal{H}_{0}} \mathbb{I}\{s_{j}^{X}\leq \tau\} }{(\sum_{j\in\mathcal{G}^{r}}\mathbb{I}\{s_{j}^{X}\leq \tau\})\vee 1} \\
    &= Q(\tau)\frac{1+\sum_{j\in \mathcal{G}^{c}\cap\mathcal{H}_{0}}\mathbb{I}\{s_{j}^{Y}\leq \tau \}}{1+\sum_{j\in\mathcal{G}^{c}}\mathbb{I}\{s_{j}^{Y}\leq \tau \}}\frac{\sum_{j\in \mathcal{G}^{r}\cap\mathcal{H}_{0}}\mathbb{I}\{s_{j}^{X}\leq \tau \}}{1+\sum_{j\in \mathcal{G}^{c}\cap\mathcal{H}_{0}}\mathbb{I}\{s_{j}^{Y} \leq \tau\}} \\
    &\leq \alpha\cdot 1 \cdot \frac{\sum_{j\in \mathcal{G}^{r}\cap\mathcal{H}_{0}}\mathbb{I}\{s_{j}^{X}\leq \tau\}}{1+\sum_{j\in \mathcal{G}^{c}\cap\mathcal{H}_{0}}\mathbb{I}\{s_{j}^{Y}\leq \tau\}}.
\end{split}
\end{equation*}
Hence, to show $\mathrm{FDR}=\mathbb{E}[\mathrm{FDP}(\tau)] \leq \alpha$, we only need to prove that 
\begin{equation}\label{aim}
    \mathbb{E}\left[ \frac{\sum_{j\in \mathcal{G}^{r}\cap\mathcal{H}_{0}}\mathbb{I}\{s_{j}^{X}\leq \tau\}}{1+\sum_{j\in \mathcal{G}^{c}\cap\mathcal{H}_{0}}\mathbb{I}\{s_{j}^{Y}\leq \tau\}} \right] \leq 1.
\end{equation}
The frequentist FDR is considered in this context, where the expectation is taken over the observed data (including the test samples and labeled null samples) while holding the hidden states $\pmb\Theta$ fixed, or equivalently, regarding $\pmb\Theta$ as a given condition to calculate the conditional expectations. The inequality presented in \eqref{aim} will be established by leveraging well-established martingale theories. 

\subsubsection{An equivalent formulation of the mirror process}

For clarity of presentation, we assume without loss of generality that there are no ties between $s_{i}^X$ and $s_i^Y$. This assumption will not change the testing result. To see this, note that for $j\in\{i\in\mathcal{G}:s_{i}^X=s_{i}^Y\}$, $H_{0,j}$ will never be rejected, and the scores $s_{j}^X$ and $s_{j}^Y$ are not involved in the FDP estimator $Q(t)$ defined by (6), therefore removing the scores in $\{i\in\mathcal{G}:s_{i}^X=s_{i}^Y\}$ has no impact on the decision.

With the constructed scores, Algorithm 1 selects the threshold as 
\begin{eqnarray*}
\tau &  = &  \max\left\{ t\in\mathbf{S}_{X}\cup\mathbf{S}_{Y}: Q(t)=\frac{1+\sum_{j\in\mathcal{G}^c}\II\{s_j^Y \leq t\}}{\left[\sum_{j\in\mathcal{G}^r}\II\{s_j^X\leq t\}\right]\vee 1} \leq \alpha \right\}.
\end{eqnarray*}
This is equivalent, for the decision rule, to select the threshold
\begin{eqnarray*}
\tau& = & \max\left\{ t\in\{s_i^X\wedge s_i^Y\}_{i\in\mathcal{G}}: Q(t)=\frac{1+\sum_{j\in\mathcal{G}^c}\II\{s_j^Y \leq t\}}{\left[\sum_{j\in\mathcal{G}^r}\II\{s_j^X\leq t\}\right]\vee 1} \leq \alpha \right\}.
\end{eqnarray*}
It follows that the search for threshold can be confined within the set $\{s_i^X\wedge s_i^Y\}_{i\in\mathcal{G}}$, because the function $Q(t)$ only jumps at the points of $\{s_i^X\wedge s_i^Y\}_{i\in\mathcal{G}}$.

Let $S_{i}=s_{i}^X\wedge{s}_{i}^Y$ and $\xi_{i}=\II\{s_{i}^X<{s}_{i}^Y\}$. Since $\PP(s_{i}^X={s}_{i}^Y)=0$, we have $\II\{s_{i}^Y<s_{i}^X\}=1-\xi_{i}$ almost surely for all $i\in\mathcal{G}$. Noete that
$\xi_j=1$ and $S_j=s_j^X$ for $j\in\mathcal{G}^{r}$. Likewise, $\xi_j=0$ and $S_j=s_j^Y$ for $j\in\mathcal{G}^{c}$. We can rewrite $Q(t)$ as
\begin{equation*}
    Q(t)= \frac{1+\sum_{j\in\mathcal{G}} (1-\xi_{j})\II\{S_j\leq t\} }{ [\sum_{j\in\mathcal{G}} \xi_{j} \II\{S_j\leq t\}]\vee1 },
\end{equation*}

Denote the ordered statistics by $S_{(1)}\leq\cdots\leq S_{(m)}$, where $m=|\mathcal{G}|$. We further have that $\tau=S_{(\hat{k})}$, where
\begin{equation}\label{anotherstop}
    \hat{k}=\max\{i\in[m]:Q(S_{(i)})\leq\alpha\}.
\end{equation}

It follows that to prove \eqref{aim}, we only need to prove the following:
\begin{equation}\label{newaim}
    \EE \left[ \frac{\sum_{j\in\mathcal{H}_0}\xi_{j}\II\{S_j\leq S_{(\hat{k})}\}}{1+\sum_{j\in\mathcal{H}_0} (1-\xi_{j})\II\{S_j\leq S_{(\hat{k})}\} } \right] \leq 1
\end{equation}

To enhance clarity in future discussions, we introduce the mapping $\rho:[m]\to\mathcal{G}$ as follows: for every $k\in[m]$, $\rho(k)=l$ if $|S_{(k)}|=|S_{l}|$. Therefore we say an ordered score $S_{(i)}$ corresponds to a null (non-null) case if $\rho(i)\in\mathcal{H}_{0}$ ($\rho(i)\notin\mathcal{H}_{0}$).

\subsubsection{Martingale arguments}

Let $\mathcal{K}=\sigma\left( \{S_{i}:i\in\mathcal{G}\}, \{\xi_{i}:i\notin\mathcal{H}_0\} \right)$. This $\sigma$-algebra encapsulates the information on score magnitudes but cannot distinguish whether the null scores are from the calibration set or test set. It will become evident that conditioning on $\mathcal{K}$ plays a crucial role in the application of the coin-flipping lemma  in \cite{barber15knockoff}. Next, consider the filtration $\mathcal{F}=(\mathcal{F}_k:k\in[m])$ generated by $\mathcal{F}_{k}=\sigma\left( \mathcal{K} \cup \sigma(V_{j}^X,{V}_{j}^Y:k\leq j \leq m) \right)$, where 
$$
V_{j}^X=\sum_{l\in\mathcal{H}_0}\xi_{l}\II\{S_{l}\leq S_{(j)}\}\; \mbox{and}\; {V}_j^Y = \sum_{l\in\mathcal{H}_0}(1-\xi_{l})\II\{S_{l}\leq S_{(j)}\}
$$
count the numbers of incorrect rejections in the two mirror processes, respectively.  
It is easy to see that $0\leq V_{j}^X\leq V_{j+1}^X \text{ and } 0\leq{V}_{j}^Y\leq{V}_{j+1}^Y$, and that $\mathcal{F}_{k}$ is decreasing in $k$ in the sense that $\mathcal{F}_{k+1}\subset\mathcal{F}_{k}$, $k\in[m-1]$. 

Next we show that the following random process
$
\left(M_{i}=\frac{V_i^X}{1+{V}_i^Y}: i\in[m]\right)
$
is a backward super-martingale with respect to the filtration $\mathcal{F}$, i.e.,
\begin{equation}\label{tower-exp}
    \EE[M_{i}|\mathcal{F}_{i+1}] \leq M_{i+1} ,\quad \forall i\in[m-1].
\end{equation}
To establish \eqref{tower-exp}, we need the following flipping-coin property. 
\begin{lemma}[Lemma 1 of \cite{barber15knockoff}]
\label{lem:barbercandes}
For any anti-symmetric function $f(x,y)$ satisfying $f(x,y)=-f(y,x)$, if the scores $(s_{i}^{X}:i\in\mathcal{G})$ and $(s_{i}^{Y}:i\in\mathcal{G})$ are pairwise exchangeable under the null, i.e., (7) holds, then $(\mathrm{sign}(f(s_{i}^{X},s_{i}^{Y})):i\in\mathcal{H}_{0})$ are i.i.d. coin flips conditional on $(|f(s_{i}^{X},s_{i}^{Y})|:i\in\mathcal{G})$.
\end{lemma}

By Lemma \ref{lem:barbercandes}, we can conclude that 
\begin{equation}\label{flipcoin}
        (\xi_{i}:i\in\mathcal{H}_0)\overset{i.i.d.}{\sim}\mbox{Binom}(1,1/2) \text{ conditional on } (S_i:i\in\mathcal{G}).
\end{equation}

We start verifying \eqref{flipcoin} by considering the sign of $s_i^X-s_i^Y$, indicated by $\{-1, 1\}$, which can also be expressed as $1-2\xi_i$. Define the following anti-symmetric function $f(x,y)$
    $$f(x,y)=\mathrm{sign}(x-y)(x\wedge y).$$
Then $S_i=|h(s_i^X,{s}_i^Y)|$, $1-2\xi_i=\mathrm{sign}(s_i^X-s_i^Y)=\mathrm{sign}(f(s_i^X,{s}_i^Y))$. 
By Lemma \ref{lem:barbercandes}, we have $\{(1-2\xi_i):i\in\mathcal{H}_0\}$ are i.i.d. coin flips conditional on $(S_i:i\in\mathcal{G})$. Thus \eqref{flipcoin} holds, i.e.
\begin{eqnarray*}
\PP(\xi_{i}=0|S_{i}:i\in\mathcal{G}) & = & \PP(1-2\xi_i=1|S_{i}:i\in\mathcal{G})=1/2; \\
\PP(\xi_{i}=1|S_{i}:i\in\mathcal{G}) & = & \PP(1-2\xi_{i}=-1|S_{i}:i\in\mathcal{G})=1/2. 
\end{eqnarray*}

To evaluate the conditional expectation $\EE[M_{k}|\mathcal{F}_{k+1}]$, we consider two situations.

Situation 1: $V_{k+1}^{Y}=0$.  In this case, we have that $V_k^Y=V_{k+1}^Y=0$, $M_{k}=V_{k}^{X}$, and $M_{k+1}=V_{k+1}^X$. It follows that $M_{k} \leq M_{k+1}$ almost surely. Therefore, we must have 
$$
\EE[M_{k}|\mathcal{F}_{k+1}] \leq \EE[M_{k+1}|\mathcal{F}_{k+1}] = M_{k+1}.
$$

Situation 2: $V^{Y}_{k+1}\neq 0$. If $S_{(k+1)}$ is a non-null score, i.e., $\rho(k+1)\notin\mathcal{H}_{0}$, we have that ${V}^Y_{k}={V}^Y_{k+1}$ and ${V}^X_{k}={V}^X_{k+1}$. Hence $M_{k}=M_{k+1}$ and \eqref{tower-exp} holds trivially. If $S_{(k+1)}$ is a null score, i.e., $\rho(k+1)\in\mathcal{H}_{0}$, only one of $V^X_{t}$ and ${V}^Y_{t}$ will decrease by one when $t$ decreases from $k+1$ to $k$. By \eqref{flipcoin}, we have that 
    \begin{equation*}
            \begin{split}
                \PP \Big(V^X_{k}=V^X_{k+1}-1,{V}^Y_{k}={V}^Y_{k+1} \Big|\mathcal{F}_{k+1} \Big) &= \frac{V^X_{k+1}}{V^X_{k+1}+{V}^Y_{k+1}}, \\
                \PP \Big(V^X_{k}=V^X_{k+1},{V}^Y_{k}={V}^Y_{k+1}-1 \Big|\mathcal{F}_{k+1} \Big) &= \frac{{V}^Y_{k+1}}{V^X_{k+1}+{V}^Y_{k+1}}.
            \end{split}
        \end{equation*}
It follows that the conditional expectation can be calculated as
        \begin{equation*}
            \begin{split}
                \EE[M_{k}|\mathcal{F}_{k+1}] 
                =& \frac{V^X_{k+1}-1}{1+{V}^Y_{k+1}} \cdot \frac{V^X_{k+1}}{V^X_{k+1}+{V}^Y_{k+1}} + \frac{V^X_{k+1}}{1+V^Y_{k+1}-1} \cdot \frac{{V}^Y_{k+1}}{V^X_{k+1}+{V}^Y_{k+1}}\\
                =& \frac{V^X_{k+1}}{V^X_{k+1}+{V}^Y_{k+1}}\left(\frac{V^X_{k+1}-1}{1+{V}^Y_{k+1}} +1 \right)
                = \frac{V^X_{k+1}}{1+{V}^Y_{k+1}} = M_{k+1}.
            \end{split}
        \end{equation*}

Therefore \eqref{tower-exp} holds in both situations, which establishes that $(M_{i}:i\in[m])$ is a super-martingale with respect to the backward filtration $\mathcal{F}$. 

\subsubsection{Stopping time arguments}
By the definition of the threshold calculated according to \eqref{anotherstop}, it is easy to see that $\hat{k}$ is an $\mathcal{F}$-stopping time, as knowing $\{\xi_{j}:j\notin\mathcal{H}_{0}\}$, $\{S_{j}:j\in\mathcal{G}\}$ and $\{V^X_{i}, {V}^Y_{i}: K\leq i\leq m\}$ is sufficient to determine whether the event $\{\hat{k}=K\}$ occurs. Therefore, we can apply Doob's stopping time theorem on $(M_{i}:i\in[m])$ and $\hat{k}$ to show that
    $$\EE \left[ \frac{\sum_{j\in\mathcal{H}_0}\xi_{j}\II\{S_j\leq S_{(\hat{k})}\}}{1+\sum_{j\in\mathcal{H}_0} (1-\xi_{j})\II\{S_j\leq S_{(\hat{k})}\} } \right]=\EE[M_{\hat{k}}]\leq\EE[M_m]=\EE\left[ \frac{\sum_{j\in\mathcal{H}_{0}} \xi_{j}}{1+\sum_{j\in\mathcal{H}_{0}} (1-\xi_{j})} \right].$$

    By \eqref{flipcoin}, we have that $\sum_{j\in\mathcal{H}_{0}} \xi_{j}\sim \mbox{Binom}(|\mathcal{H}_0|,1/2)$ conditional on $(S_i:i\in\mathcal{G})$. For a Binomial variable $B\sim \mbox{Binom}(n,1/2)$, we have that 
    \begin{equation*}
    \begin{split}
        \EE\Big[ \frac{B}{1+n-B} \Big] &= \sum_{j=1}^{n} \PP(B=i) \frac{j}{1+n-j}
        = \Big(\frac{1}{2}\Big)^{n} \sum_{j=1}^{n} \frac{n!}{(n-j)!j!} \frac{j}{1+n-j} \\
        &= \Big(\frac{1}{2}\Big)^{n} \sum_{j=1}^{n} \frac{n!}{(n+1-j)!(j-1)!} 
        = \sum_{j=1}^{n} \PP(B=i-1)\\
        &= 1-\Big(\frac{1}{2}\Big)^{n}\leq 1.
    \end{split}
    \end{equation*}

Therefore, we can conclude that
    \begin{eqnarray*}
        \EE[M_{\hat{k}}] & \leq &  \EE\left[ \frac{\sum_{j\in\mathcal{H}_{0}} \xi_{j}}{1+\sum_{j\in\mathcal{H}_{0}} (1-\xi_{j})} \right]
        \\ &  =  & \EE\left\{ \EE\Big[ \frac{\sum_{j\in\mathcal{H}_{0}} \xi_{j}}{1+\sum_{j\in\mathcal{H}_{0}} (1-\xi_{j})} \Big| S_i:i\in\mathcal{G} \Big] \right\}
        \\ &  \leq & 1,
    \end{eqnarray*}
which establishes \eqref{newaim} and completes the proof of Theorem 1.\qed

\subsection{Proof of Proposition 1}
\begin{proof} 
First, for each $\delta_{i}$, if $\delta_{i}=\II\{s_{i}^{X}\leq\tau\}\II\{s_{i}^{X}<s_{i}^{Y}\}=1$, we must have $i\in\mathcal{G}^{r}=\{i\in\mathcal{G}:s_{i}^{X}<s_{i}^{Y}\}$ and $s_{i}^X\leq\tau$. By the definition of $\tau$, we have that
$$
q_{i}=\min_{t\in\mathbf{S}_{X}\cup\mathbf{S}_{Y},t \geq s_{i}^{X}} Q(t)\leq Q(\tau)\leq \alpha.
$$
It follows that $\II\{q_{i}\leq\alpha \}=1$. 

Next, if $\II\{q_{i}\leq\alpha \}=1$, then we have that $i\in\mathcal{G}^{r}$, and $\min_{t \geq s_{i}^{X}} Q(t)\leq \alpha$. Since $Q(t)$ may only jump at anywhere in a subset of $\mathbf{S}_{X}\cup\mathbf{S}_{Y}$, it follows that 
$$
\tau=\sup\{t\in\mathbf{S}_{X}\cup\mathbf{S}_{Y}: Q(t)\leq\alpha \}\geq s_{i}^{X},
$$ 
implying that $\II\{s_{i}^{X}\leq\tau \}\II\{s_{i}^{X}<s_{i}^{Y}\}=1$. 

Combining the two arguments above, we conclude that $\delta_{i}=\II\{s_{i}^{X}\leq\tau \}\II\{s_{i}^{X}<s_{i}^{Y}\}=\II\{q_{i}\leq\alpha \}$ for $i\in\mathcal{G}$, establishing the equivalence of the two algorithms. 
\end{proof}

\section{Proofs For Results in Section 3} \label{ap:proof2}
\subsection{Proof of Theorem 3}

\begin{proof}

To prove Theorem 3, we only need to show that $\mathbb{E} \big[ \sum_{j \in \mathcal{H}_{0} }e_{j} \big]\leq m$. Following the steps outlined in the proof of Theorem 1, we have
$$    \mathbb{E} \big[ \sum_{j \in \mathcal{H}_{0} }e_{j} \big] = m \mathbb{E} \left[ \sum_{j \in \mathcal{H}_{0}} \frac{\II\{s_{j}^{X}\leq\tau\}\II\{j\in\mathcal{G}^{r}\} }{1+\sum_{j\in\mathcal{G}^{c}} \mathbb{I}\{ s_{j}^{Y}\leq \tau\}}  \right] \leq m\mathbb{E}\left[ \frac{\sum_{j\in \mathcal{G}^{r}\cap\mathcal{H}_{0}}\mathbb{I}\{s_{j}^{X}\leq \tau\}}{1+\sum_{j\in \mathcal{G}^{c}\cap\mathcal{H}_{0}}\mathbb{I}\{s_{j}^{Y}\leq \tau\}} \right] \leq m,
$$
completing the proof.
\end{proof}

\subsection{Proof of Proposition 2}

\begin{proof}
Let $R=|\mathcal{R}|$. By the definition of $\tau$, we have 
$$
\frac{1+\sum_{j\in\mathcal{G}^{c}}\mathbb{I}\{s_{j}^{Y}\leq \tau\}}{R}\leq\alpha.
$$ 
It follows that for $j\in\mathcal{R}\subset\mathcal{G}^{r}$,
$$    e_{j} = \frac{m \mathbb{I}\{ s_{j}^{X}\leq \tau\} }{1+\sum_{j\in\mathcal{G}^{c}} \mathbb{I}\{ s_{j}^{Y}\leq \tau\}} \geq \frac{m}{\alpha R}.
$$
Therefore, $\hat{k} = \max\{i:e_{(i)}\geq\frac{m}{\alpha i}\} \geq R$, which implies that $j\in \mathcal{R}_{ebh}$.

Conversely, if $j\notin \mathcal{R}$ and $e_{j}=0$, then $j$ cannot be selected by the e-BH procedure and hence $j\notin \mathcal{R}_{ebh}$.

Finally, combining the arguments in two directions, we reach our desired conclusion that $\mathcal{R}=\mathcal{R}_{ebh}$.
\end{proof}

\section{De Finetti’s theorem and semi-supervised PLIS} \label{ap:exch}

This section extends the scope of Theorem 1 to encompass the semi-supervised framework, wherein labeled null data is available in place of an explicit null distribution. Additionally, our extended theory relaxes the conditional independence assumption {given $\pmb\Theta$}, stipulated by model (2), allowing for the generation of data with possibly correlated noise.

\subsection{Exchangeability and de Finetti’s theorem}

A collection of random variables $\{\xi_{i}: i\in[n]\}$ is considered (jointly) exchangeable if for every permutation $(\tau_1, \cdots, \tau_n)$ of the indices $\{1, \cdots, n\}$,
$$
(\xi_1, \cdots, \xi_n)\overset{d}{=}(\xi_{\tau_1}, \cdots, \xi_{\tau_n}).
$$
The following de Finetti's theorem provides a powerful analytical framework for studying the properties of exchangeable random variables \citep{schervish2012theory, durrett2019probability}.
\begin{lemma}[De Finetti’s Theorem]\label{thm:definetti}
    If random variables $\{\xi_{i}: i\in[n]\}$ are exchangeable, then $\{\xi_{i}: i\in[n]\}$ are i.i.d. with respect to some conditional probability measure, i.e., there exists random variable $\eta$ such that
    \begin{equation}
        F (\xi_{1},\cdots,\xi_{n})=\int \prod_{i=1}^{n}F^{*}(\xi_{i}|\eta)dQ(\eta),
    \end{equation}
where $F(\cdot)$ stands for the joint cumulative distribution function (CDF) of $(\xi_{1},\cdots,\xi_{n})$, $F^{*}(\cdot|\eta)$ is the conditional CDF of $\xi_{i}$ given $\eta$, and $Q(\cdot)$ is the CDF of $\eta$.
\end{lemma}

Now we discuss the connection of de Finetti’s Theorem with structured probabilistic models (2). Firstly, it is easy to show that data points that are i.i.d. under the null are exchangeable in a marginal sense (without conditioning on $\pmb{\Theta}$). This simple result follows from Example 1.14 of \citet{schervish2012theory}. If the conditioning variable $\eta$ corresponds to the true states  $\pmb{\Theta}$, then the conditional independence assumption in Model (2) yields the exchangeability assumption employed in existing works in conformal inference. 
More importantly, by de Finetti's Theorem, exchangeable random elements must be i.i.d. with respect to some probability measure, or equivalently, there must exist a latent random variable $\eta$ such that the random elements are i.i.d. conditional on $\eta$. From this perspective, the assumption for Model (2) is no more stringent than those commonly found in existing works within the conformal inference literature.

\subsection{Theories on semi-supervised PLIS for exchangeable data}

This section presents theory that establishes the validity of the semi-supervised PLIS procedure (Algorithm 2) for FDR control. The proof of Theorem 2 (b) follows from the result established in Theorem 2 (a), by utilizing the same argument in proving Theorem 1. Due to the similarity and overlap of the arguments, the detailed exposition of the proof for part (b) is omitted. To establish part (a) of the theorem, we invoke de Finetti's Theorem, along with similar techniques employed in proving Properties 1-3.

\begin{proof}[Proofs of Theorem 2 (a)]

Our approach follows a similar strategy employed in the proofs of Lemma \ref{lemma1} and Properties 1-3. Accordingly, we adopt the  notations there and refrain from reiterating identical arguments. As the frequentist FDR is adopted, the following arguments are established regarding the true states $\pmb\Theta$ as a given condition.

According to de Finetti's theorem, if $\{U_{1},\cdots,U_{n},X_{i},i\in\mathcal{H}_{0}\}$ are exchangeable conditional on $\{X_{i}:i\notin\mathcal{H}_{0}\}$, then there exists a random $\eta$ such that $\{U_{1},\cdots,U_{n},X_{i},i\in\mathcal{H}_{0}\}$ are i.i.d. conditional on $(\eta, X_{i}:i\notin\mathcal{H}_{0})$.
Let $\mathbf{Y}_{0}=(Y_{i}:i\in\mathcal{H}_{0})\subset\mathbf{Y}\subset\mathbf{U}$ be the calibration data assigned to the null test units. Denote $\mathcal{C}=\sigma(\mathbf{U}\setminus\mathbf{Y}_{0},\eta,X_{i}:i\notin\mathcal{H}_{0})$. It follows that the null data points $\{X_{i}:i\in\mathcal{H}_{0}\}\cup\{Y_{i}:i\in\mathcal{H}_{0}\}$ are i.i.d. conditional on $\mathcal{C}$. 
The process in constructing $\mathbf{W}$ implies that $\{W_{j}:j\notin\mathcal{H}_{0}\}\in\mathcal{C}$, and $(X_{i},Y_{i},W_{i})\overset{d}{=}(X_{i},Y_{i},W_{i})$ conditional on $\mathcal{C}$ for $i\in\mathcal{H}_{0}$. 
Denote $\mathcal{C}^{+}=\mathcal{C}\cup\sigma(\mathbf{W})$ and $Z_{j}=\{X_{j},Y_{j}\}$. Following similar arguments in the proof of Lemma \ref{lemma1}, we can show that $\{Z_{i}:i\in\mathcal{H}_{0}\}$ are mutually independent conditional on $\mathcal{C}^{+}$. 

Upon the examination of the process of constructing the conformity scores, it is evident that $s_{j}^{X},s_{j}^{Y}\in\mathcal{C}^{+}$ for all $j\notin\mathcal{H}_{0}$. Meanwhile, given the information in $\mathcal{C}^{+}$, the randomness of  $s_{i}^{X}$ and $s_{i}^{Y}$  originates from the randomness of  $X_{i}$ and $Y_{i}$ for $i\in\mathcal{H}_{0}$. Additionally, $(s_{i}^{X},s_{i}^{Y})$ is independent of $\{(s_{k}^{X},s_{k}^{Y}):k\in\mathcal{G}\setminus\{i\}\}$. If $X_{i}$ and $Y_{i}$ are exchangeable conditional on $\mathcal{C}^{+}$, then we have
\begin{equation*}
    (s_{i}^{X},s_{i}^{Y}|\mathbf{S}_{-i},\mathcal{C}^{+}) \overset{d}{=} (s_{i}^{X},s_{i}^{Y}|\mathcal{C}^{+}) \overset{d}{=} (s_{i}^{Y},s_{i}^{X}|\mathcal{C}^{+}) \overset{d}{=} (s_{i}^{Y},s_{i}^{X}|\mathbf{S}_{-i},\mathcal{C}^{+}).
\end{equation*}
By integrating out $\mathcal{C}^{+}$, the desired conclusion is established.
\end{proof}

\section{A Comparative Review of PLIS and Related Works}
\label{ap:existing_procedure}
The primary objective of this section is to provide a comprehensive understanding of the existing works in this area, while highlighting the novel contributions that our work brings to the field. To achieve this goal, we begin with a review of the SeqStep+, Selective SeqStep+, and knockoff+ procedures, drawing from the work of \citet{barber15knockoff}. Subsequently, we present an alternative derivation of PLIS as a modified BH procedure utilizing conformal $p$-values \citep{bates21testing}, building on the insights gained from \cite{marandon22mlmeetfdr}. Finally, we conclude the section with a discussion of the unique features that distinguish PLIS from related works. To simplify the notations, we make the assumption that $\mathcal{G}=[m]$ throughout this section.
 
\subsection{The Selective SeqStep+ algorithm and its variations}
    
{\bf SeqStep+:} Consider a set of valid $p$-values $p_{1},\cdots,p_{m}$. Assuming that the null $p$-values are i.i.d. among themselves, and are independent from the non-null $p$-values, the SeqStep+ algorithm can be described as follows. For a fixed value of $c\in(0,1)$ and a subset $K\subset[m]$, define the threshold
\begin{equation}
\hat{k} = \max \left\{ k\in K:\frac{1+\#\{j\leq k: p_{j}>c\}}{1+k}\leq(1-c)\alpha \right\}.
\end{equation}
The algorithm then proceeds by rejecting  $H_{0,j}$ for all $j\leq\hat{k}$.

\medskip
        
 {\bf Selective SeqStep+:} With the same notations for SeqStep+, the Selective SeqStep+ defines the threshold as
        \begin{equation}
        \hat{k} = \max \left\{ k\in K:\frac{1+\#\{j\leq k: p_{j}>c\}}{\#\{j\leq k:p_{j}\leq c\}\vee1}\leq\frac{1-c}{c}\alpha \right\}, 
        \end{equation}
        and rejects $H_{0,j}$ for all $j\leq\hat{k}$ such that $p_{j}\leq c$. It is shown in \citet{barber15knockoff} that both {Selective SeqStep+} and {SeqStep+} control the FDR. Furthermore, the Selective SeqStep+ algorithm serves as the prototype algorithm for both the knockoff filter \citep{barber15knockoff} and the AdaPT procedure \citep{lei18adapt}, exemplifying its significance in covariate-adaptive testing.

\medskip

{\bf Knockoff+:} The knockoff+ algorithm \citep{barber15knockoff} for variable selection can be viewed as a specific instance of the Selective SeqStep+ algorithm, with $c=1/2$, and 1-bit $p$-values calculated by 
        \begin{equation}\label{1-bitpv}
           p_{j}:=\left\{\begin{matrix}
           1/2& \text{if $T_{j}>0$}\\
           1& \text{if $T_{j}<0$} 
               \end{matrix}\right. ,
        \end{equation}
        
where $T_j = T_j(Z_j,  \widetilde{Z}_j)$ is the feature importance statistic, constructed via an anti-symmetric function of the pair of scores $Z_j$ and $\widetilde{Z}_j$, a typical choice is $T_{j}=Z_j - \widetilde{Z}_j$. If $Z_j$ and $\widetilde{Z}_j$ are pairwise exchangeable under the null, \citet{barber15knockoff} proved that the null 1-bit $p$-values, as defined in \eqref{1-bitpv}, are i.i.d. conditional on $\{|T_j |:j\in[m]\}$, revealing that they are jointly exchangeable according to de Finetti’s Theorem. The implementation of Selective SeqStep+ in this case is equivalent to setting
\begin{equation}\label{ko+}
T=\inf\left\{ t\in\{|T_{i}|\}_{i=1}^{m}:\frac{1+\#\{j:T_{j}\leq -t\}}{\#\{j:T_{j}\geq t\}\vee1}\leq\alpha \right\},
\end{equation}
and rejecting for $T_{j}\geq T$, which precisely recovers the operation for the Knockoff+ algorithm. The proof for FDR control is based on the joint exchangeability between  $p_{j}$s under the null.
        
\subsection{Conformal p-values and the BH algorithm}
   
To understand how PLIS can be derived as a conformalized BH procedure, we first revisit the definition of conformal $p$-values \citep{bates21testing}, which serves as the first key component in revealing the connection of PLIS to BH. Using the notations in our paper, let $s_i^X$ and $s_i^Y$ represent scores for test data and calibration data, respectively. The $m=|\mathcal{G}|$ conformal $p$-values can be constructed as
\begin{equation}\label{cp}
p_{i}:= \frac{1+\#\{j\in[m]:s_{i}^{X}> s_{j}^{Y}\}}{1+m}, \mbox{ for } i\in[m]. 
\end{equation}
According to \citet{bates21testing}, if the null scores $\{s_1^Y,\cdots,s_m^Y,s_i^X:i\in\mathcal{H}_0\}$ are exchangeable conditional on $\{s_i^X:i\notin\mathcal{H}_0\}$, then the conformal $p$-values \eqref{cp} are super-uniform and fulfill the PRDS property. This ensures that implementing BH with these conformal $p$-values controls the FDR at the nominal level.

The second key component in revealing the connection between PLIS and BH is provided by \citet{mary22semi} and \citet{marandon22mlmeetfdr}. Concretely, following the arguments in \citet{mary22semi}, it is easy to see that the BH procedure with conformal $p$-values, as defined in \eqref{cp}, is equivalent to the following algorithm, which works by first computing
\begin{equation}\label{bhconfp}
\hat{t} = \sup\left\{ t\in\{s_{i}^{X}\}_{i=1}^{m}: \frac{\frac{1}{1+m} [1+\sum_{j=1}^{m}\II\{s_{j}^{Y}\leq t\}]}{\frac{1}{m} \sum_{j=1}^{m}\II\{s_{j}^{X}\leq t\} } \leq \alpha \right\},
\end{equation}
and then rejecting for $s_{j}^{X}\leq\hat{t}$. 

\subsection{PLIS vs conformal BH}\label{subsec:PLIS-CBH}

Upon examining formulation \eqref{bhconfp}, the connection between our PLIS algorithm and the BH procedure becomes evident. We assume, without loss of generality, that there are almost surely no ties between $s_{j}^{X}$ and $s_{j}^{Y}$, i.e. $\PP(s_{j}^{X}=s_{j}^{Y})=0$ for all $j\in\mathcal{G}$. We proceed by explaining how \eqref{bhconfp} leads to our proposal in a step-by-step manner. First note that merely following the construction in \eqref{bhconfp} can be problematic. The primary issue is that the methodology requires the null scores to be jointly exchangeable. This makes it impossible to deploy asymmetric rules, which defeats the purpose of our original goal in developing the PLIS procedure. Specifically, conformity scores learned from structured models are no longer jointly exchangeable under the null, rendering the conformal $p$-values in \eqref{cp} invalid. To address this issue, we suggest a modification to the method by eliminating the step involving conformal $p$-values and instead constructing a mirror process to make decisions directly. Specifically, we reject $H_{0,i}$ if $s_{j}^{X}\leq \tilde\tau$, where 
\begin{equation} \label{confq_origin}
           \tilde \tau = \sup\left\{ t\in\{s_{i}^{X}\}_{i=1}^{m}: \frac{1+\sum_{j=1}^{m}\II\{s_{j}^{Y}\leq t\}}{1\vee(\sum_{j=1}^{m}\II\{s_{j}^{X}\leq t\}) } \leq \alpha \right\}.
\end{equation}

This method has been mentioned in Section 3.3 and referred to as $\mbox{PLIS}_{\rm cbh}$ due to its close connection with the conformal BH procedure. Unlike \eqref{bhconfp}, $\mbox{PLIS}_{\rm cbh}$ eliminates the factor $\frac{m}{1+m}$ on the left-hand side of the inequality. The procedure in \eqref{confq_origin} can be roughly seen as equivalent to the BH procedure with improper $p$-values calculated as follows:
$$
p_{i}^{+} := \frac{1+\#\{j\in[m]:s_{i}^{X}> s_{j}^{Y}\}}{m}, \quad \mbox{for } i\in[m].
$$

When all null scores $\{s_j^X:j\in\mathcal{H}_0\}$ and $\{s_{j}^Y:j\in[m]\}$ are exchangeable conditional on non-null scores $\{s_j^X:j\notin\mathcal{H}_0\}$, the conformal BH method \eqref{bhconfp} controls FDR around $\frac{|\mathcal{H}_0|}{m}\alpha=\pi_0\alpha$ (cf. Corollary 1 of \citet{bates21testing} and Corollary 3.5 of \citet{marandon22mlmeetfdr}). And it is valid to employ the adaptive BH procedure as a way of mitigating this conservativeness, which firstly estimates the null proportion $\hat{\pi}_0$ and then implements the conformal BH method at nominal level $\alpha/\hat{\pi}_0$ (cf. Theorem 3 of \citet{bates21testing} Corollary 3.7 of \citet{marandon22mlmeetfdr}).

However, a significant limitation of the mirror process \eqref{confq_origin} is its tendency to be excessively conservative. Since the scores constructed by PLIS are no longer jointly exchangeable under the null, it is challenging to estimate $\pi_0$ to overcome the conservativeness. Moreover, when the working models incorporate structural information, the conservativeness issue is exacerbated. To see this, consider the numerator in \eqref{confq_origin}
$$
1+\sum_{j\in\mathcal{H}_0}\II\{s_{j}^Y\leq t\}+\sum_{j\notin\mathcal{H}_0}\II\{s_{j}^Y\leq t\}.
$$
By the pairwise exchangeability between $s_i^X$ and $s_i^Y$ for $i\in\mathcal{H}_0$, the term $\sum_{j\in\mathcal{H}_0}\II\{s_{j}^Y\leq t\}$ can estimate the number of false discoveries $\sum_{j\in\mathcal{H}_0}\II\{s_{j}^X\leq t\}$, while the term $\sum_{j\notin\mathcal{H}_0}\II\{s_{j}^Y\leq t\}$ results in the over-estimation. When the working model considers dependencies, such as HMM dependence, the calibration scores $s_j^Y$ will be shrunk towards zero for $j\notin\mathcal{H}_0$ as the model and algorithm leverage information from relevant observations. The shrinkage phenomenon occurs even when all calibration data points follow the null distribution. Consequently, the count $\sum_{j\notin\mathcal{H}_0}\II\{s_{j}^Y<t\}$ can become very large, resulting in an overly conservative estimate of the FDP.

To reduce the impact of the undesirable term $\sum_{j\notin\mathcal{H}_0}\II\{s_{j}^Y<t\}$, we propose to restrict the rejections to the subset $\mathcal{G}^r=\{i\in\mathcal{G}:s_i^X<s_i^Y\}$ only, with the corresponding calibration set modified to $\mathcal{G}^c=\{i\in\mathcal{G}:s_i^Y<s_i^X\}$. This adjustment can (a) remove a substantial portion of calibration scores at non-null nodes in the numerator of $Q(t)$, while (b) minimally impacting the power. The goals of (a) and (b) can be effectively achieved if we assume that, for each $j\notin\mathcal{H}_0$, $s_{j}^X<s_j^Y$ holds with high probability; this assumption tends to be reasonable with the proper choice of baseline data and working model and considering the fact that $Y_j$ is a null sample. Specifically, the assumption implies that (a) $\sum_{j\in\mathcal{G}^c}\II\{s_{j}^{Y}\leq t\}$ is much smaller than $\sum_{j\in\mathcal{G}}\II\{s_{j}^{Y}\leq t\}$, and (b) the count $\sum_{j\in\mathcal{G}^r}\II\{s_{j}^{X}\leq t \}$ would be similar to the count $\sum_{j\in\mathcal{G}}\II\{s_{j}^{X}\leq t \}$ for a suitable $t$ that is not too large. This leads to our proposed PLIS procedure, which rejects $H_{0,j}$ only when $s_{j}^{X}\leq \tau$ and $j\in\mathcal{G}^r$, with 
\begin{equation} 
\label{confq}
\tau = \sup\left\{t\in\{s_{i}^{X}\}_{i\in\mathcal{G}}\cup\{s_{i}^{Y}\}_{i\in\mathcal{G}}: \frac{1+\sum_{j\in\mathcal{G}^c}\II\{s_{j}^{Y}\leq t\}}{1\vee(\sum_{j\in\mathcal{G}^r}\II\{s_{j}^{X}\leq t \}) } \leq \alpha \right\}. 
\end{equation}

\begin{remark}\rm{
It is important to note that the PLIS procedure in \eqref{confq} can adaptively achieve the target FDR level with increased signal magnitude, as demonstrated in Figure \ref{pic:knockoff} of Section \ref{simu:sym-PLIS}. This adaptability is particularly appealing as it eliminates the need to estimate the null proportion in existing conformal methods \citep{yang21bonus,marandon22mlmeetfdr,bates21testing}, a task that can be challenging in scenarios where the joint exchangeability between null scores fails to hold. As we shall explain shortly, the two conformal methods outlined in \eqref{confq_origin} and \eqref{confq} respectively employ BH and Barber-Cand{\`e}s (BC, or Selective SeqStep+) as their base algorithms. Storey's adjustment for correcting $\pi_0$ can be applied with \eqref{confq_origin} but is no longer needed in \eqref{confq}, as PLIS enjoys the adaptivity of the BC algorithm in attaining the nominal FDR level [the property was initially noted in Appendix B of \citet{barber15knockoff}]. }
\end{remark}

\subsection{PLIS vs knockoff filters}\label{PLIS-knockoff}

PLIS is designed for a distinct scenario compared to the knockoff filter used in variable selection within regression settings, where the dependency between the constructed scores $(Z_1,\cdots,Z_p,\Tilde{Z}_1,\cdots,\Tilde{Z}_p)$ is determined by the correlation between the original and knockoff features. In our semi-supervised setup, the initial null samples are assumed to be jointly exchangeable. Upon integrating structural patterns through user-defined working models, the null scores become only pairwise exchangeable. 

As established in the next subsection, the FDP process in PLIS can be derived as a symmetrized procedure with carefully designed anti-symmetrized statistics. We give several remarks before proving the result. Firstly, when constructing the baseline data $\mathbf{W}=(W_{j})$, it is crucial that each $W_{j}$ is a symmetric function of $X_{j}$ and $Y_{j}$ to guarantee the pairwise exchangeability between null scores. By using the novel bivariate function (5), the structural information of the test data can be effectively preserved in the baseline data. 
Secondly, in the construction of the mirror process, the candidate rejection set $\mathcal{G}^{r}=\{i\in\mathcal{G}:s_{i}^{X}<s_{i}^{Y}\}$ and the mirror calibration set $\mathcal{G}^{c}=\{i\in\mathcal{G}:s_{i}^{Y}<s_{i}^{X}\}$ are disjoint sets that possess symmetry properties, i.e.  $\PP(s_{i}^{X}<s_{i}^{Y})=\PP(s_{i}^{Y}<s_{i}^{X})$ for $i\in\mathcal{H}_{0}$. The introduction of $\mathcal{G}^{r}$ and $\mathcal{G}^{c}$ allows for less conservative estimation of the FDP, while preserving the ranking of the original conformity scores $s_i^X$.

\subsection{Connection of the FDP process \eqref{confq} to the Selective SeqStep+ algorithm}\label{PLIS-symmetrized}

In this section, we prove that the FDP process \eqref{confq} is equivalent to the Selective SeqStep+ algorithm in the form of \eqref{ko+} with a carefully designed anti-symmetric statistic suggested by an insightful referee. 

\begin{proof} To start with, define the following class of anti-symmetric statistics:
\begin{equation}\label{S_xy}
       T^S_{j}= T^S(s_{j}^{X},s_{j}^{Y}) = \mathrm{sign}(s_{j}^{Y}-s_{j}^{X})\cdot [g(s_{j}^{X})\vee g(s_{j}^{Y})],\quad \forall j\in\mathcal{G},
    \end{equation}
where $g(\cdot):\mathbb{R}_{\geq0}\to\mathbb{R}_{\geq0}$ is a non-random strictly decreasing function. Consider the following mirror process
\begin{equation*}
    Q^{S}(t)=\frac{1+\sum_{j\in\mathcal{G}} \II\{T^S_{j}\leq -t\}  }{(\sum_{j\in\mathcal{G}} \II\{T^S_{j}\geq t\})\vee1}, \quad t>0.
\end{equation*}
Define $\tau'= \inf\{t\in\mathcal{T}^S:Q^{S}(t)\leq \alpha\}$, where $\mathcal{T}^S=\{|T^S_{j}|:j\in\mathcal{G}\}$.
Consider a decision rule ${\pmb\delta}^\prime=\{\delta_j^\prime: j\in\mathcal G\}$, where $\delta_j^\prime=\II\{T^S_{j}\geq \tau'\}$, then ${\pmb\delta}^\prime$ is equivalent to $\pmb{\delta}=\{\delta_j: j\in\mathcal G\}$ output by Algorithm 1.

For a non-random strictly decreasing function $g(\cdot)$ defined on $\mathbb{R}_{\geq0}\to\mathbb{R}_{\geq0}$, the value $g(s_{i}^{X})$ can be interpreted as a non-conformity score, with a higher value indicating stronger evidence against  $H_{0,i}$.  As such $g(\cdot)$ is bijective, we have 
\begin{eqnarray*}
\mathcal{G}^{r} & = & \{i\in\mathcal{G}:s_{i}^{X}<s_{i}^{Y}\}=\{i\in\mathcal{G}:g(s_{i}^{X})>g(s_{i}^{Y})\}, \mbox{ and} \\
\mathcal{G}^{c} & = & \{i\in\mathcal{G}:s_{i}^{Y}<s_{i}^{X}\}=\{i\in\mathcal{G}:g(s_{i}^{Y})>g(s_{i}^{X})\}.
\end{eqnarray*}
Consider the decision $\delta_{i}$ output by Algorithm 1. We have 
$\delta_{i}=\II\{s_{i}^{X}\leq\tau\}\II\{i\in\mathcal{G}^r\} =\II\{g(s_{i}^{X})\geq\tau'\}\II\{g(s_{i}^{X})>g(s_{i}^{Y})\}$, where
$$
\tau'=\inf\left\{t\in\{g(s_{i}^{X})\}_{i\in\mathcal{G}^{r}}\cup\{g(s_{i}^{Y})\}_{i\in\mathcal{G}^{c}}: \frac{1+\sum_{j\in\mathcal{G}} \II\{g(s_{j}^{Y})\geq t\}\II\{s_{j}^{Y}<s_{j}^{X}\} }{(\sum_{j\in\mathcal{G}} \II\{g(s_{j}^{X})\geq t\}\II\{s_{j}^{X}<s_{j}^{Y}\})\vee1 } \leq \alpha\right\}.
$$
This holds because $g(\cdot)$ is strictly decreasing, and that the function 
$$
\frac{1+\sum_{j\in\mathcal{G}} \II\{g(s_{j}^{Y})\geq t\}\II\{s_{j}^{Y}<s_{j}^{X}\} }{(\sum_{j\in\mathcal{G}} \II\{g(s_{j}^{X})\geq t\}\II\{s_{j}^{X}<s_{j}^{Y}\})\vee1 }
$$ 
only jumps at points within the set $\{g(s_{i}^{X})\}_{i\in\mathcal{G}^{r}}\cup\{g(s_{i}^{Y})\}_{i\in\mathcal{G}^{c}}$. By the definition of $T^S_{j}$ in \eqref{S_xy}, we have that, for any $t>0$:
\begin{eqnarray*}
T^S_{j}\geq t & \Longleftrightarrow  & s_{j}^{X}<s_{j}^{Y}  \mbox{ and }  g(s_{j}^{X})\geq t, \\
T^S_{j}\leq -t & \Longleftrightarrow & s_{j}^{Y}<s_{j}^{X}  \mbox{ and } g(s_{j}^{Y})\geq t.
\end{eqnarray*}
It follows that
$$
\frac{1+\sum_{j\in\mathcal{G}} \II\{g(s_{j}^{Y})\geq t\}\II\{s_{j}^{Y}<s_{j}^{X}\} }{(\sum_{j\in\mathcal{G}} \II\{g(s_{j}^{X})\geq t\}\II\{s_{j}^{X}<s_{j}^{Y}\})\vee1 } = \frac{1+\sum_{j\in\mathcal{G}} \II\{T^S_{j}\leq -t\}  }{(\sum_{j\in\mathcal{G}} \II\{T^S_{j}\geq t\})\vee1} = Q^{S}(t), \quad t>0.
$$
It is easy to see that $\mathcal{T}^S=\{g(s_{i}^{X})\}_{i\in\mathcal{G}^{r}}\cup\{g(s_{i}^{Y})\}_{i\in\mathcal{G}^{c}}$. Therefore we have $\tau'=\inf\{t\in\mathcal{T}^S:Q^{S}(t)\leq\alpha\}$ and
$\delta_{i}=\II\{s_{i}^{X}\leq\tau\}\II\{i\in\mathcal{G}^r\}=\II\{T^S_{i}\geq\tau'\}=\delta_i^\prime,$ completing the proof.

\end{proof}

\subsection{PLIS vs AdaDetect}

In our comparison of PLIS and AdaDetect, we concentrate on the situation where Efron's two-groups model (3) is employed as the working model. 

In the second example of Section 2.5, we discuss the implementation details of PLIS, which involves estimating conformity scores to emulate the oracle procedure  by \citet{sc07}. Specifically, we calculate the density ratio $r(\cdot)=f_0(\cdot)/\hat f(\cdot)$, where $f_0$ is  known, and $\hat f$ is obtained as a standard kernel density estimator. Likewise, the AdaDetect procedure \citep{marandon22mlmeetfdr} also first employs Efron's two-groups model as the working model, and then develops a novel score function, which remains permutation-invariant across both the test and calibration data, to emulate the oracle procedure by \citet{sc07}. These scores are finally utilized to construct conformal $p$-values and implement the BH procedure. 

Simulation results, as shown in Figure 2 in Section 4.1 and Figure \ref{pic:hmmmu} in Section \ref{simu-HMM:mu} of the Supplement, demonstrate the closely comparable numerical performance of PLIS (denoted as $\rm PLIS_{TG}$) and AdaDetect. However, it is important to note that, while the null scores in AdaDetect exhibit joint exchangeability, the scores in PLIS only maintain pairwise exchangeability under the null hypothesis. This distinction, although inconsequential when employing the two-group model as the working model, becomes more prominent when the working model is an HMM. We elaborate the difference between pairwise exchangeability and joint exchangeability in the next subsection.  

Finally, it is crucial to highlight that the similarity in numerical performance between PLIS and AdaDetect is limited to this particular two-group working model. In other numerical studies, PLIS demonstrates fundamental differences from other methods when alternative working models, such as an HMM, are utilized. 

\subsection{Pairwise exchangeability vs. joint exchangeability}\label{subsec:pw-jt-exch}

Due to the way in which the baseline data $\mathbf{W}$ is constructed, the pair of scores $s_{i}^{X}$ and $s_{i}^{Y}$ are exchangeable under the null, i.e., $s_{i}^{X}\overset{d}{=}s_{i}^{Y}$ conditional on other scores $\mathbf{S}_{-i}$ if $i\in\mathcal H_0$. However, this pairwise exchangeability does not imply the joint exchangeability of all null scores 
$\{s_{i}^X, i\in \mathcal H_0; s_{j}^Y, j\in\mathcal D^{cal}\}.$

The failure of joint exchangeability can be established through an argument by contradiction. Concretely, we start with two assertions: (a) $s_{i}^{X}$ and $s_{i}^{Y}$ are conditionally correlated given $(\mathbf{W},\pmb\Theta)$ due to the fact that the same data $h(X_i, Y_i)$ has been used to compute both scores, and (b)  the pair $(s_{i}^{X}, s_{i}^{Y})$ is conditionally independent of other scores $\{(s_{j}^{X}, s_{j}^{Y}): j\neq i\}$ given $(\mathbf{W},\pmb\Theta)$, as established by Property 3 in our paper. We conclude that the null scores $\{s_{i}^X, i\in \mathcal H_0; s_{j}^Y, j\in\mathcal D^{cal}\}$ cannot be jointly exchangeable, as accepting such a joint exchangeability condition implies that the null scores within the pair and across the pair must be permutation-invariant, which contradicts the established assertions (a) and (b). Specifically, (a) and (b) together demonstrate that the null scores within the pair and across the pair behave differently, in the sense that the interdependency structures in the two situations are different.

We present a toy example to further explain why the joint exchangeability may fail. Consider four independent identically distributed (i.i.d.) random variables, $\xi_{1},\xi_{2},\xi_{3},$ and $\xi_{4}$, where $z_{1}=\max(\xi_{1},\xi_{2}),z_{2}=\max(\xi_{3},\xi_{4})$ are the ``baseline data". Considering four scores $s_{1}=\xi_{1}+z_{2}, s_{2}=\xi_{2}+z_{2}, s_{3}=\xi_{3}+z_{1},$ and $s_{4}=\xi_{4}+z_{1}$, we demonstrate that although the scores are marginally identically distributed, they are not jointly exchangeable. To see this, note that conditional on $(z_{1},z_{2})$, $(s_{1},s_{2})$ and $(s_{3},s_{4})$ are independent. However, $s_{1}$ and $s_2$ are not independent since $z_{1}$ is a function of both $\xi_{1}$ and $\xi_2$. Furthermore, $Cov(s_{1},s_{2})=Var(z_{2})$, while $Cov(s_{1},s_{3})=Cov(\xi_{1},z_{1})+Cov(\xi_3,z_{2})=2Cov(\xi_{1},z_{1})$. This indicates that $(s_{1},s_{2},s_{3},s_{4})$ cannot be jointly exchangeable. 

A useful approach, as suggested by \citet{marandon22mlmeetfdr}, to verify the joint exchangeability of scores involves determining whether the score function is permutation-invariant regarding all test data. This can also be accomplished by imagining a ``parallel universe'', as proposed by \cite{liang2022integrative},  in which we can examine whether the scores obtained from the original dataset and a shuffled dataset remain the same. When permuting the test data $X_{i}$ while keeping the calibration data $Y_{i}$ unchanged, it is important to note that this alteration affects the baseline data due to the fact that $W_{i}=h(X_{i},Y_{i})\neq h(X_{j},Y_{i})$ for $X_{i}\neq X_{j}$. Since the estimation of PLIS relies on the baseline data $\mathbf{W}=(W_{i})$, exchanging $X_{i}$ and $X_{j}$ results in a change in the estimated scores. As a consequence, this failure to maintain joint exchangeability becomes evident.

\section{Additional Numerical Results}\label{ap:simu}

This section presents supplementary simulation results comparing the performance of various methods under different scenarios. Specifically, we first compare PLIS with the LIS pocedure \citep{sc09} under under model misspecification in Section \ref{simu-mis-specify}, and then investigate the behavior of the methods under Hidden Markov Models in Section \ref{simu-HMM:mu}, under general dependence structures in Section \ref{simu-more-dep}. {Numerical results for semi-supervised PLIS beyond Model (2) are provided in Section \ref{simu-puplis}-\ref{simu-puplis2}.} Numerical results on the Derandomized PLIS procedure are provided in Section \ref{simu-dePLIS}. Section \ref{ap:simu-baseline} compares different baseline data constructions in PLIS. And Section \ref{simu:sym-PLIS} contains numerical comparisons for different ways to evaluate the FDP with the same constructed PLIS scores.

\subsection{Comparisons with LIS under model misspecification}\label{simu-mis-specify}
\begin{figure}[!htbp]
    \centering
    \includegraphics[width=1\linewidth]{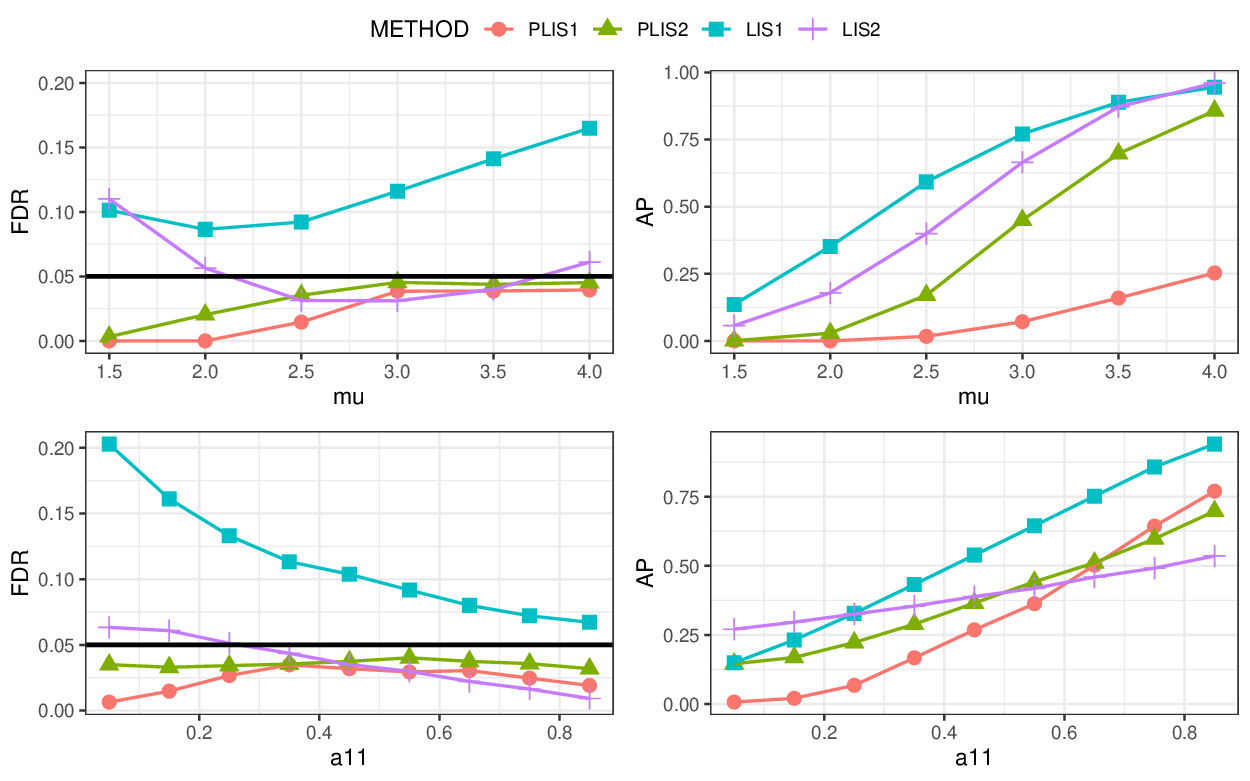}
    \caption{FDR and AP comparison when the HMM is misspecified. The null density $f_{0}$ is of $\mathcal{N}(0,1)$ and the signal's density $f_{1}$ is of $\mathcal{N}(\mu,1)$. }\label{pic:hmmmis}
\end{figure}

In this section, we examine the robustness of PLIS and LIS. The latter has been shown to be asymptotically optimal for controlling the FDR in HMMs. We consider data generated from an underlying HMM in which $\pmb\Theta$ forms a Markov chain with a transition matrix $\mathbf{A}=(a_{ij})_{i,j=0,1}$, where $a_{00}=0.95$, and the alternative distribution $F_1$ is $\mathcal N(\mu, 1)$. We consider two situations where the HMM parameters are mis-specified as follows: (a) $a_{11}=0.7$ and $F_1=\mathcal{N}(1.8,1)$, and (b) $a_{11}=0.3$ and $F_1=\mathcal{N}(3.6,1)$. We apply both PLIS and LIS with the mis-specified parameters, resulting in procedures labeled as PLIS1, PLIS2, LIS1, and LIS2. The average results from 200 replications are summarized in Figure \ref{pic:hmmmis}. In the top row, we fix $a_{11}=0.5$ and let $\mu$ vary. In the bottom row, we fix $\mu=2.5$ and let $a_{11}$ vary. Our results demonstrate that PLIS effectively controls the FDR in all situations, whereas LIS fails to control the FDR in many cases. This is because the validity of LIS requires the estimated HMM parameters to be strongly consistent, whereas PLIS is capable of controlling the FDR even in cases where the HMM parameters are mis-specified or estimated poorly.  

\subsection{HMMs with varying $\mu$}\label{simu-HMM:mu}

\begin{figure}[htbp]
    \centering
    \includegraphics[width=1\linewidth]{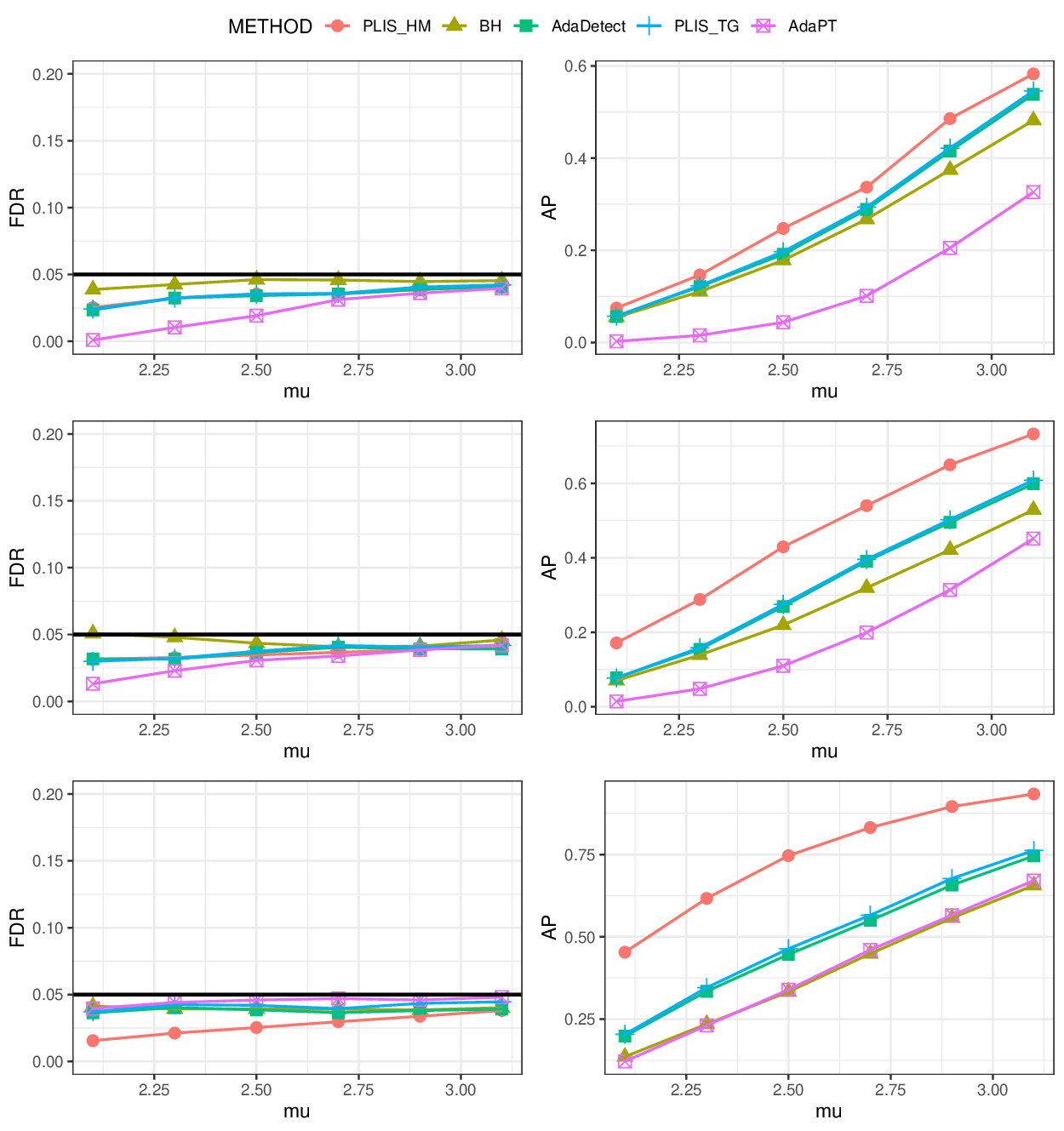}
    \caption{FDR and AP comparison for HMMs with varying $\mu$. }\label{pic:hmmmu}
\end{figure}

Consider the HMMs where $(\theta_{i})_{i=1}^{m}$ is a binary Markov chain with transition matrix $\mathbf{A}=(a_{ij})_{i,j=0,1}=(\PP(\theta_{t+1}=j|\theta_{t}=i))_{i,j=0,1}$, where $a_{00}=0.95$ is fixed and the initial state of the latent chain is $\theta_{1}=0$. We fix $a_{11}$ and vary $\mu$. We apply $\rm PLIS_{HM}$, BH, AdaDetect, $\rm PLIS_{TG}$ and AdaPT to the simulated data. The simulation results are summarized in Figure \ref{pic:hmmmu}.  

In Figure \ref{pic:hmmmu}, the top, middle, and bottom rows correspond to different values of $a_{11}$ chosen from the set $\{0.3, 0.5, 0.8\}$, respectively. We observe that all the methods under consideration control the FDR at the nominal level. However, it is worth noting that the $\rm PLIS_{HM}$ procedure exhibits conservative in FDR levels for small $\mu$, while the conservativeness is eliminated adaptively as $\mu$ grows. Nonetheless it still has the highest average power among all the methods by effectively exploiting the structural patterns in the data. 

\subsection{More results on general dependence structures}\label{simu-more-dep}

In this section, we extend the discussion in Section 4.2 of the main text to include additional examples of structured multiple testing under different dependence structures. Notably, each of the three models considered here belongs to the class of structured probabilistic models (2), but deviates from the HMM in distinct ways. Despite these differences, all the models exhibit clustering patterns that can be approximated by HMMs to some extent. In practice, we employ the HMM as the working model to capture the underlying structural patterns and utilize the PLIS framework, which is guaranteed to make valid inference under the conditions specified in (2). 

In the following, we describe the experimental setups for each of the three examples, summarize the results using figures, and finally draw observations and conclusions based on the patterns observed in the figures.

\begin{figure}[!htbp]
    \centering
    \includegraphics[width=1\linewidth]{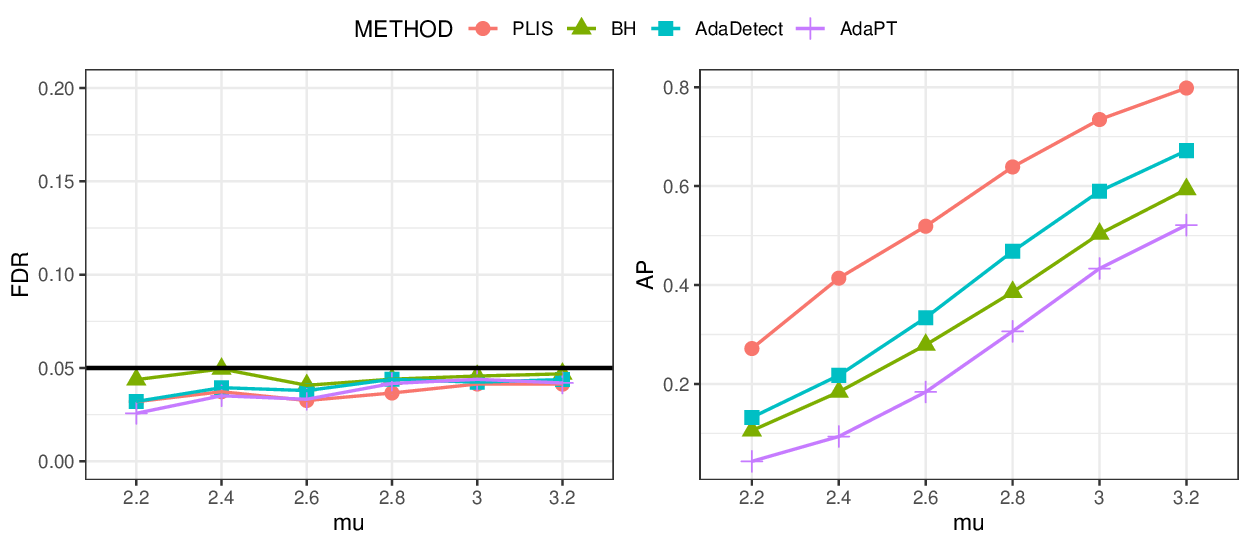}
    \caption{FDR and AP comparison when the model is a periodic heterogeneous HMM. Here $a_{11}^{(k)}=0.4\left(1+\sin\frac{k}{100} \right)$.}\label{pic:period}
\end{figure}

\begin{example}\rm{
\textbf{Heterogeneous HMMs}. The setup is similar to that of Section \ref{simu-HMM:mu}, which considers a conditional independence model with hidden states that follow a binary Markov chain: $X_{i}|\theta_{i}\overset{ind.}{\sim}(1-\theta_{i})\mathcal{N}(0,1)+\theta_{i}\mathcal{N}(\mu,1)$. However, we introduce a key modification by allowing the transition probability to vary over time. Specifically, we set $a_{11}^{(k)}=0.4\left(1+\sin\frac{k}{100} \right)$, where $k$ denotes the time index. This choice of transition probability results in a cyclical pattern in the data, which is often observed in meteorological analysis. In this context, signals tend to appear periodically, with varying patterns in the sizes of signal clusters. We apply PLIS ($\rm PLIS_{HM}$), BH and AdaDetect to the simulated data and summarize the results from 200 replications in Figure \ref{pic:period}.  \qed
}
\end{example}

\begin{figure}[!htbp]
    \centering
    \includegraphics[width=1\linewidth]{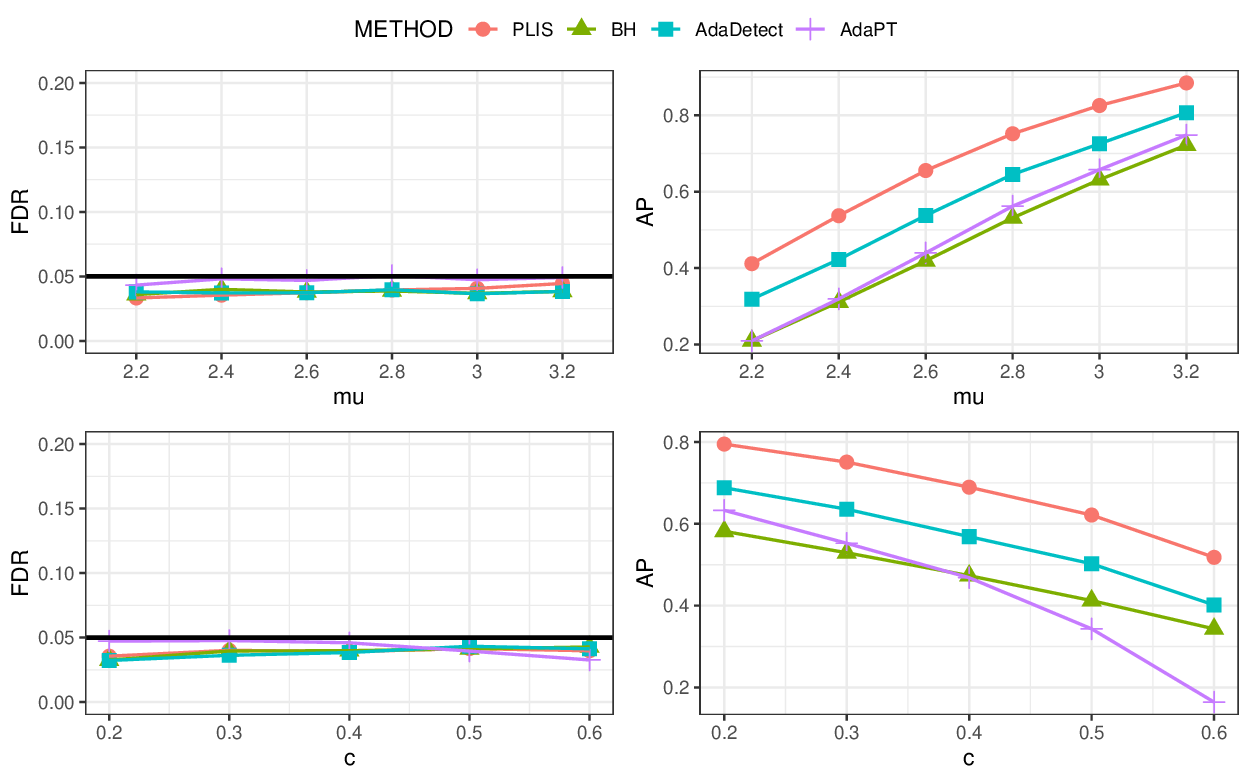}
    \caption{FDR and AP comparison under a two-layer dynamic model.}\label{pic:2layer}
\end{figure}

\begin{example}\rm{
\textbf{Two-layer dynamic models.} 
This example presents a complex scenario where it is not possible to characterize the hidden states $\pmb{\Theta}=(\theta_{i})_{i=1}^{m}$ in a closed form using a set of parameters, as is the case with HMMs. Instead, the state sequence $\Theta$ is determined by another dynamic sequence $\mathbf{Z}=(Z_{t})_{t=1}^{m}$. In turn, the state sequence determines the distribution of each observation in $\mathbf{X}$. This model, which describes the relationship between two different sequences $\mathbf{Z}$ and $\mathbf{X}$, is particularly useful for extracting common trends across different datasets.
We now describe the data generating process for this example. Let $(Z_{t})_{t=1}$ be an autoregressive moving average (ARMA) sequence satisfying 
$$
Z_{t}=c+Z_{t-1}-0.5Z_{t-2}+\varepsilon_{t}+0.1\varepsilon_{t-1},c>0, 
$$
where the innovation $\{\varepsilon_{t}\}$ are i.i.d. $\mathcal{N}(0,0.5^{2})$ variables, and the hidden states are generated according to $\theta_{i}=\mathbb{I}\{Z_{i}<0\}$. The observed data $X_i$ are generated according to the structured probabilistic model: $X_{i}|\theta_{i}\overset{ind.}{\sim}(1-\theta_{i})\mathcal{N}(0,1)+\theta_{i}\mathcal{N}(\mu,1)$. We apply PLIS, BH, and AdaDetect to the simulated data and summarize the results from 200 replications in Figure \ref{pic:2layer}. In the top row, we fix $c=0.3$ and vary $\mu$, whereas in the bottom row, we fix $\mu=2.8$ and vary $c$. Note that the average power of different methods would decrease with increasing $c$, as a larger value of $c$ causes $\mathbf{Z}$ to move in the positive direction, leading to fewer non-null hypotheses. \qed
}
\end{example}

\begin{figure}[!htbp]
    \centering
    \includegraphics[width=1\linewidth]{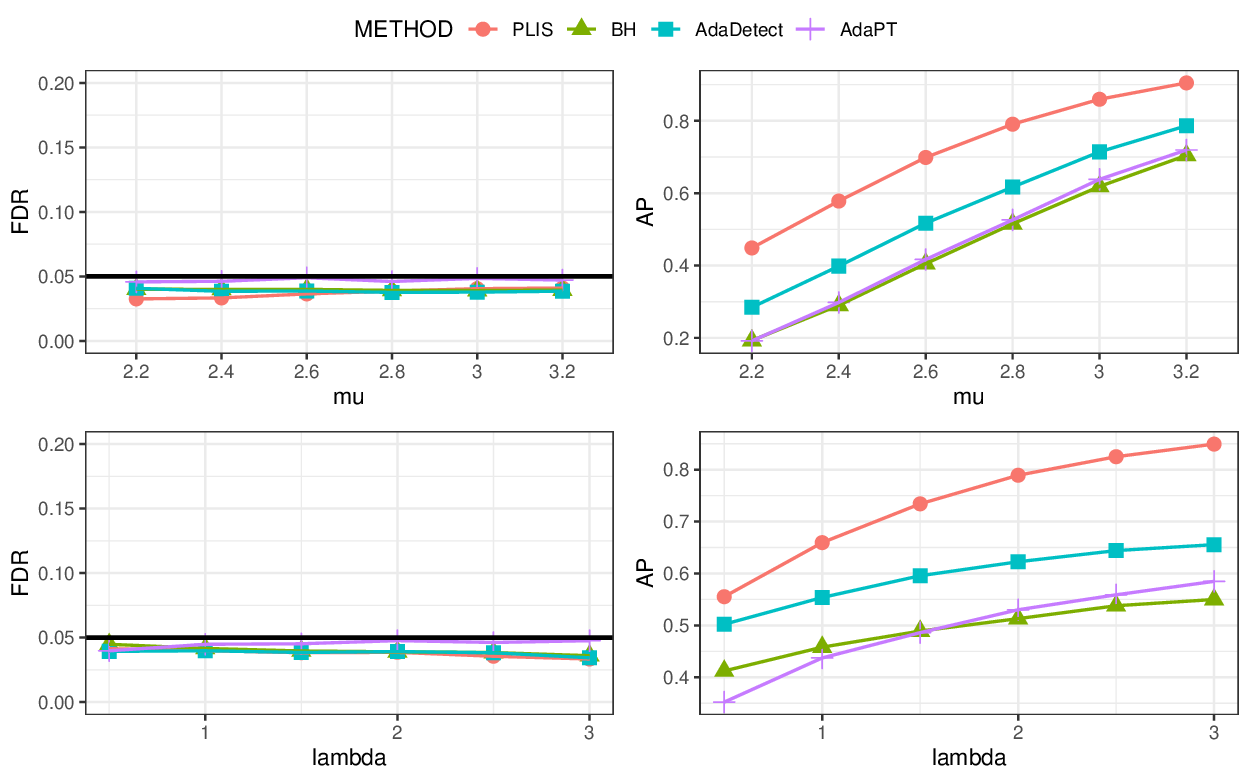}
    \caption{FDR and AP comparison in structured models generated from a renewal process.}\label{pic:renew}
\end{figure}

\begin{example}\rm{
\textbf{Structured models generated from a renewal process.}  
The data in this example are generated in three steps based on a renewal process. First, we generate $Z_{i}$ using the following scheme: 
\begin{eqnarray*}
\mbox{$Z_{i}\overset{i.i.d.}{\sim}Unif\{2,3,\cdots,20\}$ for $i=2k-1$} \\ \mbox{$Z_{i}\overset{i.i.d.}{\sim}Z$ with $Z-1\sim Poi(\lambda)$ for $i=2k$}
\end{eqnarray*}
where $k$ takes values of positive integers. Second, we generate $\theta_{j}$ as follows: 
$$
\mbox{$\theta_{j}=1$ for $j\in[1+\sum_{i=1}^{K}Z_{i}, \sum_{i=1}^{K+1}Z_{i}]$ if $K$ is odd, otherwise $\theta_{j}=0$.} 
$$
Finally, the observed data $X_i$ are generated according to the structured probabilistic model: $X_{i}|\theta_{i}\overset{ind.}{\sim}(1-\theta_{i})\mathcal{N}(0,1)+\theta_{i}\mathcal{N}(\mu,1)$. The length of the sequence is 2000. We apply PLIS, BH, and AdaDetect to the simulated data and summarize the results from 200 replications in Figure \ref{pic:renew}. In the top row, we fix $\lambda=2$ and vary $\mu$, whereas in the bottom row, we fix $\mu=2.8$ and vary $\lambda$. \qed 
}
\end{example}

It is worth noting that the three models discussed in this section pose significant challenges for analysis under the conventional ``model-based'' false discovery rate framework. In particular, this framework requires that the model be correctly specified, the model parameters be estimated consistently, and efficient algorithms be available to compute the conformity scores. However, due to the complexity of the models in above examples, meeting these requirements can be difficult, if not infeasible.

The results presented in Figures \ref{pic:hmmmu}-\ref{pic:renew} are promising. They demonstrate that although the models cannot be estimated perfectly, it is still possible to use a working model to capture the structural patterns in the data and then employ the robust PLIS framework for inference. From the simulation results, we can see that the PLIS procedure remains valid across all setups. In terms of the ability to detect non-null signals, PLIS outperforms competing methods that employ symmetric rules by a substantial margin.

\subsection{Numerical results for exchangeable data beyond Model (2)}\label{simu-puplis}
This section reports on simulation studies conducted to compare different methods under the semi-supervised setup, where only labeled null samples are available without knowledge about the null distribution. We consider the situation where the observations are exchangeable but not conditionally independent given $\pmb{\Theta}$. Specifically, the test data are generated using the following model:
$$
X_{i}=\theta_{i}\mu+\varepsilon_{1,i}+\varepsilon_{2,i}, \quad  i\in[m],
$$
where $\{\varepsilon_{1,i}\}$ are i.i.d. noise obeying $\mathcal{N}(0,1/2)$, and $\{\varepsilon_{2,i}\}$ are correlated noise with a marginal distribution of $\mathcal{N}(0,1/2)$, and a correlation of $corr(\varepsilon_{2,i}, \varepsilon_{2,i^\prime})=\rho$ for $i\neq i^\prime$.  
The labeled null samples $\mathbf{U}=\{U_{i}\}$ are generated as follows:
$$
U_{i}=\varepsilon_{1,m+i}+\varepsilon_{2,m+i},\quad i\in[2m].
$$ 
In our simulation, we set $m=2000$, and the hidden binary variables $\pmb{\Theta}=(\theta_{i})$ correspond to the hidden states in HMMs that represent the true states of the hypotheses.  

\begin{figure}[htbp]
    \centering
    \includegraphics[width=1\linewidth]{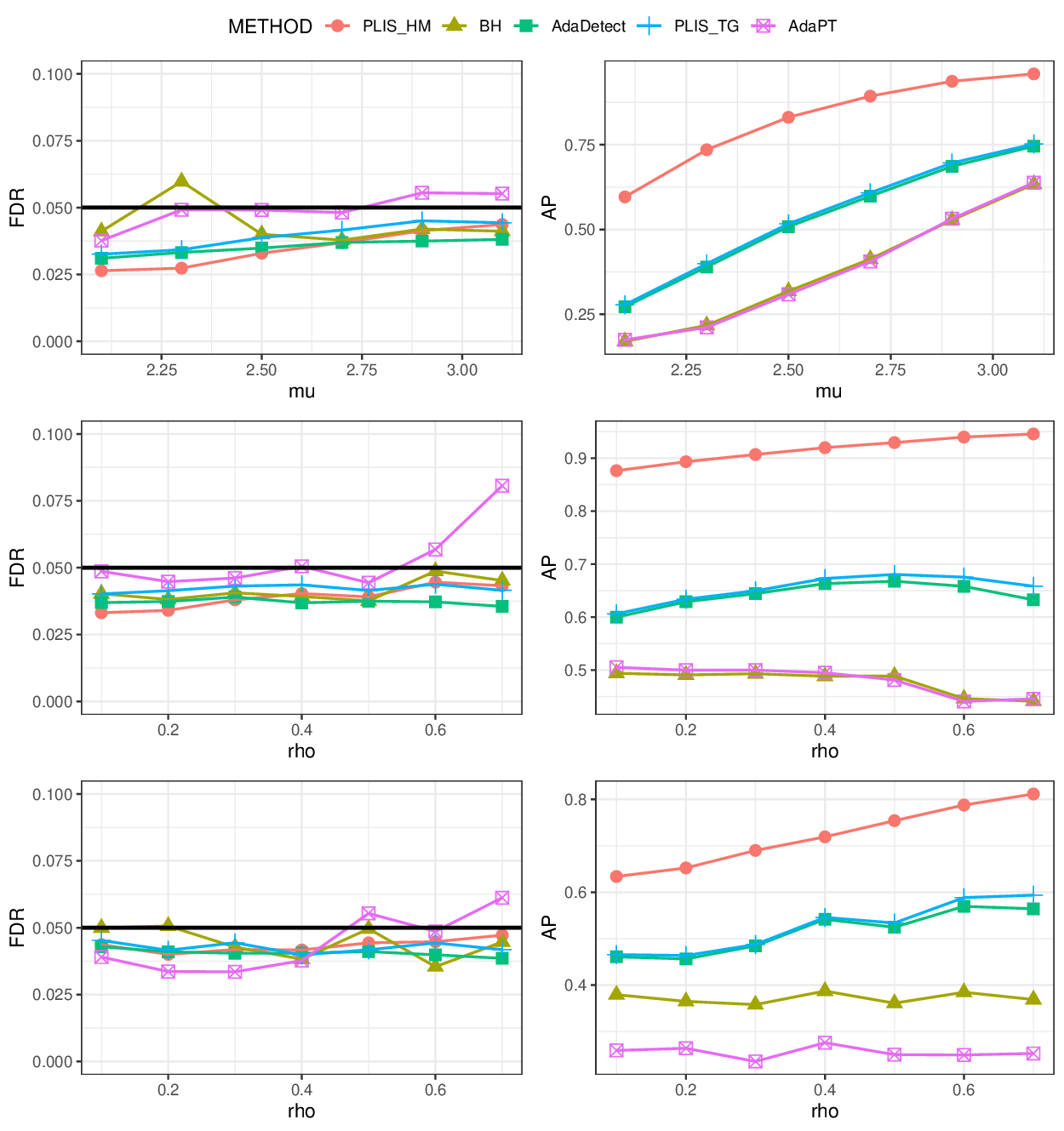}
    \caption{FDR and AP comparison when data points are exchangeable under the null but not independent conditional on $\pmb\Theta$.}\label{pic:exch}
    \end{figure}

Following the semi-supervised PLIS procedure (Algorithm 2), the first $m$ observations in $\mathbf{U}$ constitute the calibration set $\mathbf{Y}$, while the remaining data in $\mathbf{U}$ are used as training data. We compare the five methods at $\alpha=0.05$ under different settings, and obtain the FDR and AP levels by averaging the results from 200 replications. The simulation results are summarized in Figure \ref{pic:exch}, where we fix $a_{00}=0.95$, $a_{11}=0.8$, and $\rho=0.4$, and vary $\mu$ in the top row. In the middle and bottom rows, we set $a_{11}=0.8$ and $a_{11}=0.5$, respectively, and compare the different methods for varying values of $\rho$.

The results in Figure \ref{pic:exch} show that $\rm PLIS_{HM}$, $\rm PLIS_{TG}$, and AdaDetect effectively control the FDR at the nominal level in all settings. However, AdaPT shows inflated FDR levels. The proposed method $\rm PLIS_{HM}$ displays the highest average power among all the methods.

\subsection{Numerical results for non-exchangeable data}
\label{simu-puplis2}
In this section, we conduct numerical studies to evaluate the performance of the semi-supervised PLIS method (Algorithm 2) when the exchangeability assumption is violated. This situation is of particular interest as it is difficult to establish rigorous theoretical guarantees on FDR control. 
The test data are generated using the following model:
$$
X_{i}=\theta_{i}\mu+\varepsilon_{1,i}+\varepsilon_{2,i},\quad i\in[m].
$$
The labeled null samples $\mathbf{U}=\{U_{i}\}$ are generated as follows:
$$
U_{i}=\varepsilon_{1,m+i}+\varepsilon_{2,m+i},\quad i \in [2m],
$$
where $\{\varepsilon_{1,i}\}$ are i.i.d. noise obeying $\mathcal{N}(0,1/2)$, and $\{\varepsilon_{2,i}\}$ are correlated noise with a marginal distribution given by $\mathcal{N}(0,1/2)$. The correlation structure can be described as an auto-regressive model [AR(1)], i.e., the correlation coefficient is given by $\EE[\varepsilon_{2,i}\varepsilon_{2,j}]=\rho^{|i-j|}$ for some $\rho\neq0$. As before, we set $m=2000$, and the hidden binary variables $\pmb{\Theta}=(\theta_{i})$ correspond to the hidden states in HMMs that represent the true states of the hypotheses. 

\begin{figure}[htbp]
    \centering
    \includegraphics[width=1\linewidth]{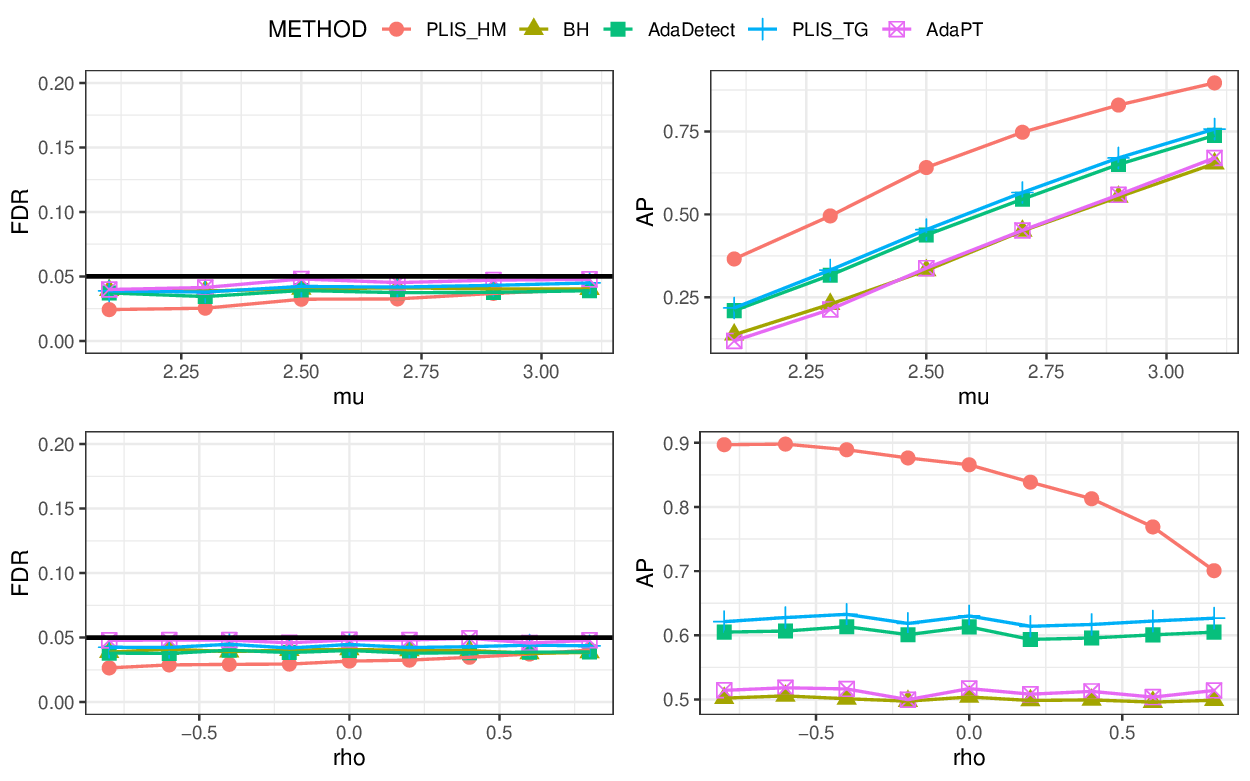}
    \caption{FDR and AP comparison for nonexchangeable data with AR(1) structure.}\label{pic:ar1}
    \end{figure}

We compare the previously mentioned  methods at $\alpha=0.05$ under different settings and obtain the FDR and AP levels by averaging the results from 200 replications. The simulation results are summarized in Figure \ref{pic:ar1}, where in the top row, we fix $a_{00}=0.95$, $a_{11}=0.8$, and $\rho=0.5$, and vary $\mu$. In the bottom row, we set $a_{11}=0.8$ and $\mu=2.8$ and compare the different methods for varying values of $\rho$.

The numerical results demonstrate that although it is difficult to establish a theoretical guarantee for PLIS when the null data are not exchangeable, the FDR can be effectively controlled under the nominal level. Compared to the results presented in Figure \ref{pic:exch}, the performance of BH, AdaDetect, $\rm PLIS_{TG}$, and AdaPT is robust against the violation of exchangeability assumption when there is an AR(1) dependence structure. This may be attributed to the weakly dependent nature of the AR(1) process, wherein the correlation coefficient decreases exponentially. The other possible explanation is that the AR(1) Gaussian process belongs to the $\alpha$-mixing processes; the kernel density estimators employed by AdaDetect and $\rm PLIS_{TG}$ are consistent under such mixing conditions. 

\subsection{Derandomized PLIS}\label{simu-dePLIS}

In this section, we investigate issues related to Derandomized PLIS, which involves averaging multiple e-values. A critical aspect is the effective selection of hyper-parameters, including the number of replicates $N$ and the FDR levels for every realized PLIS: $(\alpha_{j})_{j=1}^{N}$. This remains an open question that goes beyond the scope of this work. We provide some preliminary results for Derandomized PLIS with some pairs of $(N,\alpha_{PLIS})$, which are presented in Table \ref{tab:derandom}. We set $\alpha_{PLIS}$ to be proportional to the target FDR level $\alpha=0.05$ and choose $\alpha_{1}=\cdots=\alpha_{N}\equiv\alpha_{PLIS}$.

\begin{figure}[!htbp]
    \centering
    \includegraphics[width=1\linewidth]{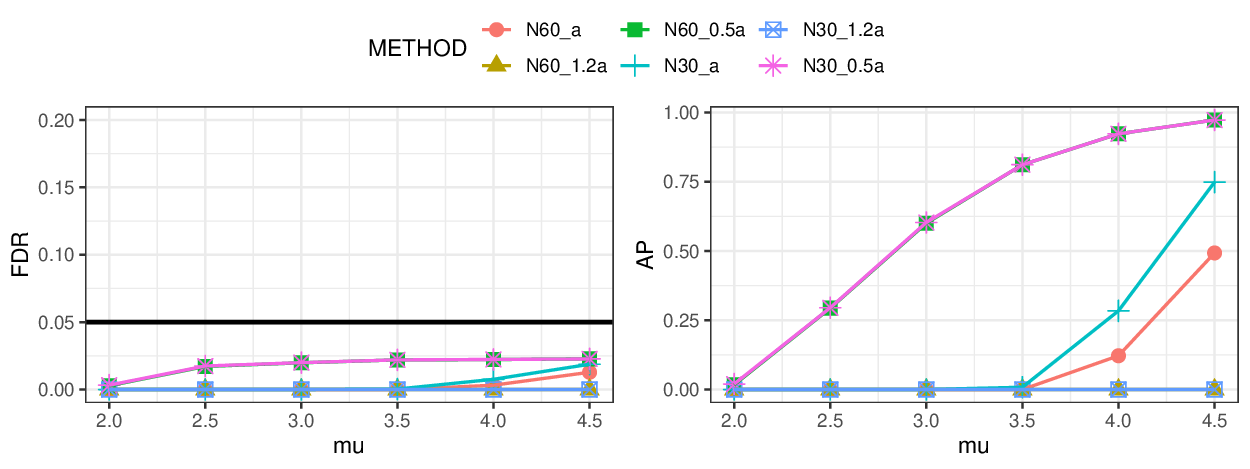}
    \caption{This figure compares the FDR and AP for the mixture model (3) with varying $\mu$, while controlling the FDR level at $\alpha=0.05$. The experiment fixes $\pi=0.2$, and all implemented methods are Derandomized PLIS with different hyper-parameters (Table \ref{tab:derandom}).}\label{pic:indepebh}
\end{figure}

In our numerical study, we generate the data using the two-group model (3), which is also utilized as the working model for PLIS:
\begin{equation*}
    X_{i}|\theta_{i}\overset{ind.}{\sim}(1-\theta_{i})\mathcal{N}(0,1)+\theta_{i}\mathcal{N}(\mu,1),
\end{equation*}
where $\theta_{1},\cdots,\theta_{m}$ are i.i.d. binary variables, $\PP(\theta_{1}=1)=\pi$. 
The density ratio (DR) statistic serves as an ideal score function. We compare the above methods on simulated data by fixing $\pi=0.2$. The average results from 200 replications are displayed in Figure \ref{pic:indepebh}.

Figure \ref{pic:indepebh} presents the comparison results for Derandomized PLIS with different hyper-parameters. It is observed that all methods are valid for FDR control, as supported by the theory of the e-BH procedure, which is employed as the final step to determine the rejections in Derandomized PLIS. Notably, the highest power is achieved when $\alpha_{PLIS}=0.5\alpha$, while no discovery is made when $\alpha_{PLIS}=1.2\alpha$. By contrast, the multiplication parameter $N$ has little effect on the performance of Derandomized PLIS procedure. 

\begin{table}[!htbp]
\centering
\caption{Methods corresponding to the pair of hyper-parameters in the experiment for Derandomized PLIS.}
\label{tab:derandom}
\begin{tabular}{|c|c|c|c|}
\hline
\textbf{METHOD} & $\alpha_{PLIS}=\alpha$ & $\alpha_{PLIS}=0.5\alpha$ & $\alpha_{PLIS}=1.2\alpha$ \\ \hline
$N=60$          & N60\_a                 & N60\_0.5a                 & N60\_1.2a                 \\ \hline
$N=30$          & N30\_a                 & N30\_0.5a                 & N30\_1.2a                 \\ \hline
\end{tabular}
\end{table}

As highlighted by \citet{ren22derandomized}, it is intuitively preferable to choose $\alpha_{PLIS}$ to be less than $\alpha$ when the sample size $N>1$. To illustrate this, consider a scenario where the data contains $m_{1}$ non-nulls, each exhibiting remarkably strong signals, resulting in the high likelihood of selecting all $m_{1}$ non-nulls in a single run of PLIS. In each run $j$, we would anticipate an approximate false discovery proportion of $\alpha_{PLIS}$. This implies that we would expect to observe around $\frac{\alpha_{PLIS}}{1-\alpha_{PLIS}}m_{1}$ false discoveries along with the $m_{1}$ true discoveries. Since each non-null $i$ is selected in nearly every run of PLIS, while it is possible for a null $k\in\mathcal{H}_{0}$ to be chosen only in a small proportion of the runs, we would expect the e-values to behave as $e_{j}\approx m/(\frac{\alpha_{PLIS}}{1-\alpha_{PLIS}}m_{1})$ for $j\notin\mathcal{H}_{0}$, and approximately $e_{j}\approx 0$ for $j\in\mathcal{H}_{0}$.

Applying the e-BH procedure at level $\alpha$ to these e-values, we observe that the power of the procedure can be high only if $m/(\frac{\alpha_{PLIS}}{1-\alpha_{PLIS}}m_{1})\geq m/(\alpha m_{1})$. Otherwise, if this condition is not met, the power will be zero, as no discoveries can be made. Consequently, selecting $\alpha_{PLIS}\leq \frac{\alpha}{1+\alpha}$ may result in a powerful procedure. However, it is important to note that our examination of the relationship between the $\alpha_{k}$s and $\alpha$ is still at a preliminary stage and has not reached a conclusive point. We believe that further investigation into this topic is warranted.

\begin{figure}[!htbp]
    \centering
    \includegraphics[width=1\linewidth]{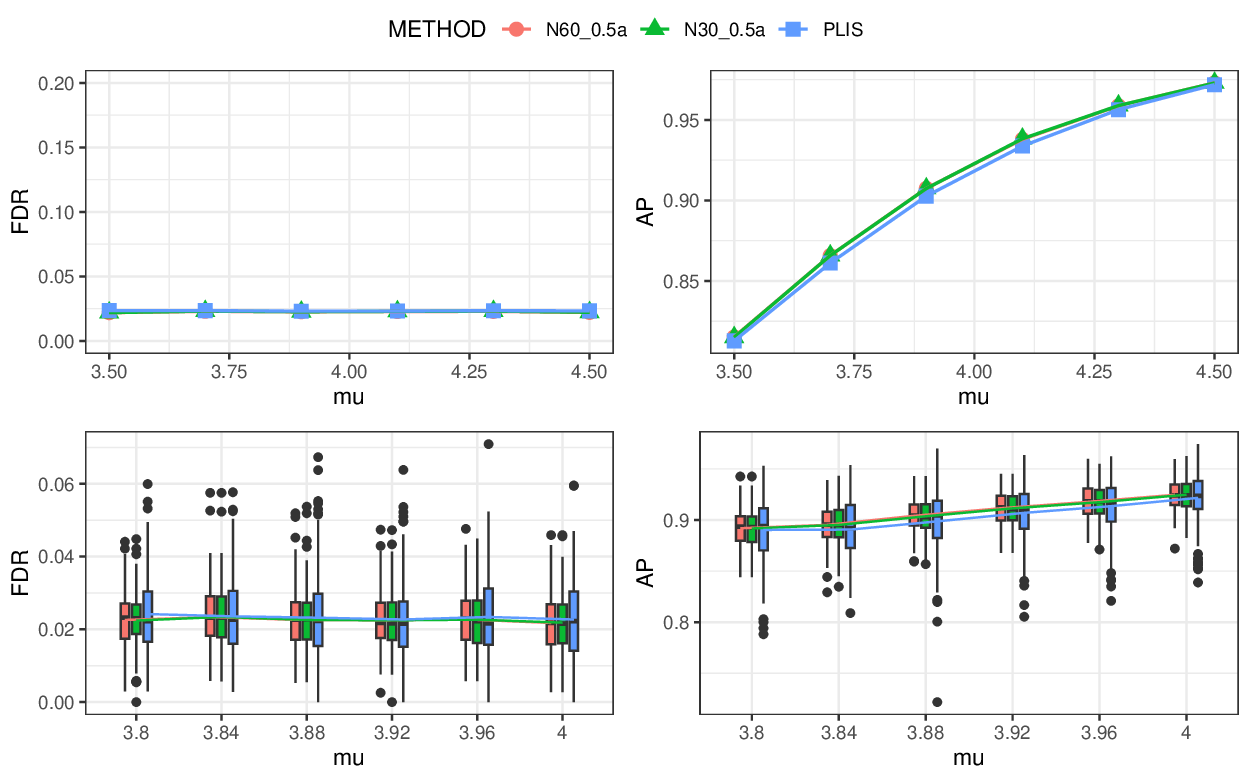}
    \caption{Comparison for Derandomized PLIS with $\alpha_{k}=0.5$ and PLIS. The top row is the line chart representing the trends of FDR and AP for varying $\mu$, while the bottom row is the boxplots for comparison. }\label{pic:derandom}
\end{figure}

To demonstrate the effectiveness of derandomization, we compare the FDR and AP levels between PLIS and Derandomized PLIS, utilizing a mixture model (3) with a mixing proportion of $\pi = 0.2$. The nominal FDR level is set at $\alpha = 0.05$. A summary of the simulation results is presented in Figure \ref{pic:derandom}. It becomes apparent that employing Derandomized PLIS with $\alpha_{k}=0.5\alpha$ yields improvement in power. Furthermore, the boxplots displayed in the bottom row of the figure visually reinforce the notion that derandomization effectively reduces both the variance and number of outliers. As a result, derandomization provides a useful tool for enhancing the replicability and reliability of the discoveries made.

\subsection{Comparison for different baseline data constructions in PLIS}\label{ap:simu-baseline}

In this section, we provide numerical results to compare the efficacy of different baseline data construction methods. The pairwise exchangeability properties (Properties 1-3) hold as long as the bivariate function is symmetric, i.e., $h(x,y)=h(y,x)$. Among the baseline data construction methods, we recommended using the function
    \begin{equation*}
       h(x,y)=\left\{\begin{matrix}
       x& \text{if $|x|\geq |y|$}\\
        y& \text{otherwise} 
         \end{matrix}\right.. 
    \end{equation*}
This choice of function provides two significant benefits. First, it guarantees pairwise exchangeability, which our theoretical analysis requires. Second, it maintains large non-null effects and hence essential structural information. However, alternative functions such as $h^\prime(x,y)=x+y$ also guarantee pairwise exchangeability and can lead to a valid PLIS procedure. 

We demonstrate that using alternative functions, such as $h(x,y)=x+y$, can lead to a dilution of non-null effects in the test data, thereby decreasing the power of the PLIS procedure. Denote the PLIS procedure with $h^\prime(x,y)=x+y$ as PLIS.2. We generate test data using the process 
$$
X_{i}|\theta_{i}\overset{ind.}{\sim} (1-\theta_{i})\mathcal{N}(0,1)+\theta_{i}\mathcal{N}(0,1), i\in[m],
$$
with $m=2000$. The calibration data $Y_{1},\cdots,Y_{m}$ are i.i.d. $\mathcal{N}$(0,1). 
\begin{figure}[htbp]
    \centering
    \includegraphics[width=1\linewidth]{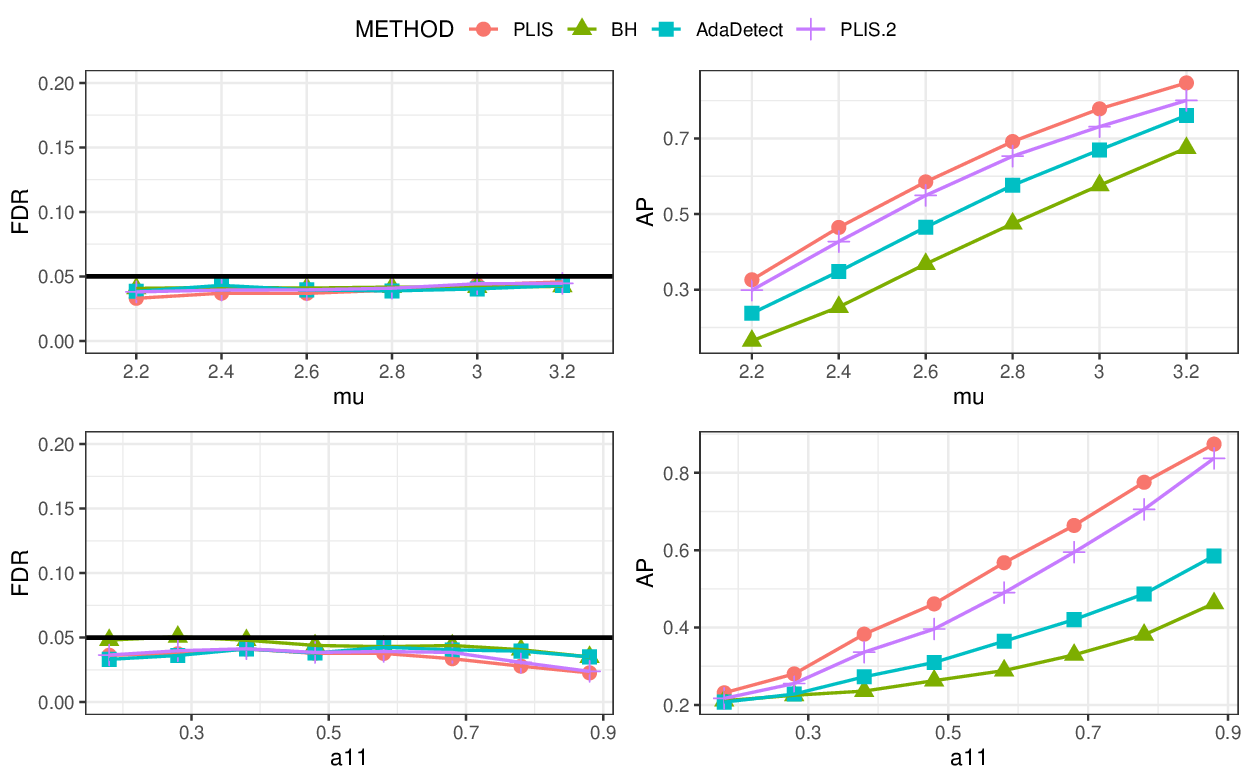}
    \caption{FDR and AP comparison for PLIS with different baseline data constructions. }\label{pic:baseline}
    \end{figure}

We consider two simulation setups. In the first setup, the underlying states $\pmb\Theta$ are generated from a two-layer model. Specifically, let $(Z_{t})_{t=1}$ be an autoregressive moving average (ARMA) sequence satisfying $Z_{t}=0.4+Z_{t-1}-0.5Z_{t-2}+\varepsilon_{t}+0.1\varepsilon_{t-1}$, where the innovations $\{\varepsilon_{t}\}$ are i.i.d. $\mathcal{N}(0,0.5^{2})$ variables, and the hidden states are generated according to $\theta_{i}=\mathbb{I}\{Z_{i}<0\}$. In the second setup, the hidden states $\pmb\Theta=(\theta_{i})$ are generated as a Markov chain with $a_{00}=0.95$, $\mu=2.6$, and varying $a_{11}$. 

We apply various methods to the simulated data at FDR level $\alpha=0.05$. The simulation results for the two setups are summarized in the first and second rows of Figure \ref{pic:baseline}, respectively. We can see that PLIS.2 control the FDR at the nominal level and outperforms BH and AdaDetect in power. However, PLIS.2 is not as efficient as the proposed PLIS in terms of power. This corroborates the merits of the proposed bivariate function employed for construction of the baseline data in our paper.

\subsection{Comparisons with the variations of PLIS}\label{simu:sym-PLIS}

This section presents a comparison of the numerical performance between PLIS and its two variations, ${\rm PLIS}_{\rm cbh}$ and ${\rm PLIS}_{\rm sym}$. The primary objective is to illustrate the following  points: 
(a) ${\rm PLIS}_{\rm cbh}$ exhibits excessive conservativeness under strong dependence;
(b) the rankings utilized by ${\rm PLIS}_{\rm sym}$ are suboptimal due to the information loss during the symmetrization step; and 
(c) the proposed PLIS procedure effectively overcomes these limitations.

\begin{figure}[!htbp]
    \centering
    \includegraphics[width=1\linewidth]{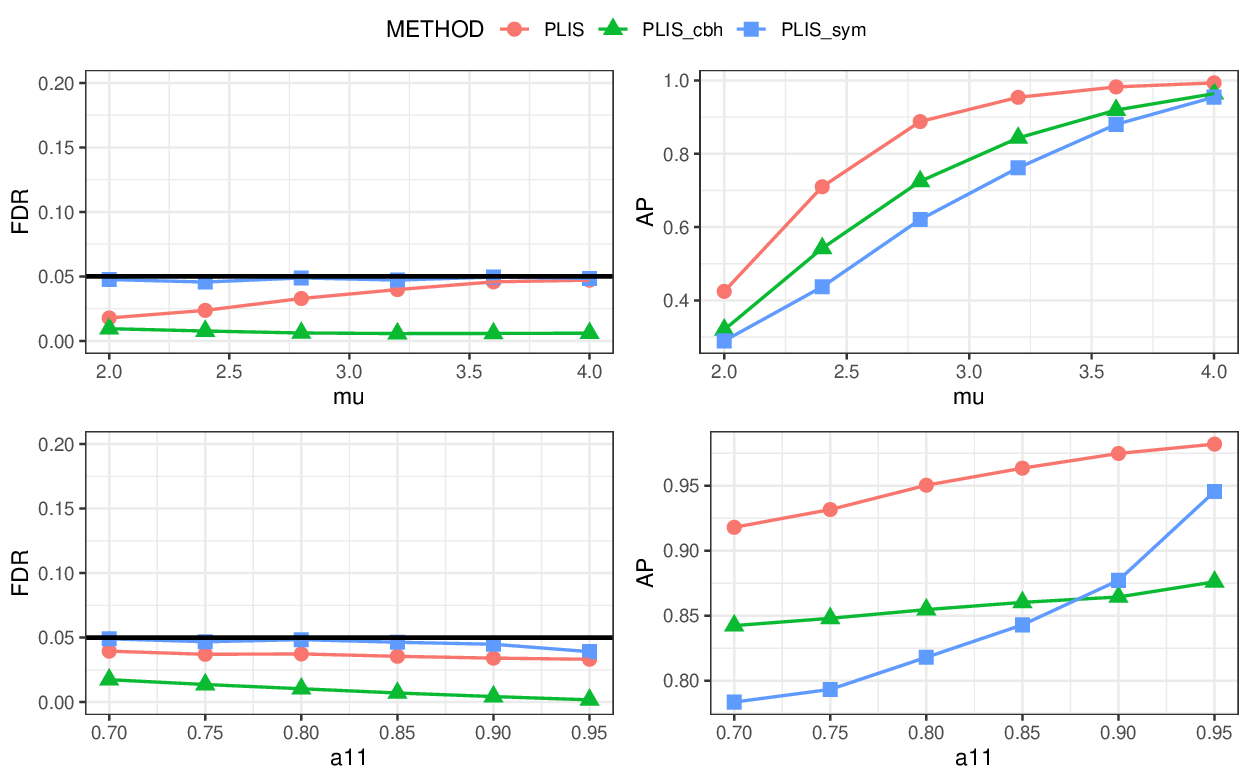}
    \caption{FDR and AP comparison for PLIS, ${\rm PLIS}_{\rm cbh}$ and ${\rm PLIS}_{\rm sym}$.}\label{pic:knockoff}
\end{figure}

The data are generated from the HMM discussed in Section 4.1. For evaluating the performance of three methods, namely PLIS, ${\rm PLIS}_{\rm cbh}$, and ${\rm PLIS}_{\rm sym}$ in Section 3.3, we employ the same scores that are calculated based on the HMM working model $\mathcal {M}_{\rm HM}$ and the forward-backward algorithm $\mathcal{A}_{\rm FB}$, as described in the first example of Section 2.5. We apply the methods at FDR level $\alpha=0.05$. The simulation results are summarized in Figure \ref{pic:knockoff}, where the top row fixes the values of $a_{00}=0.9$ and $a_{11}=0.8$, while varying $\mu$. The bottom row, on the other hand, fixes $\mu=3$ and $a_{00}=0.6$, while varying $a_{11}$.

In the top row, we can see that the FDR of ${\rm PLIS}_{\rm sym}$ consistently remains close to the nominal level, while ${\rm PLIS}_{\rm cbh}$ exhibits an overly conservative behavior. Our proposed method, PLIS, demonstrates a certain degree of conservativeness when $\mu$ is small but adaptively achieves the desired nominal FDR level as $\mu$ increases. Despite the conservativeness observed in ${\rm PLIS}_{\rm cbh}$, it outperforms ${\rm PLIS}_{\rm sym}$ in terms of average power, indicating that the rankings obtained from $s_j^X$ are superior to those derived from the contrast statistics $T_j=s_j^Y-s_j^X$. 
In the bottom row, however, as the dependence becomes stronger, the conservativeness of ${\rm PLIS}_{\rm cbh}$ becomes increasingly prominent, eventually leading to a decrease in power compared to ${\rm PLIS}_{\rm sym}$. 

Remarkably, our proposed method, PLIS, consistently outperforms both variants, ${\rm PLIS}_{\rm cbh}$ and ${\rm PLIS}_{\rm sym}$, across all scenarios. The simulation results highlight two key advantages of PLIS. Firstly, PLIS effectively addresses the sub-optimal ranking issue observed in ${\rm PLIS}_{\rm sym}$ by maintaining the efficient ranking established from $s_j^X$, a pseudo score that aims to mirror the ranking of $\PP(\theta_j=0|\mathbf{X})$. Secondly, PLIS mitigates the conservativeness of ${\rm PLIS}_{\rm cbh}$ by refining the candidate rejection set from $\mathcal G$ to $\mathcal G^r$. It adaptively attains the nominal FDR level as the signal effect increases.

\end{document}